\documentclass[twocolumn,showpacs,preprintnumbers,amsmath,amssymb,prb]{revtex4}

\usepackage{graphicx}
\usepackage{dcolumn}
\usepackage{bm}

\newcommand{\bol}[1]{\boldsymbol #1}
\def\rnum#1{\expandafter{%
\romannumeral #1}}
\def\Rnum#1{\uppercase\expandafter{%
\romannumeral #1}}

\begin{document}

\title{
Low-energy properties 
of two-leg spin-1 antiferromagnetic ladders 
with commensurate external fields and their extensions}

\author{Masahiro Sato}

\affiliation{Department of Physics, Tokyo Institute of Technology, 
Oh-okayama, Meguro-ku, Tokyo 152-8550, Japan}


\begin{abstract}
This study addresses low-energy properties of 
2-leg spin-1 ladders with antiferromagnetic (AF) intrachain
coupling under a uniform or staggered external field $H$, and a few 
of their modifications. The generalization to spin-$S$ ladders 
is also discussed.
In the strong AF rung (interchain)-coupling $J_\perp$ region, degenerate
perturbation theory applied to spin-$S$ ladders predicts $2S$
critical curves in the parameter space $(J_\perp, H)$
for the staggered-field case, in contrast to $2S$ finite critical 
regions for the uniform-field case. All critical areas belong to a 
universality with central charge $c=1$. 
On the other hand, we employ Abelian and non-Abelian bosonization 
techniques in the weak rung-coupling region. They show that in the 
spin-1 ladder, a sufficiently strong uniform field engenders a $c=1$ 
critical state regardless of the sign of $J_\perp$, whereas the 
staggered field is expected not to yield any singular phenomena.
From the bosonization techniques, new field-theoretical expressions of 
string order parameters in the spin-1 systems are also proposed.
\end{abstract}


\maketitle

\section{\label{sec1} Introduction}
Spin ladder systems have been investigated theoretically for more than a
decade. Several real magnets corresponding to them have been synthesized 
and observed.~\cite{ladder1,SrCuO,Cu2--(NO3)2,Cu2--Cl4,DT-TTF,BIP-TENO} 
A trigger of such studies may be the discovery of 
high-$T_{\rm c}$ materials and the connection between the 
Hubbard model (one of the high-$T_{\rm c}$ models) and 
the antiferromagnetic (AF) Heisenberg model.~\cite{Rev-Heis} 
Recently, it has been recognized that spin ladders themselves can 
provide several theoretically interesting phenomena: for example, 
quantum critical phenomena, nontrivial magnetization processes 
(plateaux and cusps), 
the connection with field theories and integrable models, 
topological or exotic orders, etc. 
Especially, the intensive studies have largely developed 
the physics of $2$-leg spin-$\frac{1}{2}$ ladders. 

In the spin ladder systems, like other magnetic systems, their responses
to external magnetic fields have received theoretical and experimental 
attentions. Recently, in addition to the standard uniform magnetic
field, staggered fields, which have an alternating component along a
direction of the system, have been in the 
spotlight.~\cite{O-A,A-O,Nomu,Z-R-M,Mas-Zhe,Erco,Wang,M-O}  
Actually, several phenomena induced by them have been observed,
and some mechanisms generating them in real magnets are known.~\cite{A-O,Wang}   
As will be discussed in the next section, uniform- or staggered-field
effects in the $2$-leg spin-$\frac{1}{2}$ ladder with AF intrachain
coupling have been understood well.
In the uniform-field magnetization process, a massless phase exists
between the saturated state and a massive spin-liquid state. A staggered
field yields a quantum phase transition.  

Here, the following natural question would arise: 
how the external fields influence high-spin or $N$-leg ladder systems?
To answer this (partially), this study specifically addresses 
low-energy properties of $2$-leg spin-$1$ ladders with external fields 
through the use of several analytical tools. 
Our main target is the simple Hamiltonian,
\begin{eqnarray}
\label{eq1}
\hat {\cal H} &=& J\sum_{l,j}\vec S_{l,j}\cdot\vec S_{l,j+1}
+J_{\perp}\sum_{j}\vec S_{1,j}\cdot\vec S_{2,j}+\hat{\cal H}_{\rm Z},
\end{eqnarray}
where $\vec S_{l,j}$ is the spin-$1$ operator on the site
$(l,j)$; $l$ ($=1$ or $2$) is the chain-number index and the integer $j$
runs along each chain. The intrachain coupling  
$J$ is positive. We will refer to such ladders as
``AF'' ladders. The interchain coupling $J_{\perp}$ 
is called the rung coupling. The Zeeman term 
$\hat{\cal H}_{\rm Z}$ here is chosen as two types 
\begin{subequations}
\label{eq2}
\begin{eqnarray}
\hat{\cal H}_{\rm u}&=&  -H\sum_{j}(S_{1,j}^z+S_{2,j}^z), \label{eq2-1}\\
\hat{\cal H}_{\rm s}&=&  -H\sum_{j}(-1)^j(S_{1,j}^z+S_{2,j}^z),\label{eq2-2}
\end{eqnarray}
\end{subequations}
where $H(\geq 0)$ is the strength of external fields.
The latter (\ref{eq2-2}) has a staggered field along the chain $(-1)^jH$.
Obviously, the AF rung coupling and the staggered field compete with
each other.

\begin{figure}
\scalebox{0.5}{\includegraphics{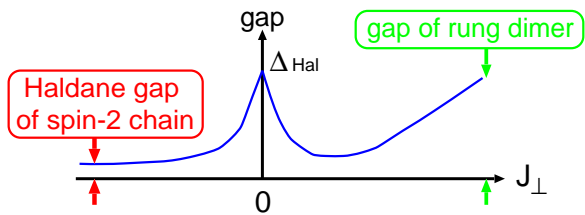}}
\caption{Schematic gap profile in the spin-$1$ AF ladder. The symbol
 $\Delta_{\rm Hal}$ denotes the Haldane gap ($\simeq 0.41J$). For
 details, see Ref.~\onlinecite{Todo}.}
\label{fig11}
\end{figure}

\begin{figure}
\scalebox{0.6}{\includegraphics{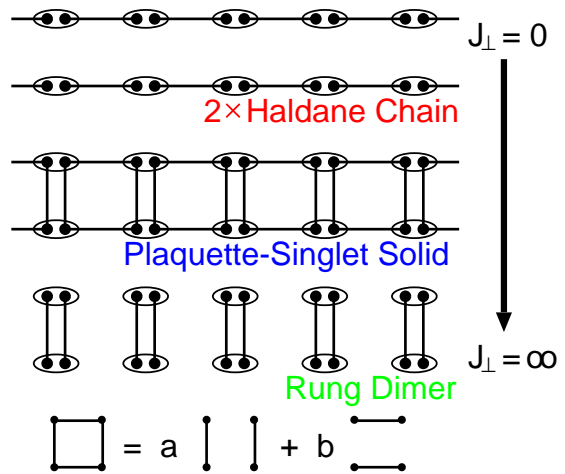}}
\caption{RVB pictures in the GS of the spin-$1$ AF ladder (\ref{eq1})
without external fields. The black point denotes a spin-$\frac{1}{2}$ state. 
The ellipse including two black points indicates the symmetrization of 
two spin-$\frac{1}{2}$ states and recovers an original spin-$1$ site. 
The black line represents for the singlet bond, which consists of 
a spin-$\frac{1}{2}$ pair. 
For $0< J_{\perp}< \infty$, the GS is well described by 
the PSS state. With increasing the rung-coupling $J_{\perp}$, 
the weight $a$ ($b$) increases (decreases) in the GS wave function, 
and it approaches the tensor product of singlet dimers 
[singlet of the spin-$1$ pair]. For details, see Ref.~\onlinecite{Todo}.}
\label{fig4}
\end{figure}

Regarding the case without external fields, some preceding 
theoretical studies~\cite{Sene,Allen,Todo} 
of the spin-$1$ ladder (\ref{eq1}) exist (although, as will see in
Sec.~\ref{sec2}, there are also a few studies dealing with 
$\hat{\cal H}_{\rm u}$). Their results deserve to be 
summarized here for our consideration in later sections. 
These studies explain how the first-excitation gap varies in dependence
upon the rung coupling $J_\perp$. At the decoupled point $J_\perp=0$, 
the model (\ref{eq1}) is reduced to two spin-$1$ AF chains. As known
well, the chain has a finite first-excitation gap (Haldane gap).~\cite{NLSM}
Around the decoupled point, 
the gap reduction takes place with $|J_\perp|$ increasing. 
Namely, the gap has a cusp structure around $J_\perp=0$. 
Far from the decoupled point, in the AF-rung side, the gap increases
together with the growth of $J_\perp$. It approaches the gap of the 
rung dimer (two spins along the rung) which is the strong AF
rung-coupling limit of the model (\ref{eq1}). 
On the other hand, for the ferromagnetic (FM)-rung side, even away from
the decoupled point, the gap decreases monotonically. It reaches the
Haldane gap of the spin-$2$ AF chain with the bond $J'=J/2$, which is the
strong FM rung-coupling limit. (In general, 
the $2$-leg spin-$S$ AF ladder with the intrachain coupling $J$ is
reduced to the spin-$2S$ AF chain with $J'=J/2$ in the strong FM
rung-coupling limit.)~\cite{M-comm,Matsu}
The first-excitation-gap profile is summarized as Fig.~\ref{fig11}.
A recent quantum Monte Carlo analysis~\cite{Todo} quantitatively surveys
the ground-state (GS) properties of the model (\ref{eq1}) 
without Zeeman terms. It estimates 
the first-excitation gap and the spin-spin correlation length. Moreover,  
it proposes a new string-type parameter, which is discussed in a later
section. The authors conclude that the GS 
is always massive from $J_{\perp}=0$ to $\infty$, 
and is characterized by the ``plaquette-singlet'' solid
(PSS) state, one of short-range resonating-valence-bond 
(RVB) states,~\cite{note1} which is explained in Fig.~\ref{fig4}.
It smoothly connects two limiting states: 
the Haldane state ($J_{\perp}=0$) and 
the rung-dimer state ($J_{\perp}\rightarrow \infty$).
The new string parameter is able to capture a feature of the PSS state. 

In addition to the external-field effects,
we will reexamine and revisit some of the results without external fields.
 
Our analysis specifically addresses two regions: the strong AF rung-coupling region
($J_\perp\gg J$) and the weak rung-coupling one ($J\gg |J_\perp|$). 
For the former (latter) region, we mainly employ degenerate
perturbation theory (DPT) (field-theoretical methods). 
In the former region, we can also investigate $2$-leg spin-$S$
ladders. It contributes to a systematic understanding of 
spin ladder systems. Furthermore, the comparison of our results and
existing results of spin-$\frac{1}{2}$ ladders will help to elucidate the
spin-ladder systems.

The organization of the remainder of this paper is as follows. 
First, we give a brief review of $2$-leg spin-$\frac{1}{2}$ ladders in
Sec.~\ref{rev-lad}. As mentioned previously, it is useful in 
comparing our spin-$1$ case with the spin-$\frac{1}{2}$ case 
and highlighting the features of the spin-$1$ case.
Section~\ref{sec2} specifically addresses the strong AF rung-coupling region 
($J_{\perp}\gg J$). For the spin-$1$ ladder (\ref{eq1}), the DPT provides 
effective Hamiltonians and predicts that there are two critical
areas (points) in the magnetization process applying the uniform
(staggered) field. We also apply the DPT to the $2$-leg spin-$S$ ladders, 
to find that in the spin-$S$ case there are 
$2S$ critical regions (points) through the uniform(staggered)-field 
magnetization process. 
In addition, we propose RVB-type pictures of noncritical phases 
in the uniform-field case with $\hat {\cal H}_{\rm u}$.
In Sec.~\ref{sec3}, we consider the weak rung-coupling region 
($J\gg |J_{\perp}|$) and employ some field-theoretical approaches. In
particular, to map 1D spin-$1$ systems onto a field theory, 
we utilize a non-abelian bosonization (NAB),~\cite{Witt,Aff,A-H} 
i.e., a Wess-Zumino-Novikov-Witten (WZNW) model description.
The first two subsections are devoted to an explanation of the NAB. 
Through it, the spin-$1$ ladder (\ref{eq1}) is
described by using a fermion field theory, 
which was originally proposed by Tsvelik.~\cite{Tsv}  
Subsequently, we will consider the uniform-field case in Sec.~\ref{sec3-3}. 
In this case, the NAB is considerably effective because it
can treat the uniform Zeeman term nonperturbatively.~\cite{Tsv} 
From consequences of the DPT in Sec.~\ref{sec2}, 
the NAB here, and the gap profile in Fig.~\ref{fig11}, 
we can determine the whole GS phase diagram in Sec.~\ref{sec3-3-A}. 
The critical regime is ``simply connected'' 
and has a $c=1$ criticality except for the decoupled point $J_{\perp}=0$.  
In Sec.~\ref{sec3-4}, we devote our attention to spin-$1$ ladders 
without external fields. 
Section~\ref{sec3-4-1} is assigned to the evaluation of string-type
parameters in one-dimensional (1D) spin-$1$ systems within our 
field-theoretical framework. We propose their field-theoretical expressions. 
Using a RG analysis, we consider the GS phase diagram of a spin-1 ladder 
extended from the model (\ref{eq1}) in Sec.~\ref{sec3-4-2}.   
Field-theoretical approaches used here are not sufficiently 
efficient for the staggered-field case with $\hat {\cal H}_{\rm s}$. 
A brief discussion on it is in Sec.~\ref{sec3-5}. 
In Sec.~\ref{sec4}, we summarize all results and briefly discuss them. 
The Appendix offers readers supplements of field-theoretical 
techniques and calculations in Sec.~\ref{sec3}.

\section{\label{rev-lad}Review of Spin-1/2 Ladders}
In this section, we review low-energy properties of the
spin-$\frac{1}{2}$ AF ladder which is equivalent to the model
(\ref{eq1}), in which spin-$1$ operators are replaced with 
spin-$\frac{1}{2}$ operators.

\begin{figure}[h]
\scalebox{0.6}{\includegraphics{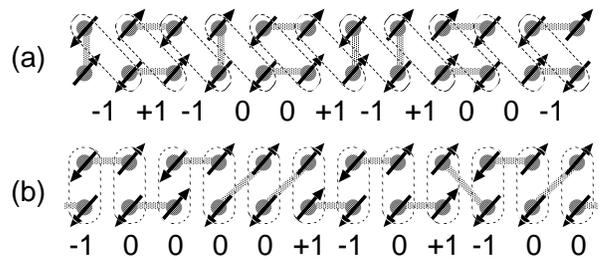}}
\caption{RVB pictures of the GS of the spin-$\frac{1}{2}$ ladder. 
Panels (a) and (b) are typical spin configurations in the AF-rung and 
FM-rung spin-liquid phases, respectively. 
The gray line denotes a singlet bond. Up and down arrows 
represent $S_{l,j}^z=+\frac{1}{2}$ and $-\frac{1}{2}$, respectively. 
Each number under the dashed loop encircling two sites 
shows the value of $\tilde{S}_j^z=S_{1,j}^z + S_{2,j+1}^z$ in (a), 
and one of $S_{j}^z=S_{1,j}^z + S_{2,j}^z$ in (b). 
Removing all sites of $\tilde{S}_j^z=0$ or $S_{j}^z=0$, 
one can see a ``hidden'' N\'eel order ($+1,-1,+1,-1,\cdots$).
For more details, see Refs.~\onlinecite{Nishi} and \onlinecite{Kim}.}
\label{fig3}
\end{figure}

For the case without external fields, a finite rung coupling engenders a
gapped spin-liquid (no long-range orders occur) 
GS irrespective of the sign of $J_\perp$. This is true because 
the GS of the spin-$\frac{1}{2}$ AF Heisenberg chain is critical
(massless) and the rung coupling is relevant from the standpoint 
of the perturbative renormalization group (RG) picture.~\cite{Sel}
The spin liquid can be illustrated using the short-range RVB
picture~\cite{White1,White2,Nishi,S-D,Kim} 
as in Fig.~\ref{fig3}. The figure indicates that the excitation gap in
the spin-liquid phase has the same order as the energy 
required to cut a singlet bond. In the AF-rung side, singlet bonds tend
to occur along both chain and rung directions. In contrast,
two spins on the rung tend to construct a triplet state on the
FM-rung side. The tendency removes singlet bonds along rungs, and 
allows those along the diagonal direction. The FM-rung spin liquid 
must connect with the GS of the spin-$1$ AF chain
($J_\perp\to -\infty$) smoothly. Therefore, we call it the Haldane phase.
In Refs.~\onlinecite{Nishi} and \onlinecite{Kim}, the authors show that 
these two kinds of massive spin liquids [(a) and (b) in Fig.~\ref{fig3}] 
can be detected by the two string-type parameters,
\begin{subequations}
\label{eq3}
\begin{eqnarray}
{\cal O}_{\rm odd}^a= -\lim_{|j-k|\rightarrow\infty}       
\left\langle S_j^a\,e^{i\pi\sum_{n=j+1}^{k-1}S_n^a} 
\,S_k^a\right\rangle,\label{eq3-1}\\
{\cal O}_{\rm even}^a = -\lim_{|j-k|\rightarrow\infty}   
\left\langle \tilde{S}_j^a\,
e^{i\pi\sum_{n=j+1}^{k-1}\tilde{S}_n^a} 
\,\tilde{S}_k^a \right\rangle ,\label{eq3-2}
\end{eqnarray}
\end{subequations}
where $a=x,y,z$, $\langle \cdots\rangle$ represents the
expectation value of the GS. We defined two new operators 
$S_j^a= S_{1,j}^a + S_{2,j}^a$ and $\tilde{S}_j^a= S_{1,j}^a + S_{2,j+1}^a$. 
Indeed, from Fig.~\ref{fig3}, one can confirm that
${\cal O}_{\rm even}^a\neq 0$ and ${\cal O}_{\rm odd}^a=0$ 
(${\cal O}_{\rm odd}^a\neq 0$ and ${\cal O}_{\rm even}^a= 0$) are 
realized in the AF-rung spin-liquid (Haldane) phase. 

\begin{figure}[h]
\scalebox{0.6}{\includegraphics{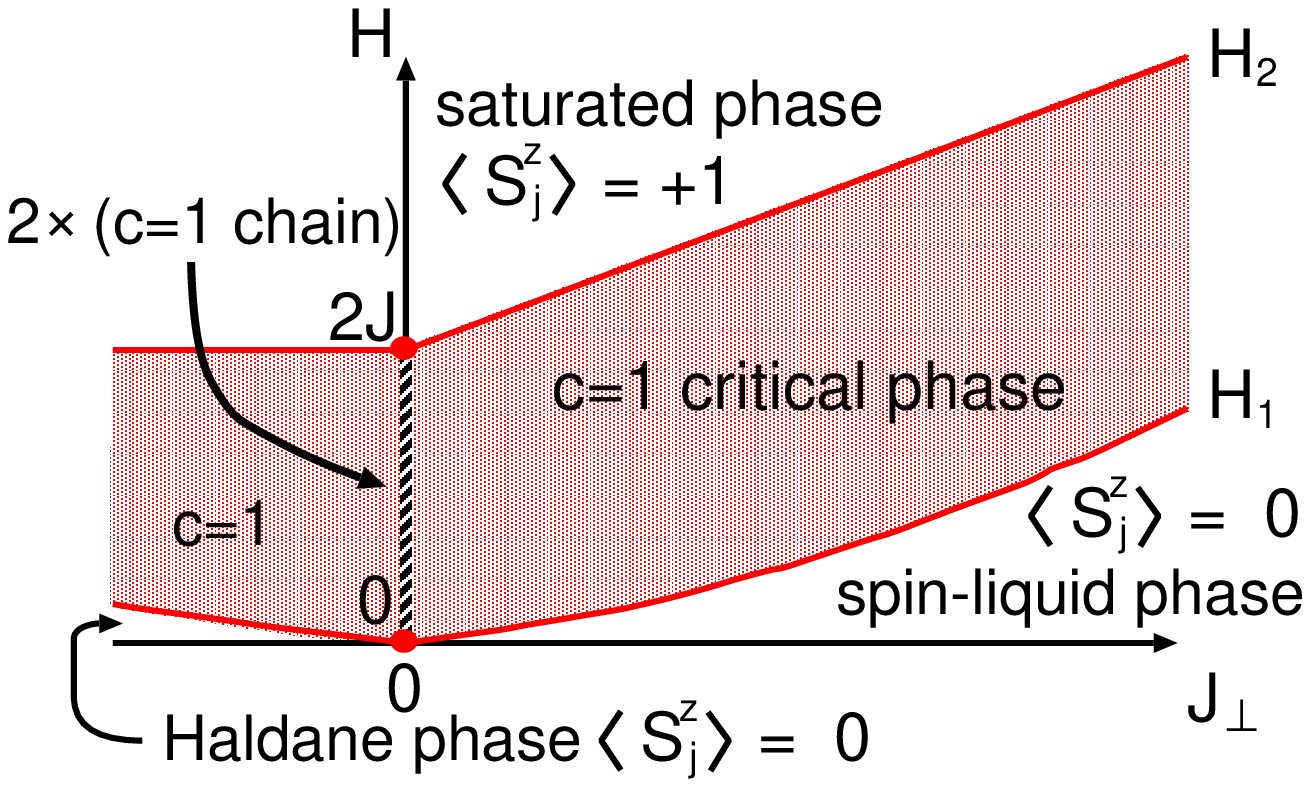}}
\caption{Schematic GS phase diagram of the spin-$\frac{1}{2}$ ladder 
with the uniform Zeeman term (\ref{eq2-1}).}
\label{fig1}
\end{figure}

For the uniform-field case, low-energy properties 
have become well understood, but the quantitative GS
phase diagram has not been constructed yet as far as we know. 
For details, see e.g., Refs.~\onlinecite{Cab1,Cab2,Mila,Fu-Zh,Hi-Fu} 
and references therein. 
The schematic GS phase diagram is expected as in Fig.~\ref{fig1}. 
The spin-liquid, Haldane and saturated phases correspond to massive plateau
regions in the uniform-field magnetization process. 
The critical phase, except for the decoupled line $J_\perp=0$, 
can be regarded as an one-component Tomonaga-Luttinger liquid 
(TLL),~\cite{Gogo} which is identical to a conformal field theory 
(CFT)~\cite{CFT} with the central charge $c=1$. 
In this area, the magnetization per rung $\langle S_{j}^z\rangle$ 
changes continuously.    
When the field is increased, quantum transitions take place 
at lower and upper critical fields ($H_{1}$ and $H_{2}$). 
They are of a commensurate-incommensurate 
(C-IC) type.~\cite{P-T,Sch,Ch-Gi,B-M} The upper critical field $H_2$
can be determined by calculating the exact spin-wave excitation energy 
in the saturated (perfect ferromagnetic) state: $H_{2}=2J+J_{\perp}$ 
for AF-rung side, and $H_2=2J$ for the FM-rung side.  
The same logic shows that the upper critical field of 
the spin-$1$ AF chain with $J'=J/2$ (strong FM rung-coupling limit) 
is $2J$, which is consistent with the result $H_2=2J$.

\begin{figure}
\scalebox{0.6}{\includegraphics{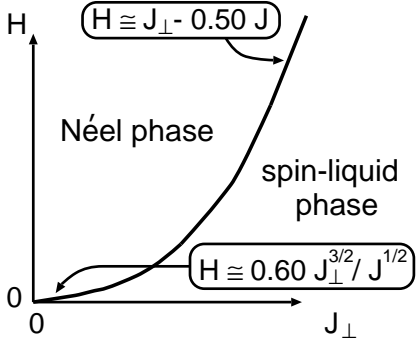}}
\caption{Schematic GS phase diagram of the spin-$\frac{1}{2}$ ladder 
with the staggered Zeeman term (\ref{eq2-2}). 
In weak and strong rung-coupling regions, 
the transition curve follows $h\simeq 0.60\times J_{\perp}^{3/2}/J^{1/2}$ 
and $h\simeq J_{\perp}-0.50\times J$, respectively, where the 
factors ($0.60$ and $0.50$) are determined by our numerical 
calculations.~\cite{Ide} These results are consistent with 
analytical ones in Ref.~\onlinecite{Wang}.}
\label{fig2}
\end{figure}

Wang {\em et al}. have investigated the spin-$\frac{1}{2}$
staggered-field case with the term (\ref{eq2-2}) 
by employing the DPT and Abelian 
bosonization.~\cite{Hal,Lec-Aff,Boso,CFT,D-S,Gogo,Lec-Sene,LowD-QFT,2D-QFT} 
They predict that the competition between the AF rung
coupling and the field $H$ creates a second-order quantum phase transition. 
It belongs to a Gaussian type with $c=1$, and separates a
N\'eel phase and the massive spin-liquid phase, which continuously 
connects with the spin-liquid (a) in Fig.~\ref{fig3}.
Furthermore, Ide, Nakamura, and Sato~\cite{Ide} recently determined the transition 
curve with high accuracy using the level-crossing method~\cite{level} and 
a new twisted operator method.~\cite{Naka} 
The curve starts from the origin in the space
$(J_{\perp},H)$ because both the rung-coupling term and 
the staggered Zeeman term are relevant to the single chain. 
The GS phase diagram is given in Fig.~\ref{fig2}.


\section{\label{sec2} Strong Rung-Coupling Limit}
This section employs the DPT for the strong rung-coupling region: 
$J_{\perp}\gg J$. Frequently in 1D systems, 
theories of the strong coupling limit such as the DPT provide 
visualizations of GSs and low-lying excitations. 
This is the case in the strong rung-coupling limit of the model
(\ref{eq1}), as we discuss in this section.

\subsection{\label{sec2-1} Spin-$\bol 1$ ladders}
First, we investigate the spin-$1$ AF-rung ladder with the uniform field.
The DPT~\cite{Mila} for this model has already been performed, for example in 
Ref.~\onlinecite{Oka}. However, we reproduce it here as a preparation
for later discussions.

\begin{figure}
\scalebox{0.6}{\includegraphics{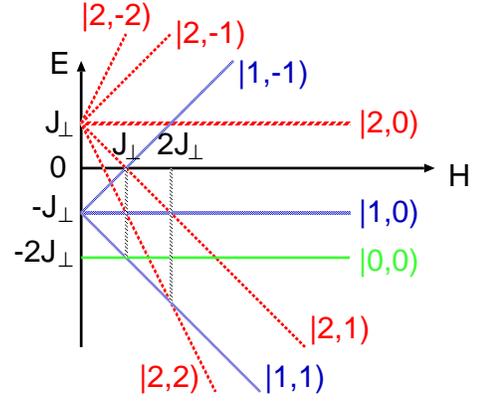}}
\caption{Eigenstates and eigenenergies of $\hat {\cal H}_j$ with the strength of 
the uniform field $H$ varying. The thick dotted line corresponds to the GS. 
Each vector $|\cdots,\cdots)$ represents a state 
$|\mathbb{S},\mathbb{S}^z\rangle_j$.}
\label{fig5}
\end{figure}

  
Under condition $J_{\perp}\gg J$, the $0$-th approximation takes 
the following Hamiltonian:
\begin{subequations}
\label{eq4}
\begin{eqnarray}
\hat {\cal H}^{0\text{-th}}&=&\sum_{j}\hat {\cal H}_j,\label{eq4-1}\\
\hat {\cal H}_j&=& J_{\perp}\vec S_{1,j}\cdot\vec S_{2,j}
-H(S_{1,j}^z+S_{2,j}^z),\label{eq4-2}
\end{eqnarray}
\end{subequations} 
where the original model is reduced to a set of 
two-body problems on each rung. Because the $j$-th rung Hamiltonian 
$\hat {\cal H}_j$ has two conserved quantities, $\vec S_j^2$ and
$S_j^z$, one can easily solve it. Each eigenstate has a one-to-one 
correspondence to a state $|\mathbb{S},\mathbb{S}^z\rangle_{j}$ where 
$\mathbb{S}$ and $\mathbb{S}^z$ are magnitudes of the rung spin 
$\vec S_j$ and its $z$ component, respectively. 
The solution is given in Fig.~\ref{fig5}.
The GS of the rung encounters two level crossings in the magnetization
process. In parameter space $(J_{\perp},H)$, we 
concentrate on vicinities of these two level-crossing lines; $H=J_{\perp}$
and $H=2J_{\perp}$. Near one line, $H=2J_{\perp}$, 
two low-lying states on the $j$-th rung are
\begin{eqnarray}
\label{eq5}
|+\rangle_{1,j} \equiv |1,1\rangle_{j} , && |-\rangle_{1,j} \equiv |2,2\rangle_{j}. 
\end{eqnarray}
Low-lying states near another line are 
\begin{eqnarray}
\label{eq6}
|+\rangle_{2,j} \equiv |0,0\rangle_{j} , && |-\rangle_{2,j} \equiv |1,1\rangle_{j}. 
\end{eqnarray}
The first-order calculation of the DPT is equivalent to projecting the
total Hamiltonian (\ref{eq1}) onto the subspace which consists of the
set of two states (\ref{eq5}) [or (\ref{eq6})] over all rungs.
Two projection operators can be defined as $\hat P_1 =\prod_j\hat
P_{1,j}$ and $\hat P_2 =\prod_j\hat P_{2,j}$, where 
\begin{eqnarray}
\label{eq7}
\hat P_{1,j} &=& |+\rangle_{1,j}{}_{1,j}\langle +| 
\,\,+\,\,|-\rangle_{1,j}{}_{1,j}\langle -|,\nonumber\\
\hat P_{2,j} &=& |+\rangle_{2,j}{}_{2,j}\langle +| 
\,\,+\,\,|-\rangle_{2,j}{}_{2,j}\langle -|.
\end{eqnarray}
For convenience, we define new spin-$\frac{1}{2}$ operators as
\begin{eqnarray}
\label{eq8}
U_j^z = \frac{1}{2}\bigg[ |+\rangle_{1,j}{}_{1,j}\langle +| 
\,\,-\,\,|-\rangle_{1,j}{}_{1,j}\langle -|\bigg],\nonumber\\
U_j^+ = |+\rangle_{1,j}{}_{1,j}\langle -|,\,\,\,\,
U_j^- = |-\rangle_{1,j}{}_{1,j}\langle +|.
\end{eqnarray}
Similarly, another spin-$\frac{1}{2}$ operator $\vec T_j$ is defined by 
replacing the subscript $(1,j)$ to $(2,j)$ in Eq.~(\ref{eq8}).
The relation between these pseudo-spin operators and original spin-$1$
ones is
\begin{subequations}
\label{eq9}
\begin{eqnarray}
\hat P_{1,j}S_{l,j}^z\hat P_{1,j}=-\frac{1}{2}U_j^z +\frac{3}{4},\,
\hat P_{1,j}S_{l,j}^{\pm}\hat P_{1,j}=(-1)^l U_j^{\mp},
\,\,\,\,\,&& \label{eq9-1}\\
\hat P_{2,j}S_{l,j}^z\hat P_{2,j}=-\frac{1}{2}T_j^z +\frac{1}{4},\,
\hat P_{2,j}S_{l,j}^{\pm}\hat P_{2,j}=\frac{2(-1)^l}{\sqrt{3}}T_j^{\mp}.
\,\,\,\,\,&&\label{eq9-2}
\end{eqnarray}
\end{subequations}
We read that $\langle S_j^z\rangle = 0$, $1$ and $2$
correspond to $\langle T_j^z\rangle = \frac{1}{2}$, 
$\langle T_j^z\rangle = -\frac{1}{2}$ (or $\langle U_j^z\rangle=\frac{1}{2}$) 
and $\langle U_j^z\rangle=-\frac{1}{2}$, respectively.
Under these preparations, we can obtain the following effective
Hamiltonian near $H=2J_{\perp}$: 
\begin{eqnarray}
\label{eq10}
\hat{\cal H}_{\text{u},1}^{\text{eff}}&\equiv & \hat P_1 \hat {\cal H}\hat P_1
=\sum_{i,j}\hat P_{1,i}\hat {\cal H} \hat P_{1,j}\nonumber\\
&=&\sum_j {\cal J}_1\left[U_j^x U_{j+1}^x + U_j^y U_{j+1}^y 
+\Delta_1 U_j^z U_{j+1}^z \right] \nonumber\\
&&-H_{\text{u},1}\sum_j U_j^z+\text{const},
\end{eqnarray}
where ${\cal J}_1=2J$, $\Delta_1=\frac{1}{4}$ and 
$H_{\text{u},1}=2J_{\perp}-H+\frac{3}{2}J$. 
Similarly, near $H=J_{\perp}$, we obtain 
\begin{eqnarray}
\label{eq11}
\hat{\cal H}_{\text{u},2}^{\text{eff}}
&=&\sum_j {\cal J}_2\left[T_j^x T_{j+1}^x + T_j^y T_{j+1}^y 
+\Delta_2 T_j^z T_{j+1}^z \right] \nonumber\\
&&-H_{\text{u},2}\sum_j T_j^z + \text{const} ,
\end{eqnarray}
where ${\cal J}_2=\frac{8}{3}J$, $\Delta_2=\frac{3}{16}$ and 
$H_{\text{u},2}=J_{\perp}-H+\frac{J}{2}$.
Both (\ref{eq10}) and (\ref{eq11}) are 
a spin-$\frac{1}{2}$ XXZ chain with a uniform field.~\cite{note2} 
Known exact results on the spin-$\frac{1}{2}$ XXZ chain immediately
lead to the following predictions.
From the model (\ref{eq10}), in the region 
$|H_{\text{u},1}|\leq {\cal J}_1(1+\Delta_1)$ the system has 
a $c=1$ criticality, in which the uniform magnetization and the 
critical exponents of pseudo-spin correlation functions vary 
continuously with $H_{\text{u},1}$ varying. 
Otherwise [i.e., $|H_{\text{u},1}| > {\cal J}_1(1+\Delta_1)$], the magnetization
is saturated. Similarly, from (\ref{eq11}), another massless 
region with $c=1$ exists in $|H_{\text{u},2}| \leq {\cal J}_2(1+\Delta_2)$.
By interpreting these in the context of the original ladder model in the
uniform field, we predict that two $c=1$ critical areas exist: 
$J_{\perp}-\frac{8}{3}J \leq H \leq J_{\perp}+\frac{11}{3}J$ and 
$2J_{\perp}-J \leq H \leq 2J_{\perp}+4J$. 
Therefore, in the strong AF rung-coupling region, 
the spin-$1$ AF ladder has an intermediate plateau region with 
$\langle S_j^z\rangle = 1$ in addition to two trivial plateau regions: 
the saturated state and the PSS state described 
in Fig.~\ref{fig4}. The presence of the intermediate plateau is 
contrasted with the spin-$\frac{1}{2}$ case (Fig.~\ref{fig1}).
From effective models (\ref{eq10}) and (\ref{eq11}), we also see 
that all critical phenomena in the magnetization process 
do not involve any spontaneous symmetry breakings. 
While effective models also predict 
the intermediate plateau vanishes at the point 
$(J_{\perp},H)= (\frac{14}{3}J,\frac{25}{3}J)$, where
the lower and upper boundary curves $H=J_{\perp}+\frac{11}{3}J$ 
and $H=2J_{\perp}-J$ cross each other.  
However, this estimation is too rough because the lowest-order DPT 
is probably valid only in the sufficiently strong rung-coupling cases.
A recent numerical study~\cite{Sakai,Oka2} 
evaluates the vanishing point $(J_{\perp},H)\approx(1.44J,2.7J)$. 
It claims that in the subspace fixing the total magnetization, 
the transition between the $c=1$ region and the plateau is of a 
Beresinski-Kosterliz-Thouless (BKT) type.~\cite{note4}

\begin{figure}
\scalebox{0.6}{\includegraphics{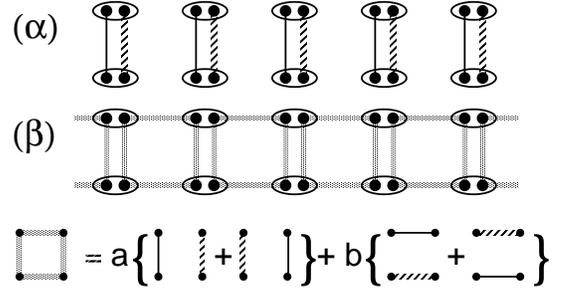}}
\caption{Intermediate plateau state in the spin-$1$ 
ladder (\ref{eq1}) with the uniform field (\ref{eq2-1}): 
the case with $J\rightarrow 0$ ($\alpha$) and the case where $J$ is 
finite ($\beta$). Thin black bonds mean singlet bonds of 
spin-$\frac{1}{2}$ pair. The dotted bonds mean triplet bonds 
where both $z$ components of two spins are $+\frac{1}{2}$. 
In the panel ($\beta$), gray plaquettes represent 
a super-position state drawn in the lowest part of the figure.}
\label{fig6}
\end{figure}

Imitating the RVB picture in Fig.~\ref{fig4}, we attempt to regard the 
intermediate plateau state as a state comprising bonds of
nearest-neighbor spin-$\frac{1}{2}$ pairs. 
One should note that the pairs contain triplet bonds as well as singlet
ones because the plateau has $\langle S_j^z\rangle = 1$. 
From Ref.~\onlinecite{OYA}, the necessary condition for a plateau state is 
\begin{eqnarray}
\label{eq-OYA}
S_u - M_u &=& {\rm an \,\,integer}, 
\end{eqnarray}
where $S_u$ and $M_u$ are, respectively, the sum of spin magnitudes and
magnetizations within the unit cell of the plateau state.
According to this condition, relations~(\ref{eq9}), and two effective models, 
the plateau state should be invariant under the one-site translation 
along the chain and the exchange of two chains.
These allow us to illustrate the bond picture of the intermediate plateau 
as Fig.~\ref{fig6}. 
The state ($\beta$) connects with the state ($\alpha$) smoothly by
decreasing the weight $b$.
Hereafter, we refer to bond pictures including singlet and triplet bonds 
as ``RVB'' pictures.

Next, we turn to the staggered-field case, which has never been discussed
in previous studies.
In the $0$-th order DPT, the even-$j$ rung Hamiltonian 
has the same form as Eq.~(\ref{eq4-2}), but the odd-$j$ one has  
the uniform field pointing to the opposite direction to Eq.~(\ref{eq4-2}).
Therefore, low-lying states in odd rungs must be modified as
follows:
\begin{subequations}
\label{eq12}
\begin{eqnarray}
|+\rangle_{1,j=\text{odd}}\rightarrow |1,-1\rangle_j, &
|-\rangle_{1,j=\text{odd}}\rightarrow |2,-2\rangle_j, &\label{eq12-1}\\
|+\rangle_{2,j=\text{odd}}= |0,0\rangle_j, & 
|-\rangle_{2,j=\text{odd}}\rightarrow |1,-1\rangle_j. &\label{eq12-2}
\end{eqnarray}
\end{subequations}
Consequently, projection and pseudo-spin operators are redefined: 
new pseudo-spins in odd-$j$ sites satisfy 
the same-type relation as Eq.~(\ref{eq9}), in
which $-\frac{1}{2}U_j^z+\frac{3}{4}$ ($-\frac{1}{2}T_j^z+\frac{1}{4}$)
and $U_j^{\mp}$ ($T_j^{\mp}$) are replaced with 
$\frac{1}{2}U_j^z-\frac{3}{4}$ ($\frac{1}{2}T_j^z-\frac{1}{4}$) and 
$-U_j^{\pm}$ ($-T_j^{\pm}$), respectively.
Using these new tools, we can also construct effective models for
the staggered-field cases in a similar way deriving (\ref{eq10}) or (\ref{eq11}).   
In the vicinity of the line $H=2J_{\perp}$, 
the effective Hamiltonian is
\begin{eqnarray}
\label{eq13}
\hat{\cal H}_{\text{s},1}^{\text{eff}}
&=&\sum_j {\cal J}_1\left[\tilde U_j^x \tilde U_{j+1}^x + 
\tilde U_j^y \tilde U_{j+1}^y 
+\Delta_1 \tilde U_j^z \tilde U_{j+1}^z \right] \nonumber\\
&&-H_{\text{s},1}\sum_j (-1)^j\tilde U_j^z + \text{const} ,
\end{eqnarray}
where $\tilde U_j^{x,y}=(-1)^j U_j^{x,y}$, $\tilde U_j^{z}= U_j^{z}$ 
and $H_{\text{s},1}=2J_{\perp}-H-\frac{3}{2}J$. Similarly, near the line 
$H=J_{\perp}$, the effective one $\hat{\cal H}_{\text{s},2}^{\text{eff}}$
is the same type as $\hat{\cal H}_{\text{s},1}^{\text{eff}}$, in which 
$({\cal J}_1,\Delta_1,\tilde U_j^{\alpha})\rightarrow
({\cal J}_2,\Delta_2,\tilde T_j^{\alpha})$ and 
$H_{\text{s},1}\rightarrow H_{\text{s},2}=J_{\perp}-H-\frac{1}{2}J$.
Unlike the uniform-field cases, the above two models have 
effective staggered fields $H_{\text{s},1}$ and $H_{\text{s},2}$, 
respectively. Because such alternating terms are relevant for the 
spin-$\frac{1}{2}$ critical chain, infinitesimal values of staggered 
fields immediately yield a finite excitation gap. In other words, only
when the field $H_{\text{s},1(2)}$ vanishes, the GS is critical.
Therefore, there exist two critical lines (not areas) with $c=1$: 
$H=J_{\perp}-\frac{1}{2}J$ and $H=2J_{\perp}-\frac{3}{2}J$. 
The former (latter) critical line satisfies 
$\langle \tilde U_j^z\rangle =0$ and $\langle S_j^z\rangle=\frac{3}{2}(-1)^j$
($\langle \tilde T_j^z\rangle =0$ and 
$\langle S_j^z\rangle=\frac{1}{2}(-1)^j$). Both lines are of a $c=1$
Gaussian-type transition in common with the spin-$\frac{1}{2}$ case
(Fig.~\ref{fig2}). On the critical lines, the staggered susceptibility 
$(-1)^j\frac{\partial \langle S_j^z\rangle}{\partial H}$ diverges 
[see the next subsection]. Like the uniform-field case, any symmetry
breakings do not occur at the transitions. 
Numerical~\cite{Nomu} and analytical~\cite{A-O,Alca} works show 
that the staggered magnetization in $\hat {\cal H}_{\text{s},1}^{\text{eff}}$ 
(or $\hat {\cal H}_{\text{s},2}^{\text{eff}}$) changes continuously with
the staggered field varying. Thereby, we can confirm that the 
original staggered magnetization $(-1)^j\langle S_j^z\rangle$ 
has no plateau, contrary to the uniform-field cases. 

We should discuss effects of the higher-order terms of $J$ in the DPT, 
although they have already been explained in Ref.~\onlinecite{Wang}. 
Especially, let us consider whether any mechanisms 
varying the properties of above $c=1$ criticalities emerge or not, 
from higher-order effects. In the vicinity of each transition, 
both two low-lying states in the rung 
and the perturbative intrachain coupling part are 
invariant under spin rotations around 
the $z$ axis of the total spin. Therefore, the U(1) symmetry can not be broken 
by higher-order terms. For instance, an anisotropic XY exchange 
interaction, which brings a mass generation, does not occur. 
The continuous U(1) symmetry is one of the characteristic natures 
in the $c=1$ criticality. Because the original model (\ref{eq1}) 
has a site-parity symmetry, we can also say that bond-alternating terms, 
which produce a mass too, do not appear.  
As long as we focus on the strong rung-coupling cases ($J_{\perp}\gg J$), 
anisotropy parameter $\Delta_{1(2)}$ will not 
exceed the value of the BKT transition point $\Delta_{\text{BKT}}=1$.
These considerations imply that higher-order perturbations can not make 
$c=1$ criticalities of transitions change, 
although they will modify parameters 
$({\cal J}_{1(2)},\Delta_{1(2)},H_{\text{u(s)},1(2)})$ slightly 
and generate several new but small terms (e.g., next nearest neighbor
interaction terms) in effective models. 

Using analytical tools, we can also evaluate 
several physical quantities around each transition. 
We will discuss them in the next subsection along with
general spin-$S$ cases.

\subsection{\label{sec2-2} Generalization to spin-$\bol S$ ladders}
Viewing results of DPTs in spin-$\frac{1}{2}$ cases 
(Figs.~\ref{fig1},~\ref{fig2}, Refs.~\onlinecite{Wang} and \onlinecite{Cab2}) 
and above spin-$1$ cases, we find it possible to generalize them to
$2$-leg spin-$S$ AF ladders with external fields (\ref{eq2-1}) or (\ref{eq2-2}). 
Even for the spin-$S$ case, in which $\vec S_{l,j}$ 
means a spin-$S$ operator, the rung Hamiltonian is solvable through
states $|\mathbb{S},\mathbb{S}^z\rangle_j$, in which $\mathbb{S}$ runs from
$0$ to $2S$. The number of independent states 
$|\mathbb{S},\mathbb{S}^z\rangle_j$ is $\sum_{m=0}^{2S}(2m+1)=(2S+1)^2$, 
and the table of Clebsch-Gordan (CG) coefficients 
(or the Wigner-Eckart theorem) can determine the relation between 
them and the states of original two spins $(\vec S_{1,j},\vec S_{2,j})$.  
The eigen energy of the state $|\mathbb{S},\mathbb{S}^z\rangle_j$ is 
\begin{eqnarray}
\label{eq14}
\frac{J_{\perp}}{2}\mathbb{S}(\mathbb{S}+1)-J_{\perp}S(S+1)-H\mathbb{S}^z,
\end{eqnarray}
where we consider the even-$j$ rung.
In a subspace with a fixed value of $\mathbb{S}$, the GS of the rung is  
$|\mathbb{S},\mathbb{S}\rangle_j$, although at $H=0$, all $(2\mathbb{S}+1)$
states are degenerate. Because the GS energy of the above subspace 
goes down with the slope $-\mathbb{S}$ 
when $H$ is increased, the GS of the full space has $2S$ level
crossings: $H=J_{\perp},2J_{\perp},\dots , 2SJ_{\perp}$.
(This is a general result of Fig.~\ref{fig5}.) Hence, one readily expects
that the spin-$S$ case has $2S$ critical phenomena in the
uniform(staggered)-field magnetization process. 
The first-order DPT near each level-crossing line gives 
the effective model, which is the same type as the spin-$1$ case:
\begin{table*}
\caption{Effective models around the highest-field lines $H=2SJ_\perp$. 
Parameters $({\cal J}_1,\Delta_1,H_{\text{u},1}, H_{\text{s},1})$ are 
values of the effective exchange coupling constant, the anisotropy
 parameter, the effective uniform field and the effective staggered
 field, respectively. The sign $\Gamma$ denotes the width of the $c=1$
 critical region along the $H$-axis in the original parameter space 
$(J_{\perp},H)$, $M_{\text{u},1}$ represents the possible values of
 $\langle S_{l,j}^z\rangle$ in the critical region $\Gamma$, and  
$M_{\text{s},1}$ is the value of $|\langle S_{l,j}^z\rangle|$ 
at the transition line $H_{\text{s},1}=0$.}
\label{tab1}
\begin{ruledtabular}
\begin{tabular}{cccccccc}
spin & ${\cal J}_1$ &$\Delta_1$ & $H_{\text{u},1}$& $\Gamma$ & $M_{\text{u},1}$ 
& $H_{\text{s},1}$ & $M_{\text{s},1}$ \\
\hline
$1/2$ &  $J$ & $1/2$ & $J_{\perp}-H+\frac{1}{2}J$ & $3 J$ 
& $0\rightarrow \frac{1}{2}$ & $J_{\perp}-H-\frac{1}{2}J$  &  $1/4$\\
$1$  & $2J$ & $1/4$ & $2J_{\perp}-H+\frac{3}{2}J$ & $5J$  
& $\frac{1}{2}\rightarrow 1$ & $2J_{\perp}-H-\frac{3}{2}J$ &  $3/4$\\
$3/2$ & $3J$ & $1/6$ & $3J_{\perp}-H+\frac{5}{2}J$ & $7J$
& $1\rightarrow \frac{3}{2}$ & $3J_{\perp}-H-\frac{5}{2}J$ &  $5/4$\\
$2$ & $4J$ & $1/8$ & $4J_{\perp}-H+\frac{7}{2}J$ & $9J$
& $\frac{3}{2}\rightarrow 2$ &$4J_{\perp}-H-\frac{7}{2}J$ &  $7/4$\\
\hline
$S$ & $2SJ$ & $1/(4S)$ & $2SJ_{\perp}-H+\frac{4S-1}{2}J$  
& $(4S+1)J$& $S-\frac{1}{2}\rightarrow S$ &
 $2SJ_{\perp}-H-\frac{4S-1}{2}J$ &  $S-\frac{1}{4}$\\
\end{tabular}
\end{ruledtabular}
\caption{Effective models around the second-highest-field lines $H=(2S-1)J_\perp$. 
Parameters $({\cal J}_2,\Delta_2,H_{\text{u},2},\Gamma,M_{\text{u},2}, H_{\text{s},2},
M_{\text{s},2})$ have the same roles as 
$({\cal J}_1,\Delta_1,H_{\text{u},1},\Gamma,M_{\text{u},1}, H_{\text{s},1},
M_{\text{s},1})$ in Table \ref{tab1}, respectively. Note that the
 spin-$\frac{1}{2}$ case does not possess the effective model.}
\label{tab2}
\begin{ruledtabular}
\begin{tabular}{ccccccccc}
spin & ${\cal J}_2$ &$\Delta_2$ & $H_{\text{u},2}$ & $\Gamma$ &
 $M_{\text{u},2}$ & $H_{\text{s},2}$ & $M_{\text{s},2}$\\
\hline
$1$ &  $\frac{8}{3}J $ & $3/16$ & $J_{\perp}-H+\frac{1}{2}J$ 
& $\frac{19}{3}J$ & $0\rightarrow \frac{1}{2}$  
& $J_{\perp}-H-\frac{1}{2}J$  & $1/4$ \\
$3/2$ & $\frac{24}{5}J$ & $5/48$ & $2J_{\perp}-H+\frac{3}{2}J$ 
& $\frac{53}{5}J$ & $\frac{1}{2}\rightarrow 1$  
& $2J_{\perp}-H-\frac{3}{2}J$ & $3/4$ \\
$2$ & $\frac{48}{7} J$ & $7/96$ & $3J_{\perp}-H+\frac{5}{2}J$ 
&  $\frac{103}{7}J$ & $1\rightarrow \frac{3}{2}$  
& $3 J_{\perp}-H-\frac{5}{2}J$ & $5/4$ \\
$5/2$ & $\frac{80}{9} J$ & $9/160$ & $4 J_{\perp}-H+\frac{7}{2}J$ 
&  $\frac{169}{9}J$ & $\frac{3}{2}\rightarrow 2$  
& $4J_{\perp}-H-\frac{7}{2}J$ &  $7/4$ \\
\hline
$S$ &  $\frac{8S(2S-1)}{4S-1}J$ & $\frac{4S-1}{16S(2S-1)}$ 
&  $(2S-1)J_{\perp}-H+\frac{4S-3}{2}J$ 
& $\frac{32S^{2}-12S-1}{4S-1}J$ & $S-1\rightarrow S-\frac{1}{2}$
& $(2S-1)J_{\perp}-H-\frac{4S-3}{2}J$  & $S-\frac{3}{4}$\\
\end{tabular}
\end{ruledtabular}
\end{table*}
we always obtain a spin-$\frac{1}{2}$ XXZ chain with a uniform 
(staggered) field for the uniform(staggered)-field cases.
Therefore, even in the spin-$S$ cases, 
similar conclusions as the spin-$1$ cases can be available. 
In principle, one can obtain effective models 
in all vicinities of level-crossing lines. 
However, derivations of the models associated with lower-field crossing lines 
must calculate numerous CG coefficients (syntheses of spins).
Tables~\ref{tab1} and \ref{tab2} depict only the effective
models near the highest-field and second-highest-field lines. 
Nevertheless, these two will be sufficient to conclude that the spin-$S$
case has $2S$ critical regions (points) in the uniform(staggered)-field 
magnetization process. Note that in these two tables, we use the same
symbols $({\cal J}_{1(2)},\Delta_{1(2)},H_{\text{u(s)},1(2)})$ in 
the spin-$1$ cases to represent effective models. 

From these tables, we can extract the following information on the
GS of the spin-$S$ case. 
(\rnum{1}) For uniform-field cases,
the critical areas around the highest-field line and
second-highest-field line, respectively, are
\begin{subequations}
\label{eq14'}
\begin{eqnarray}
2SJ_{\perp}-J\leq H\leq 2SJ_{\perp}+4SJ,\label{eq14'-1}\\
(2S-1)J_{\perp}-\frac{8S^2+2S-2}{4S-1}J  \leq H\nonumber \\ 
\leq (2S-1)J_{\perp}+\frac{24S^2-14S+1}{4S-1}J,\label{eq14'-2}
\end{eqnarray}
\end{subequations}
which determine the width of the critical region $\Gamma$ in the tables.
The plateau regions exist outside the regions $\Gamma$, 
as with the spin-$1$ case.  
(\rnum{2}) The larger the magnitude of spin $S$ becomes,
the more the width $\Gamma$ increases as a result of the growth of 
the effective coupling ${\cal J}_{1(2)}$.
The effective model approaches a spin-$\frac{1}{2}$ XY chain. 
Because plateaux are characteristic in ``quantum'' spin systems, 
the growth of $\Gamma$, or the decrease of plateau regions,        
means the system approaches the classical vector spin system.
On the other hand, in the case fixing $S$, the highest-field critical
region $\Gamma$ is smaller than the second highest-field one.
Therefore we expect that the lower-field critical regions are larger.
(\rnum{3}) Like the spin-1 case, all intermediate plateaux must vanish
at a sufficiently weak rung-coupling region.
(\rnum{4}) For the staggered-field cases, the transition lines with the 
highest field and the second-highest field are, respectively,
\begin{subequations}
\label{eq14''}
\begin{eqnarray}
H&=&2SJ_{\perp}-(4S-1)J/2,\label{eq14''-1}\\
H&=&(2S-1)J_{\perp}-(4S-3)J/2.\label{eq14''-2}
\end{eqnarray}
\end{subequations}
(\rnum{5}) All critical phenomena do not involve spontaneous 
symmetry breakings as with the spin-1 case.


Now, let us investigate the physical quantities 
around each criticality in more detail.
If we represent the effective model near each transition with the 
(second) highest field using the pseudo-spin operator 
$U_j^{\alpha}$ ($T_j^{\alpha}$) in a similar manner of the preceding subsection, 
the original spins are projected out as follows:
\begin{eqnarray}
\label{eq15}
S_{l,j}^z\rightarrow -\frac{1}{2}U_j^z+S-\frac{1}{4},&&
S_{l,j}^{\pm}\rightarrow (-1)^l\sqrt{S}U_j^{\mp},\\ 
S_{l,j}^z\rightarrow -\frac{1}{2}T_j^z+S-\frac{3}{4},&&
S_{l,j}^{\pm}\rightarrow (-1)^l 2\sqrt{\frac{S(2S-1)}{4S-1}}T_j^{\mp},\nonumber
\end{eqnarray} 
where we consider even-$j$ sites, again. For the staggered-field
cases, as Eqs.~(\ref{eq12}) and (\ref{eq13}), we must redefine 
$U_j^{\alpha}$ ($T_j^{\alpha}$) in all odd-$j$ rungs
and define ${\tilde U}_j^{\alpha}$ (${\tilde T}_j^{\alpha}$). 
The relation (\ref{eq15}), of course, contains Eq.~(\ref{eq9}), 
and provides the magnetic relation between the effective and
original models. As mentioned earlier, 
the low-energy properties of the effective models 
are elucidated well by several analytical methods. 
Therefore, the combination of the relation (\ref{eq15}) and such methods 
can provide accurate predictions of the original spin-$S$ ladders.

First we discuss the uniform-field cases. 
According to Bethe ansatz~\cite{Cab2,K-B-I,M-Taka} and Abelian 
bosonization,~\cite{Hal,Lec-Aff,Boso,CFT,D-S,Gogo,Lec-Sene,LowD-QFT,2D-QFT} 
the low-energy physics of the massless area 
$|H_{\text{u},1(2)}|\leq {\cal J}_{1(2)}(1+\Delta_{1(2)})$
is governed by the free boson field theory with the spin-wave [massless
excitation] velocity $v_{1(2)}$ and 
the compactification radius $R_{1(2)}$. (We refer the reader to
Appendix~\ref{app1} for an explanation of Abelian bosonization.) 
Utilizing their knowledge, 
one can derive the uniform susceptibility formula~\cite{A-O,E-A-T}
\begin{eqnarray}
\label{eq16}
\chi_{1(2)}
\equiv
\frac{\partial \langle U_j^z\rangle}{\partial H_{\text{u},1}}\,\,
\left(\text{or}\,\,
\frac{\partial \langle T_j^z\rangle}{\partial H_{\text{u},2}}\right) 
&=& \frac{a_0}{(2\pi)^2 R_{1(2)}^2 v_{1(2)}},\,\,\,\,\,\,\,\,\,\,\,\,
\end{eqnarray}
where $a_0$ is the lattice constant.
Because the effective field $H_{\text{u},1(2)}$ varies 
linearly with the original field $H$ (see Tables~\ref{tab1} and
\ref{tab2}), the relation $\chi_{1(2)}=\frac{1}{2}
\frac{\partial \langle S_{l,j}^z\rangle}{\partial H}$ 
is realized for the magnetization process varying only
$H$,~\cite{note00} at least in the first-order DPT.
One thus can regard all the behavior of $\chi_{1(2)}$ as those of the
original susceptibility, except for the difference of the factor $\frac{1}{2}$. 
The linear relation between the effective
and original fields is often used below. 
The Bethe ansatz can determine the radius $R_{1(2)}$ and the
velocity $v_{1(2)}$ as functions of ${\cal J}_{1(2)}$, $\Delta_{1(2)}$
and $H_{\text{u},1(2)}$.
Especially for $H_{\text{u},1(2)}=0$, the radius and
the velocity are represented analytically as 
\begin{eqnarray}
\label{eq17}
R_{1(2)}&=&\frac{1}{\sqrt{2\pi}}
\left(1-\frac{1}{\pi}\arccos\Delta_{1(2)}\right)^{1/2},\nonumber\\
v_{1(2)}&=&\frac{\pi}{2}\,\,\frac{\sqrt{1-\Delta_{1(2)}^2}}{\arccos\Delta_{1(2)}}
\,\,{\cal J}_{1(2)}a_0.
\end{eqnarray}
Inserting Eq.~(\ref{eq17}) to Eq.~(\ref{eq16}), 
we see that when $S$ goes from one half (one) to $\infty$, 
the pseudo-spin magnetization curve slope 
[i.e., the susceptibility (\ref{eq16})] at the midpoint
of the (second) highest-field critical regime 
[i.e., at $H_{\text{u},1(2)}=0$] decreases monotonically from 
$\frac{1}{\sqrt{3}\pi J}\cong 0.184\times J^{-1}$ 
($0.0954\times J^{-1}$) to $\frac{1}{2\pi S J}$ ($\frac{1}{4\pi S J}$). 
Fixing $S$, we also find that $\chi_1(H_{\text{u},1}=0)$ 
is larger than $\chi_2(H_{\text{u},2}=0)$. 
For example, in the spin-$1$ case, 
$\chi_{1}(H_{\text{u},1}=0)\cong 0.119\times J^{-1}$ and 
$\chi_{2}(H_{\text{u},2}=0)\cong0.0954\times J^{-1}$.
These imply a reasonable fact that  
the larger $\Gamma$ is, the smaller is the magnetization slope 
at $H_{\text{u},1(2)}=0$.  
Subsequently, let us consider how magnetization
approaches the value of the plateau (saturation). 
Without utilizing formula (\ref{eq16}), 
studies of the C-IC transition~\cite{P-T,Sch,Ch-Gi,B-M} 
have shown that near the saturation, the magnetization behaves as     
\begin{eqnarray}
\label{eq18}
\langle U_j^z\rangle\,\,(\text{or}\,\,\langle T_j^z\rangle)&\sim&
|H_{\text{u},1(2)}^{\text{cr}}-H_{\text{u},1(2)}|^{1/2},
\end{eqnarray}
where $|H_{\text{u},1(2)}^{\text{cr}}|$ is the critical value 
${\cal J}_{1(2)}(1+\Delta_{1(2)})$.~\cite{note5} 
Hence, in the magnetization process, the original magnetization
$\langle S_{l,j}^z\rangle$ also behaves, near each plateau, 
$\sim |H^{\text{cr}}-H|^{1/2}$, where $H^{\text{cr}}$ is the critical
field of each $c=1$ region. This power law is a universal property.
That is, it is independent of the spin magnitude $S$ and 
the level-crossing line we choose in the two-spin problem of the rung. 
The spin-wave analysis can exactly calculate the gap in
the saturated state of each effective model. 
For $|H_{\text{u},1(2)}|> |H_{\text{u},1(2)}^{\text{cr}}|$, the excitation gap is
estimated as $|H_{\text{u},1(2)}|-|H_{\text{u},1(2)}^{\text{cr}}|$.~\cite{note6}  
Again, translating this into the original model, we see that when $H$ is
moved just outside each $c=1$ region, the gap grows as 
$|H-H^{\text{cr}}|$.~\cite{note7}  

Performing the same spin-wave analysis 
in the saturated state of the original spin-$S$ ladders, 
we can determine the upper critical uniform field $H_c$. 
The field gives the boundary between the saturation with 
$\langle S_j^z\rangle =2S$ and the $c=1$ phase just under it. 
The result is  
\begin{eqnarray}
\label{eq19}
H_c=4SJ+2SJ_{\perp}.
\end{eqnarray}
Surprisingly, $H_c$ perfectly agrees with the upper critical field derived
from the effective model around the highest-field line 
[see the region (\ref{eq14'-1})]. In other words, $H_c$ is not 
modified by the higher-order perturbation effects of the DPT. 
For the FM-rung side, the spin-wave theory implies that 
\begin{eqnarray}
\label{eq19-2}
H_c=4SJ,
\end{eqnarray}
where $H_c$ does not depend upon the rung coupling
$J_\perp$. Equations~(\ref{eq19}) and (\ref{eq19-2}) are a generalization
of the critical field $H_2$ in the spin-$\frac{1}{2}$ case (Fig.~\ref{fig1}).

Next, we shift our focus to the staggered-field cases. 
When the effective staggered field $H_{\text{s},1(2)}$ 
is vanishing [i.e., the GS is massless], 
the bosonization translates the pseudo-spin operator to the following
boson representation:
\begin{eqnarray}
\label{eq20}
\tilde{U}_j^z\,\left(\text{or} \,\tilde{T}_j^z\right) 
\approx   \frac{a_0\,\partial_x \phi}{2\pi R_{1(2)}}
+(-1)^j A_{1(2)} \sin\left(\frac{\phi}{R_{1(2)}}\right),
\end{eqnarray}
where $\phi(x)$ [$x=j\times a_0$] and $A_{1(2)}$ are 
the boson field and 
a nonuniversal constant,~\cite{L-Z,Luk,H-F} respectively. 
Here we neglect the so-called Klein factor.~\cite{D-S,Lec-Sene,LowD-QFT,2D-QFT}
From Eq.~(\ref{eq20}), a finite $H_{\text{s},1(2)}$ leads to 
a perturbation term proportional to 
$H_{\text{s},1(2)}\sin(\phi/R_{1(2)})$ for the effective boson field
theory, and it then becomes a sine-Gordon model.
The vertex operator $\sin(\phi/R_{1(2)})$ has the scaling dimension 
$x_{1(2)}=\frac{1}{4\pi R_{1(2)}^2}$ and is always relevant (i.e.,
$x_{1(2)}<2$). Therefore the scaling argument~\cite{A-O} 
tells us that any small fields 
$H_{\text{s},1(2)}$ yield an excitation gap $m_{1(2)}$ 
and the pseudo-spin staggered magnetization as:
\begin{subequations}
\label{eq21} 
\begin{eqnarray}
m_{1(2)}&\sim & |H_{\text{s},1(2)}|^{1/[2-x_{1(2)}]},\\ \label{eq21-1}
\langle \tilde{U}_j^z\rangle\,(\text{or}\,\langle \tilde{T}_j^z\rangle)
&\sim &  (-1)^j|H_{\text{s},1(2)}|^{x_{1(2)}/[2-x_{1(2)}]},\label{eq21-2}
\end{eqnarray}
\end{subequations}
where if $H_{\text{s},1(2)}<0$, we take the replacement
$(-1)^j\rightarrow (-1)^{j+1}$ in Eq.~(\ref{eq21-2}). 
Substituting $R_{1(2)}$ in Tables~\ref{tab1} and
\ref{tab2} for $x_{1(2)}$, we find that the larger the value $S$
becomes, the slower the growths of both the gap and the magnetization  
become. In other words, the singularity of the staggered susceptibility
$\frac{\partial \langle \tilde{U}_j^z\rangle}
{\partial H_{\text{s},1}}|_{H_{\text{s},1}\rightarrow 0}$ (or 
$\frac{\partial \langle \tilde{T}_j^z\rangle}{\partial
H_{\text{s},2}}|_{H_{\text{s},2}\rightarrow 0}$) decreases with $S$ increasing.
Particularly, in the limit $S\rightarrow \infty$, 
$m_{1(2)}\sim |H_{\text{s},1(2)}|$ and 
$|\langle \tilde{U}_j^z\rangle|\, (\text{or}\,\,|\langle \tilde{T}_j^z\rangle|)
\sim |H_{\text{s},1(2)}|$, which means the staggered susceptibility does
not diverge at the ``transition'' line $H_{\text{s},1(2)}=0$. 
These can be again interpreted as a sign of the approach to the classical
spin system. On the other hand, fixing $S$, one observes that 
when the fields $H_{\text{s},1}=H_{\text{s},2}$ are small enough,
$m_1$ and $|\langle \tilde{U}_j^z\rangle|$ are larger than $m_2$ and 
$|\langle \tilde{T}_j^z\rangle|$, respectively. We therefore anticipate that 
the transition with the higher field has a stronger singularity.
In the spin-$1$ case, we have $m_1\sim |H_{\text{s},1}|^{0.878}$, 
$m_2\sim |H_{\text{s},2}|^{0.903}$, 
$|\langle \tilde{U}_j^z\rangle| \sim |H_{\text{s},1}|^{0.756}$ 
and $|\langle \tilde{T}_j^z\rangle| \sim |H_{\text{s},2}|^{0.806}$. 
In common with the uniform-field cases,  
$H_{\text{s},1(2)}$ has a linear relation with $H$. 
In order to translate all consequences into ones of the model 
(\ref{eq1}) in the original staggered-field magnetization process, 
it is sufficient to replace $H_{\text{s},1(2)}$
and $\langle \tilde{U}_j^z\rangle$ (or $\langle \tilde{T}_j^z\rangle$),
respectively, with $H$ and 
$\langle S_{l,j}^z\rangle - (-1)^j M_{\text{s},1(2)}$, where 
$M_{\text{s},1(2)}$ is the staggered magnetization par site 
at each transition line (see Tables~\ref{tab1} and \ref{tab2}).

\begin{figure}
\scalebox{0.35}{\includegraphics{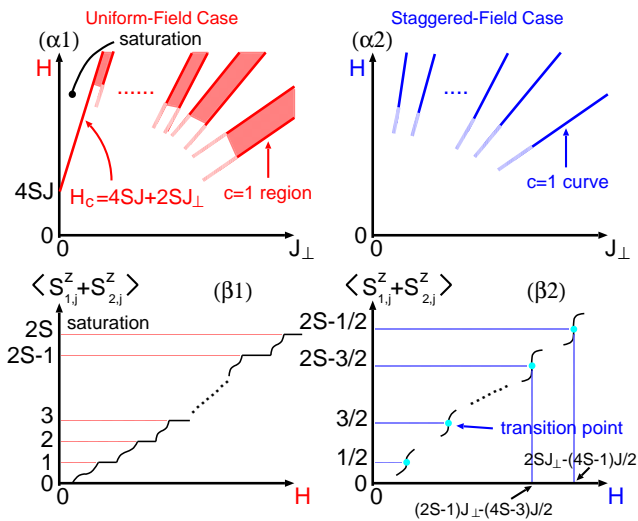}}
\caption{Panels ($\alpha 1$) and ($\alpha 2$) are, respectively, 
the GS phase diagrams in the uniform-field case (\ref{eq2-1}) 
and the staggered-field one (\ref{eq2-2}). 
Panels ($\beta 1$) and ($\beta 2$) are, respectively, 
the uniform magnetization curve in the case (\ref{eq2-1}) and 
the staggered one in the case (\ref{eq2-2}). In the panel ($\beta 2$), 
we denote the magnetization of the even-$j$ rung.}
\label{fig7}
\end{figure}

Summarizing all the above results about the spin-$S$ cases,
we can draw GS phase diagrams and the uniform and staggered 
magnetizations as in Fig.~\ref{fig7}. 
Notably, Fig.~\ref{fig7} is valid in the strong
AF rung-coupling limit.

Utilizing solutions of the Bethe ansatz integral
equations,~\cite{A-O} values of the nonuniversal constants 
$A_{1(2)}$,~\cite{Wang,H-F} etc., we can serve more quantitative predictions.      
We omit them here.

\begin{figure}
\scalebox{0.5}{\includegraphics{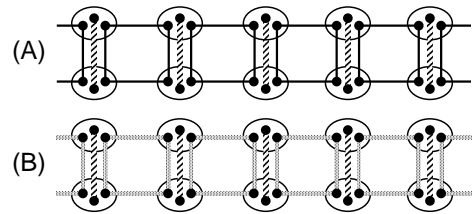}}
\caption{Expected ``RVB'' pictures for the plateaux with 
$\langle S_j^z\rangle=1$ (A) and $\langle S_j^z\rangle=2$ (B) 
in the spin-$\frac{3}{2}$ case.}
\label{fig-3/2}
\end{figure}

Finally, we speculate the short-range ``RVB'' picture of the plateau
states in the uniform-field case, without any computations. 
The spin-$S$ case has $2S+1$ plateau regions including two trivial
plateaux: the saturated state and the $H=0$ state.
The guides to guess the ``RVB'' pictures are the effective models, in
which translational symmetry does not break, the plateau
condition (\ref{eq-OYA}), and the expectation that the bonds along the
chain are subject to taking the singlet state and all plateaux vanish 
in the sufficiently weak AF rung-coupling region. 
From these, in order to build the ``RVB'' picture for the plateau with 
$\langle S_j^z\rangle=\tilde S$ in the spin-$S$ case, we should perform
just the following two procedures: (\rnum{1}) putting $\tilde S$ triplet
bonds and $(2S-\tilde S)$ singlet bonds per a rung, and (\rnum{2}) 
``joining'' two nearest-neighbor rungs by singlet or triplet bonds, not
to break the translational symmetry along the chain.
The plaquette states of the lowest panels in Figs.~\ref{fig4} and 
\ref{fig6} are available for the procedure (\rnum{2}).
For instance, two plateau states with $\langle S_j^z\rangle=1$ and 
$\langle S_j^z\rangle=2$ in the spin-$\frac{3}{2}$ case 
are described as (A) and (B) in Fig.~\ref{fig-3/2}, respectively.
Similarly, the plateau with $\langle S_j^z\rangle=0$ could be captured
by the set of a PSS state and a spin-liquid state (a) in Fig.~\ref{fig3}. 
Following the above speculations, one would easily produce ``RVB'' pictures 
for any plateaux.

Within the ``RVB'' picture, every time the GS moves from a plateau to 
the plateau just above it with $H$ increasing, one singlet bond along 
each rung is cut and exchanged for a triplet bond. This phenomena is 
reminiscent of successive transitions in the spin-$S$ bond-alternated 
chain.~\cite{Lec-Aff,A-H,Sier}

\section{\label{sec3} Weak Rung-Coupling Limit}
This section describes the spin-$1$ ladder (\ref{eq1}) under 
the opposite condition to the last section: $J\gg |J_{\perp}|,H$.
We start from the limit $J_{\perp}=H=0$. There, the ladder becomes 
two decoupled spin-$1$ AF Heisenberg chains. 
The Haldane gap of the single chain complicates the mapping to some field
theories. Nevertheless, two famous mappings
exist:~\cite{schul} Haldane's method based on 
a non-linear sigma model (NLSM)~\cite{NLSM} and the method applying a 
NAB.~\cite{Witt,Aff,A-H} We will use the latter in this section. 
It is useful for treating several additional terms, 
in first principle, by the perturbation theory and the RG picture.

\subsection{\label{sec3-1} Critical points of spin-1 AF Chain}
The application of the NAB to 1D spin-$1$ systems~\cite{Tsv,I-O,Kita,Allen} 
is supported by low-energy properties of the spin-$1$ 
Takhtajan-Babujian (TB) chain~\cite{Tak,Bab}
and the following spin-$1$ bilinear-biquadratic chain:
\begin{eqnarray} 
\label{eq24}
\hat {\cal H}_{\delta}&=& J \sum_{j}\left[\vec S_j\cdot\vec S_{j+1}
+\delta(\vec S_j\cdot\vec S_{j+1})^2\right], 
\end{eqnarray}
where $J>0$, $\vec S_j$ is the spin-$1$ operator on the site $j$, 
and the TB chain corresponds to $\delta=-1=\delta_{\text{TB}}$.
This subsection present a brief review the model (\ref{eq24}).~\cite{note9}

\begin{figure}
\scalebox{0.45}{\includegraphics{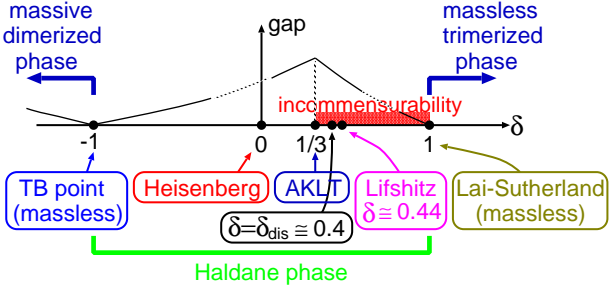}}
\caption{GS phase diagram and the schematic gap behavior in the model (\ref{eq24}).}
\label{fig8}
\end{figure}

The efforts of several 
people~\cite{Sol,Ken,F-S,B-X-G,S-J-G,SMKYM,G-J-S,F-Su,Nomu2,L-S-T} 
since the late 1980's have advanced our understanding of the model (\ref{eq24}). 
In particular, the low-energy properties in the region $J>0$ and
$|\delta|\leq 1$ have been elucidated well.
The GS phase diagram and the excitation gap in this region are summarized as 
in Fig.~\ref{fig8}. At least three special points exist aside from our target, 
the Heisenberg point $\delta=0$. The TB chain is integrable and 
has massless excitations with the wave numbers (momenta) $k=0$ and $\pi/a_0$.
The point $\delta_{\text{LS}}=1$, called the 
Lai-Sutherland (LS) model,~\cite{Lai-Sut} 
is also integrable. It has massless excitations with $k=0$ and $\pm 2\pi/3a_0$. 
The low-energy limits of TB and LS points are, respectively, 
equal to the level-$2$ SU(2) WZNW model~\cite{A-M,AGSZ,Avd,I-M} 
(a CFT) and the level-$1$ SU(3) WZNW model. 
On the $\delta$ axis, these two points are located in a quantum phase transition. 
The TB point separates the Haldane phase ($|\delta|\leq 1$) and the
massive dimerized phase ($\delta<\delta_{\text{TB}} $), 
which has twice degenerate GSs. 
On the other hand, the LS point separates the Haldane and massless
``trimerized'' phase ($\delta>\delta_{\text{LS}}$). The transition belongs to a
generalized SU(3) BKT type.~\cite{I-K} Features of the Haldane phase are the
existence of a finite gap between the unique GS and the isolated spin-$1$
magnon mode, and a ``hidden'' AF order detected by a string
order parameter:
\begin{eqnarray} 
\label{eq25}
{\cal O}^a&=& -\lim_{|j-k|\rightarrow\infty}       
\left\langle S_j^a\,\,e^{i\pi\sum_{n=j+1}^{k-1}S_n^a} 
\,\,S_k^a\right\rangle, \nonumber\\
&\equiv& \langle\hat {\cal O}^a\rangle,
\end{eqnarray}
which is the same form as the string parameter (\ref{eq3-1}) in the
spin-$\frac{1}{2}$ ladder, except that $S_j^{\alpha}$ is a spin-$1$
operator. The AKLT~\cite{AKLT} point 
$\delta_{\text{aklt}} =\frac{1}{3}$ is noteworthy because its 
GS is exactly identical to a valence-bond-solid state. Moreover,
the point $\delta_{\text{aklt}}$ is the onset of an incommensurability: 
in $(1>)$ $\delta > \delta_{\text{aklt}}$, 
the real-space spin correlations has an incommensurate fluctuation, 
that connects smoothly with the three-site period one at the LS point. 

Asides from points above, recent studies have described there are 
two characteristic points: $\delta_{\text{dis}}\cong 0.4$ 
(Ref.~\onlinecite{G-J-S}) and
Lifshitz point $\delta_{\text{Lif}}\cong 0.44$.~\cite{B-X-G,S-J-G}
In the region $-1< \delta <\delta_{\text{dis}}$, the momentum of 
the lowest magnon excitation stays at $k=\pi/a_0$. However, 
on the right side of $\delta_{\text{dis}}$, it has a deviation from $k=\pi/a_0$,
and splits into two incommensurate momenta, which smoothly reach the massless
points $+3\pi/2a_0$ and $-3\pi/2a_0$ at the LS chain, respectively.   
At the Lifshitz point, an incommensurability appears in the spin structure factor.

\subsection{\label{sec3-2} Effective field theory}
Reviewing the last paragraph, one notes that the TB point in the
model (\ref{eq24}) can be adopted as an underlying point to consider 
an effective field theory for the Heisenberg chain ($\delta=0$). 
It is likely that the LS point is also available for the field theoretical
description. However, as mentioned already, 
there are three points changing the low-energy properties between 
the LS and Heisenberg points.
Therefore, the LS point is not appropriate for deriving the effective theory of
the Heisenberg point. This subsection provides 
the effective field theory for the spin-$1$ ladder (\ref{eq1}) 
in our notation.~\cite{Allen}

The level-$2$ SU(2) WZNW model, which describes the TB
point, has two primary fields: the $2\times 2$ matrix field $g_{mn}$ 
[$m,n\in \{1,2\}$] with left and right conformal
weights $(\frac{3}{16},\frac{3}{16})$ and 
the $3\times 3$ matrix field $\Phi_{mn}$ [$m,n\in \{1,2,3\}$] with
weights $(\frac{1}{2},\frac{1}{2})$. 
This WZNW model is identical to three copies of massless Majorana 
(real) fermions, 
as the WZNW model has $c=3/2$ and the Majorana fermion theory, 
which is equivalent to a 2D critical Ising model, has $c=1/2$. 
This identification allows a discussion of low-energy properties of the
TB chain using the $c=1/2$ CFT, instead of the WZNW model.  
In imaginary time formalism where 
$\tau=i t$ ($t$ is real time), we have the fermionic Euclidean 
action for the TB chain, 
\begin{eqnarray} 
\label{eq26}
S_{\text{TB}}&=& \int d\tau dx\,\,\, 
v[\xi_{L}^a\partial_{\bar{z}}\xi_{L}^a+\xi_{R}^a\partial_z\xi_{R}^a],
\end{eqnarray}
where $\xi_{L}^a(z)$ and $\xi_{R}^a(\bar{z})$ [$a=1,2,3$] 
are, respectively, the left mover of a Majorana fermion with
weights $(\frac{1}{2},0)$ and the right mover with $(0,\frac{1}{2})$. 
The velocity of the excitations $v$ is the order of $Ja_0$.~\cite{A-M,Aff3}
Here, we introduced the ``light-cone'' coordinate:
$z=v\tau+ix$, $\bar{z}=v\tau-ix$, 
$\partial_z=\frac{1}{2}(\frac{1}{v}\partial_\tau -i\partial_x)$, and  
$\partial_{\bar{z}}=\frac{1}{2}(\frac{1}{v}\partial_\tau +i\partial_x)$.
The repeated indices are summed.
In the operator formalism, the fermions satisfy 
the following equal-time anticommutation relations: 
$\{\xi_\alpha^a(x),\xi_\beta^b(y)\}
=\delta_{ab}\delta_{\alpha\beta}\delta(x-y)$.  
On the top of the action, there exist correspondences among the
WZNW primary fields and fields of the $c=1/2$ CFT. 
In accordance with Refs.~\onlinecite{Zam-Fat}, \onlinecite{Kita} and
\onlinecite{Allen},
$\Phi_{mn}(z,\bar{z})$ comprises a bilinear form of 
$\xi_{L}^a(z)$ and $\xi_{R}^a(\bar{z})$, which indeed has the weights 
$(\frac{1}{2},\frac{1}{2})$. In addition, from Ref.~\onlinecite{Allen}, 
we have 
\begin{eqnarray} 
\label{eq28}
g_{mn}(z,\bar{z})\propto\sum_{\alpha=0}^3 (\tau_\alpha)_{mn} 
{\cal G}_\alpha(z,\bar{z}),
\end{eqnarray}
where $(\tau_0)_{mn}=\delta_{mn}$ and $\tau_{1,2,3}$ are the Pauli
matrices. The fields ${\cal G}_\alpha(z,\bar{z})$ are determined as
\begin{eqnarray} 
\label{eq29}
{\cal G}_0=\sigma_1\sigma_2\sigma_3, &&
{\cal G}_a=i\sigma_a\mu_{a+1}\mu_{a+2}, 
\end{eqnarray}
where $\sigma_a(z,\bar{z})$ and $\mu_a(z,\bar{z})$ are, 
respectively, the order and disorder fields in the critical 
Ising model ($c=1/2$ CFT). 
Here, we use the cyclic index $a+3=a$. 
Both fields $\sigma_a$ and $\mu_a$ have weights
$(\frac{1}{16},\frac{1}{16})$. An imaginary unit $i$ is embedded in
Eq.~(\ref{eq29}) to let the field $g$ be a SU(2) matrix. The SU(2) 
current operators $J_L^a(z)$ and $J_R^a(\bar{z})$ in the WZNW model 
can be defined by fermions as follows: 
\begin{eqnarray} 
\label{eq30}
J_L^a(z)=-\frac{i}{2}\epsilon_{abc}\xi_L^b\xi_L^c, &&
J_R^a(\bar{z})=-\frac{i}{2}\epsilon_{abc}\xi_R^b\xi_R^c ,
\end{eqnarray}
where $\epsilon_{abc}$ is the totally antisymmetric tensor and
$\epsilon_{123}=1$.
The currents $J_{L,R}^a(x)$ satisfy level-$2$ SU(2) Kac-Moody algebra.
Through the NAB,~\cite{Aff,A-H,C-P-R} the spin operator in the TB chain 
is translated into the following sum of the uniform and staggered parts: 
\begin{eqnarray} 
\label{eq32}
\frac{1}{a_0}S_j^a &\approx& 
J_L^a+J_R^a +i C_0 (-1)^j \text{Tr}[(g-g^{\dag})\tau_a]\nonumber\\
&=&  J_L^a+J_R^a +C_1 (-1)^j \sigma_a\mu_{a+1}\mu_{a+2},
\end{eqnarray}
where both $C_0$ and $C_1$ are nonuniversal constants. 
On the left-hand side, $a=1$, $2$, and $3$ correspond to $x$, $y$ and
$z$, respectively. This formula connects smoothly with the bosonized 
spin density of the spin-$\frac{1}{2}$ ladder.~\cite{Sel} 
From Eq.~(\ref{eq32}), one notes that the one-site translation causes
$g(x)\rightarrow -g(x+a_0)$.~\cite{Aff,A-H}

Here we must mention a subtle point. The OPEs in Appendix~\ref{app2-2}
show that a disorder field $\mu_a$ has an anticommuting
character (so far we implicitly think of it as a bosonic object). 
One solution to maintain it and the Hermitian property of 
the staggered part of the spin density (\ref{eq32}) is to 
modify the staggered part as 
\begin{eqnarray} 
\label{eq32-2}
\sigma_a\mu_{a+1}\mu_{a+2}&\rightarrow& \kappa\,\,\sigma_a\mu_{a+1}\mu_{a+2},
\end{eqnarray}
where the new parameter $\kappa$ has the same properties
as an imaginary unit: $\kappa^{*}=-\kappa$ and $\kappa^2=-1$.
We will sometimes use this modification below.

Making full use of the above relations, one obtains 
low-energy properties of the TB chain. 
In order to achieve the field theory 
for the Heisenberg chain beyond the TB chain, 
one must add the following two perturbation terms to the action
(\ref{eq26}).
\begin{eqnarray} 
\label{eq33} 
i m\xi_L^a\xi_R^a - \lambda J_L^a J_R^a
\end{eqnarray}
As long as the attached terms are restricted to relevant or marginal ones, 
only these two terms are admitted and possess spin rotational 
(see Appendix~\ref{app4}) and one-site translational symmetries.~\cite{note11}
From the forms of Eqs.~(\ref{eq32}), (\ref{eq32-2}), (\ref{eq33}) 
and the Hamiltonian for the TB action (\ref{eq26}),  
we can infer that the time reversal transformation 
$(S_j^a,i)\rightarrow (-S_j^a,-i)$ is mapped to 
$(\xi_L^a,\xi_R^a,\kappa,i)\rightarrow (\xi_R^a,\xi_L^a,-\kappa,-i)$.
Similarly, we infer that link-parity transformation 
$S_j^a\rightarrow S_{-j+1}^a$ and site-parity one 
$S_j^a\rightarrow S_{-j}^a$ ($S_0^a$ is fixed) correspond to    
$[\xi_L^a(x),\xi_R^a(x),g(x)]\rightarrow 
[\mp\xi_R^a(-x),\pm \xi_L^a(-x),g(-x)]$ and 
$[\xi_L^a(x),\xi_R^a(x),g(x)]\rightarrow 
[\mp\xi_R^a(-x),\pm \xi_L^a(-x),-g(-x)]$, respectively.~\cite{note11'}

Because the Heisenberg point ($\delta=0$) is far from 
the TB point ($\delta=-1$), $m$ and $\lambda$ are phenomenological parameters. 
It is known that one may take $m>0$ and $\lambda >0$ 
in the Haldane phase (Fig.~\ref{fig8}).~\cite{Tsv} 
The inequality $\lambda >0$ means that the term $- \lambda J_L^a J_R^a$
is marginally irrelevant. As shown in Eq.~(\ref{b5}), 
the Ising model picture tells us that $m>0$ indicates that each Ising
model is in the disordered phase $\langle \mu_a\rangle\neq 0$.
The mass parameter $m$ must contribute to the Haldane gap. 
Consequently, three bands built of $\xi_L^a$ and $\xi_R^a$ 
can be regarded as the spin-$1$ magnon modes in the chain
(\ref{eq24}). Running from $\delta=-1$ to $\delta=0$ allows the velocity
$v$ to be renormalized. However, it still has the order of $Ja_0$.~\cite{G-J-S,Taka} 
We will use the same symbol $v$ for the renormalized velocity. 
In addition to $v$, for the other parameters ($m$,
$\lambda$, etc.), hereafter we use the same symbols whether they contain
any renormalization effects or not. 
It is widely believed that the formula (\ref{eq32}) is applicable
even for the Heisenberg chain because its low-energy excitations 
still stay around the uniform point $k=0$ and the staggered one 
$\pi/a_0$ (see the previous subsection).

Heretofore, we have obtained an effective field theory for the spin-$1$ 
Heisenberg chain. This framework was first proposed by 
Tsvelik~\cite{Tsv} in 1990. 
Utilizing Eq.~(\ref{eq32}), we easily obtain 
the field theory for the spin-$1$ ladder (\ref{eq1}) without external
fields. The action is 
\begin{eqnarray} 
\label{eq34}
S_{\text{lad}}&=&S_{\text{TB}}[\xi_L^a,\xi_R^a]
+S_{\text{TB}}[\tilde{\xi}_L^a,\tilde{\xi}_R^a]\nonumber\\
&&+\int d\tau dx\Big[i m\xi_L^a\xi_R^a - \lambda J_L^a J_R^a 
+i m\tilde{\xi}_L^a\tilde{\xi}_R^a 
- \lambda \tilde{J}_L^a \tilde{J}_R^a \nonumber\\
&& + J_{\perp}a_0 (J_L^a\tilde{J}_R^a+J_R^a\tilde{J}_L^a
+J_L^a\tilde{J}_L^a +J_R^a\tilde{J}_R^a)\nonumber\\
&& + C_1^2 J_{\perp} a_0\,\,\kappa\tilde{\kappa}\sigma_a\mu_{a+1}\mu_{a+2}
\tilde{\sigma}_a\tilde{\mu}_{a+1}\tilde{\mu}_{a+2}\Big],
\end{eqnarray}
where quantities without (with) an overtilde
$\tilde{\,\,}$ represent fields of the chain $1$ (chain $2$) with $l=1$ ($l=2$).
We stress that the Hamiltonian for the action $S_{\text{lad}}$ 
is invariant under spin rotation, one-site translation, time reversal, 
two parity transformations and exchanging the chain indices.  

On top of the rung-coupling term, 
it is possible to translate the other terms into 
field-theoretical expressions. 
The uniform Zeeman term (\ref{eq2-1}) and the staggered one 
(\ref{eq2-2}) are, respectively, mapped to 
\begin{subequations}
\label{eq35}
\begin{eqnarray}
\approx &&
-H \int dx\,(J_L^3+J_R^3+\tilde{J}_L^3+\tilde{J}_R^3),\label{eq35-1}\\
\approx &&    -C_1 H \int dx\,
(\kappa\sigma_3\mu_1\mu_2 +
\tilde{\kappa}\tilde{\sigma}_3\tilde{\mu}_1\tilde{\mu}_2).\label{eq35-2}
\end{eqnarray}
\end{subequations}
An advantage of the field theory used here is that 
the uniform Zeeman term is translated
into the fermionic quadratic form (\ref{eq35-1}), which can be treated
nonperturbatively. 
Equation (\ref{eq35-2}) is not invariant under time reversal and
one-site translational operations. 

Other considerations regarding symmetries between the spin-$1$ ladder and
its effective theory (\ref{eq34}) are found in Ref.~\onlinecite{Allen}.

\subsection{\label{sec3-3} Uniform-field case}
Sections~\ref{sec3-1} and \ref{sec3-2} complete the main preparation
dealing with the spin-$1$ ladder (\ref{eq1}). This paragraph presents
a discussion of the ladder with a uniform Zeeman term (\ref{eq2-1}). 
We clarify what critical phase emerges when the uniform field is
applied. It is easy to infer that a weak rung coupling does
not collapse the Haldane gap of two decoupled spin-$1$ AF chains 
(see the Introduction and the next subsection). 
Moreover, in the single chain, 
a sufficiently strong uniform field engenders a $c=1$ critical 
state.~\cite{Tsv,S-T,Aff2,S-A,K-F,Fath}
Therefore, we take the following strategy. (\rnum{1}) We review the effective
theory for the $c=1$ critical phase appearing 
in the spin-$1$ chain with a strong uniform field.~\cite{Tsv} 
(\rnum{2}) Then, adding the rung coupling terms perturbatively, 
we investigate the low-energy physics of the ladder (\ref{eq1}) 
with a strong uniform field.
  
\begin{figure}
\scalebox{0.4}{\includegraphics{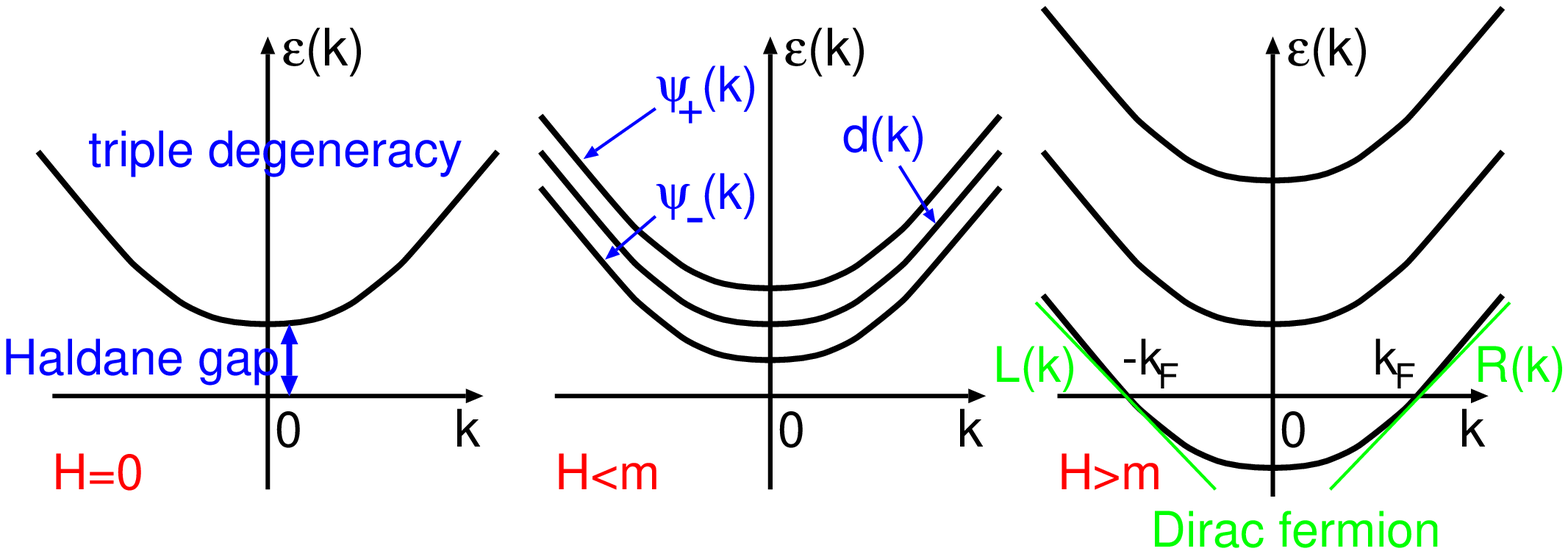}}
\caption{Band structures of the spin-$1$ AF chain with a uniform field $H$.}
\label{fig9}
\end{figure}

Following the above scenario, first we explain how the $c=1$ state is
described within the field-theoretical scheme. 
We neglect the four-body interaction, the $\lambda$ term. 
One may interpret that it vanishes via the RG procedure.
Actually, it is believed that the effective theory without the 
$\lambda$ term is sufficient to describe the low-energy physics of the
Heisenberg chain.~\cite{Tsv} In this case, the Hamiltonian for the chain
$1$ is 
\begin{eqnarray}
\label{eq36}
\hat {\cal H}_{\text{chain1}}&=&\int dx \,\, iv
(\psi_L^{\dag}\partial_x\psi_L-\psi_R^{\dag}\partial_x\psi_R)\nonumber\\
&&+i m(\psi_L^{\dag}\psi_R-\psi_R^{\dag}\psi_L)
+H(\psi_R^{\dag}\psi_R+\psi_L^{\dag}\psi_L)\nonumber\\
&& +i \frac{v}{2}(\xi_L^3\partial_x\xi_L^3 - \xi_R^3\partial_x\xi_R^3)
+im \xi_L^3\xi_R^3, 
\end{eqnarray}
where, for convenience, we introduced a Dirac (complex) fermion, 
\begin{eqnarray}
\label{eq37}
\left(
\begin{array}{lll}
\psi_R\\
\psi_L
\end{array}
\right)
&=&\frac{1}{\sqrt{2}}
\left(
\begin{array}{lll}
\xi_R^1+i\,\xi_R^2\\
\xi_L^1+i\,\xi_L^2
\end{array}
\right).
\end{eqnarray}
The Hamiltonian (\ref{eq36}) is a quadratic form. 
The uniform field mixes two species $\xi_{L,R}^1$ and $\xi_{L,R}^2$,
while it does not affect  $\xi_{L,R}^3$. Through Fourier transformations 
$\psi_{L,R}(x)=\int_{k:\text{all}} \frac{dk}{2\pi}
e^{ikx}\psi_{L,R}(k)$ and 
$\xi_{L,R}^3(x)=\int_{k>0} \frac{dk}{2\pi}\{e^{ikx}c_{L,R}(k)
+e^{-ikx}c_{L,R}^{\dag}(k)\}$ [${\xi_{L,R}^3}^{\dag}(x)=\xi_{L,R}^3(x)$], 
and Bogoliubov transformations,
\begin{eqnarray}
\label{eq38} 
\left(
\begin{array}{lll}
\psi_R(k)\\
\psi_L(k)
\end{array}
\right)
&=&{\cal U}(k)
\left(
\begin{array}{lll}
\psi_+(k)\\
\psi_-^{\dag}(k)
\end{array}
\right),\nonumber\\
\left(
\begin{array}{lll}
c_R(k)\\
c_L(k)
\end{array}
\right)
&=&{\cal U}_3(k)
\left(
\begin{array}{lll}
d(k)\\
d^{\dag}(-k)
\end{array}
\right),
\end{eqnarray}
where ${\cal U}(k)_{11,12}=m/[2\epsilon(k)(\epsilon(k)\mp kv)]^{1/2}$, 
${\cal U}(k)_{21,22}=\pm i[\epsilon(k)\mp kv]^{1/2}/[2\epsilon(k)]^{1/2}$, 
${\cal U}_3(k)_{11,12}={\cal U}(k)_{11,12}$, 
${\cal U}_3(k)_{21,22}=i [\epsilon(k)\mp kv]^{1/2}/[2\epsilon(k)]^{1/2}$ and  
$\epsilon(k)=[(kv)^2+m^2]^{1/2}$, 
we obtain the diagonalized Hamiltonian 
\begin{eqnarray}
\label{eq39-0} 
\hat {\cal H}_{\text{chain1}}=\int \frac{dk}{2\pi}\,
[\epsilon_+(k)\psi_+^{\dag}(k)\psi_+(k)\nonumber\\
+\epsilon_-(k)\psi_-^{\dag}(k)\psi_-(k)
+\epsilon(k)d^{\dag}(k)d(k)] + \text{const},
\end{eqnarray}
where $\epsilon_{\pm}(k)=\epsilon(k)\pm H$. The three bands 
$\epsilon(k)$ and $\epsilon_{\pm}(k)$ reproduce the Zeeman splitting 
of the spin-$1$ magnon modes. Their structures are given in Fig.~\ref{fig9}.
When the uniform field exceeds a critical value $m$, a Fermi surface
appears in the lowest band $\epsilon_-(k)$ and its low-energy 
excitations then can be captured as a massless Dirac fermion. This 
corresponds precisely to the $c=1$ state we are looking for.
Here, the mass parameter $m$, after including all the renormalization
effects, can be identified with the Haldane gap. 

Now, in this strong-field region ($H>m$), 
let us recover the $\lambda$ term, and take into account 
the rung coupling. Provided that we focus on low-energy physics,
expanding the modes $\psi_+(x)$ and $\psi_-(x)$, respectively, 
around $k=0$ and the Fermi momentum $k=k_F=(H^2-m^2)^{1/2}/v$ will be
allowed. Therefore, we obtain  
\begin{eqnarray}
\label{eq39} 
\psi_R(x) &\approx&
\frac{1}{\sqrt{2}}\psi(x)+e^{ik_F x}U_+ R^{\dag}(-x)\nonumber\\
&&+e^{-ik_F x}U_- L^{\dag}(-x),\nonumber\\
\psi_L(x) &\approx&
\frac{i}{\sqrt{2}}\psi(x)-i e^{ik_F x}U_- R^{\dag}(-x)\nonumber\\
&&-i e^{-ik_F x} U_+ L^{\dag}(-x),
\end{eqnarray}
where $U_\pm ={\cal U}(\pm k_F )_{12}
=m/\sqrt{2H [H+(H^2\pm m^2)^{1/2}]}$, $R(x)$ and $L(x)$ are,
respectively, the left and right movers of the Dirac fermion 
[$\psi_-(x)\approx e^{ik_F x}R(x)+e^{-ik_F x}L(x)$]. 
In addition, we defined the field $\psi(x)$ as 
$\int\frac{dk}{2\pi}\epsilon_+(k)\psi_+^{\dag}(k)\psi_+(k)\approx 
\int dx \psi^{\dag}(-\frac{v^2}{2m}\partial_x^2+m+H)\psi$.
Similarly, we introduce fields $\tilde{L}$, $\tilde{R}$ and
$\tilde{\psi}$ in chain $2$.
Substituting these new fields for the effective field theory
($\hat {\cal H}_{\text{chain1}}+\hat {\cal H}_{\text{chain2}}
+\lambda\,\,\text{term}+ \text{rung\,\,coupling}$), we obtain the 
effective Hamiltonian (\ref{c1}) under the condition $H>m$. 
(Because it is too lengthy, the explicit form is put in
Appendix~\ref{app3}.) In the process, using Eqs.~(\ref{eq39}), 
we carefully approximated mass terms, interaction ones and currents
$J_{L,R}^a$ while conserving the Hermitian property of each term. 
Moreover, we dumped all terms possessing some rapid fluctuation factors 
$\exp(\pm in k_F x)$ ($n$ is an integer).
The action toward the Hamiltonian (\ref{c1}) is written as
\begin{eqnarray}
\label{eq40} 
S_{H>m}&=& S_0^1[L,R,\tilde{L},\tilde{R}] 
+S_0^2[\xi_\alpha^3,\tilde{\xi}_\alpha^3]+S_0^3[\psi,\tilde{\psi}]\nonumber\\
&&+S_{\text{int}}^1[L,R,\tilde{L},\tilde{R}]
+S_{\text{int}}^2[L,R,\tilde{L},\tilde{R};
\xi_\alpha^3,\tilde{\xi}_\alpha^3]\nonumber\\
&&+S_{\text{int}}^3[L,R,\tilde{L},\tilde{R};\psi,\tilde{\psi}]\nonumber\\
&&+S_{\text{int}}^4[\xi_\alpha^3,\tilde{\xi}_\alpha^3;\psi,\tilde{\psi}]
+S_{\text{int}}^5[\psi,\tilde{\psi}]+S_{\text{stag}},
\end{eqnarray}
where $S_0^\alpha$, $S_{\text{int}}^\alpha$ and $S_{\text{stag}}$
denote the free part of each field, the four-body interaction terms 
and the staggered part of the rung-coupling term 
[i.e., the final term in Eq.~(\ref{eq34})], respectively. 
Integrating out massive fields 
$\xi_\alpha^3$, $\tilde{\xi}_\alpha^3$, $\psi$, and $\tilde{\psi}$
leads to the effective action containing only soft modes 
$L$, $R$, $\tilde{L}$, and $\tilde{R}$. 
Through a cumulant expansion, it can be expressed as
\begin{eqnarray}
\label{eq41} 
S_{\text{eff}}[L,R,\tilde{L},\tilde{R}]&=& 
S_0^1+S_{\rm int}^1+\langle \bar S_{\text{int}}\rangle_M \nonumber\\
&&+\frac{1}{2}[\langle \bar S_{\text{int}}^2\rangle_M 
-\langle \bar S_{\text{int}}\rangle_M^2]+\dots ,
\end{eqnarray}
where $\bar S_{\text{int}}=\sum_{\alpha=2}^5 S_{\text{int}}^\alpha
+S_{\text{stag}}$, and $\langle\cdots\rangle_M$ indicates 
the expectation value of free parts of four massive fermions. 
Because the Abelian bosonization is useful for interacting Dirac 
fermion models (see Appendix~\ref{app1}), 
we introduce the boson fields $\phi$ and $\tilde{\phi}$ 
from the Dirac fermions $(L,R)$ and $(\tilde{L},\tilde{R})$,
respectively. Applying the formula~(\ref{b9}) and the known results
of the 2-leg spin-$\frac{1}{2}$ ladder with a uniform field [refer
Eqs.~(29) and (32) in Ref.~\onlinecite{Fu-Zh}], we can bosonize the
Ising-field product of $S_{\rm{stag}}$ as well as the fermion fields.
As a result, $S_{\rm{stag}}$ is mapped as follows: 
\begin{eqnarray}
\label{eq-stag}
C_1^2 J_{\perp} a_0\kappa\tilde{\kappa}\sigma_a\mu_{a+1}\mu_{a+2}
\tilde{\sigma}_a\tilde{\mu}_{a+1}\tilde{\mu}_{a+2}\,\,\sim\hspace{2cm}
\nonumber\\
C_1^2 J_{\perp} a_0\big\{\mu_3\tilde\mu_3\cos[\sqrt{\pi}(\theta-\tilde\theta)]
+\frac{1}{2}\sigma_3\tilde\sigma_3\cos[\sqrt{\pi}(\phi-\tilde\phi)]\big\}
\nonumber\\
\,\,\,\,+\,\,\text{fluctuating or irrelevant terms},\,\,\,\,\,\,\,\,
\hspace{1cm}
\end{eqnarray}
where $\theta$ ($\tilde\theta$) is the dual of the boson field $\phi$
($\tilde\phi$), and the first (second) term on the right-hand side is
generated from the $x$ and $y$ ($z$) components of the rung-coupling 
staggered part. The result~(\ref{eq-stag}) is also supported by the NLSM
approach.~\cite{Aff2,K-F} The final fluctuating or irrelevant terms 
may be negligible in the low-energy limit. 
From Eqs.~(\ref{eq40})-(\ref{eq-stag}), up to the first cumulant, 
the effective action $S_{\text{eff}}$ yields the following bosonized Hamiltonian: 
\begin{eqnarray}
\label{eq42} 
\hat{\cal H}_{\text{1st}}&=&\int dx \Big[A\, \Pi_+^2
+B_+(\partial_x\phi_+)^2
+C\partial_x\phi_+  \\
&&+A\, \Pi_-^2+B_{-}(\partial_{x}\phi_-)^2\nonumber\\
&&+C_{\theta_-}\cos(\sqrt{2\pi}\theta_-)
+C_{\phi_-}\cos(\sqrt{8\pi}\phi_-)\Big],\nonumber
\end{eqnarray}
where we define the symmetric boson field 
$\phi_+ =(\phi + \tilde{\phi})/\sqrt{2}$ and the antisymmetric one
$\phi_- =(\phi - \tilde{\phi})/\sqrt{2}$, and 
$\Pi_\pm=\partial_t\phi_\pm/v'$ ($\theta_\pm$) is the canonical
conjugate (dual) of the boson $\phi_\pm$. 
New parameters in $\hat{\cal H}_{\text{1st}}$ are
\begin{subequations}
\label{eq43}
\begin{eqnarray}
A  &=& \frac{v'}{2} -\frac{\lambda}{2\pi}U_+^2 U_-^2
+\frac{\lambda}{4\pi}(U_+^2- U_-^2)^2 , \label{eq43-1}\\
B_{\pm} &=& \frac{v'}{2} -\frac{\lambda}{2\pi}U_+^2 U_-^2
-\frac{\lambda}{4\pi}(U_+^2- U_-^2)^2  \nonumber\\
&&\pm \frac{J_{\perp}a_0}{2\pi}(U_+^2+U_-^2)^2,\label{eq43-2}\\
C&=& \sqrt{\frac{2}{\pi}}\Big[
2\lambda\delta_\xi U_+ U_- -\frac{\lambda}{4}\delta_\psi(U_+ + U_-)^2
\nonumber\\
&&+\frac{J_{\perp}a_0}{2}(U_+^2 + U_-^2)\Big],\label{eq43-3}\\
C_{\phi_-} &=& 2\frac{J_{\perp}a_0}{\pi^2\alpha^2}U_+^2 U_-^2,\label{eq43-4}\\
C_{\theta_-} &\propto& C_1^2J_\perp a_0 \langle\mu_3\rangle_M^2,\label{eq43-5}
\end{eqnarray}
\end{subequations}
where $\alpha$ in Eq.~(\ref{eq43-4}) is the cut-off parameter in 
the bosonization formula (\ref{a3}),  
$\delta_\xi=i\langle \xi_L^3\xi_R^3\rangle_M
=i\langle \tilde{\xi}_L^3\tilde{\xi}_R^3\rangle_M>0$ 
and $\delta_\psi=\langle \psi\psi^{\dag}\rangle_M
=\langle \tilde{\psi}\tilde{\psi}^{\dag}\rangle_M>0$.
Both $\delta_\xi$ and $\delta_\psi$ are $O(1/a_0)$.
Except for these two, the averages of products of massive fermions 
vanish in the first cumulant $\langle S_{\text{int}}\rangle_M$. 
One should note 
$\langle\mu_3\rangle_M=\langle\tilde\mu_3\rangle_M \neq 0$ and 
$\langle\sigma_3\rangle_M=\langle\tilde\sigma_3\rangle_M =0$. 
In the derivation of Eq.~(\ref{eq42}), for simplicity, we assumed
that the Dirac fermion bands are always half-filled. From this,
for example, we employed the relation 
$L^{\dag}L-LL^{\dag}=2L^{\dag}L-\delta_0=2:L^{\dag}L:$, where the order
of $\delta_0$ is the inverse of the fermion wave-number cut off, 
and the symbol $:\,\,:$ means the normal-ordered product. 
Observing the Hamiltonian (\ref{c1}) carefully and using the operator
product expansion (OPE) in the $c=1$ and $c=1/2$ CFTs (Appendix~\ref{app2-2}), 
we find that the second cumulant yields new interaction 
terms $\cos(\sqrt{8\pi} \theta_-)$, 
$\cos(2\sqrt{8\pi}\phi_{-,L(R)})$ and 
$\cos[\sqrt{2\pi}(\phi_{-,L(R)}+3\phi_{-,R(L)})]$ from 
$\langle (S_{\text{int}}^2)^2 \rangle_M$ and 
$\langle S_{\text{int}}^2 S_{\rm{stag}}\rangle_M$.
[Here, $\phi_{\pm,L(R)}$ is the left (right) mover of $\phi_\pm$.] 
This means that the second cumulant does not produce
any vertex operators having the symmetric bosons $\phi_+$, $\theta_+$ 
and $\phi_{+,L(R)}$. To verify this result, 
it is sufficient to note that the presence of 
$\exp(i\sqrt{8\pi}\phi_+)$, $\exp(i\sqrt{8\pi}\theta_+)$, 
$\exp(i2\sqrt{8\pi}\phi_{+,L})$ and $\exp(i2\sqrt{8\pi}\phi_{+,R})$ 
requires, respectively, fermion four-body terms 
$L^{\dag}R\tilde{L}^{\dag}\tilde{R}$,
$L^{\dag}R^{\dag}\tilde{L}^{\dag}\tilde{R}^{\dag}$, 
$L^{\dag}L^{\dag}\tilde{L}^{\dag}\tilde{L}^{\dag}$ and 
$R R\tilde{R}\tilde{R}$.
It is expected that except for the above vertex terms, 
relevant or marginal interactions do not emerge in the higher
cumulants.

Besides the discussion related to the explicit counting of vertex operators, 
the Hamiltonian (\ref{c1}) and the cumulant expansion 
have the following four remarkable points. 
(\rnum{1}) Above vertex operators corresponding to some four-body fermions 
occur only through the rung coupling (meaning that the $\lambda$ term 
does not violate the $c=1$ phase in the decoupled chain). 
(\rnum{2}) We can apply an argument in Ref.~\onlinecite{OYA}, which
cleverly employs the bosonization and symmetries of spin systems.
From Appendix~\ref{app4}, the U(1) transformation associated with the spin
rotation around $z$ axis is given by $\psi_{L,R}\rightarrow 
\psi_{L,R}e^{i\varphi}$ ($\varphi$ is a real number) in the field
theory, which accompanies $L^{\dag}\rightarrow L^{\dag}e^{i\varphi}$, 
$R^{\dag}\rightarrow R^{\dag}e^{i\varphi}$ and 
$\psi\rightarrow \psi e^{i\varphi}$ via Eq.~(\ref{eq39}). 
Of course, fields with an overtilde $\tilde{\,\,}$ 
also receive the same transformations.  
In the boson language, they correspond to a shift of 
the symmetric dual field $\theta_+\rightarrow \theta_+ +\text{constant}$.
This U(1) symmetry hence prohibits the emergence of 
any vertex operators with the dual field $\theta_+$.
Similarly, let us consider the one-site translation.
As one can see from Eq.~(\ref{eq39}), it causes 
$L^{\dag}(-x)\rightarrow L^{\dag}(-x-a_0)e^{-ik_F a_0}$ and 
$R^{\dag}(-x)\rightarrow R^{\dag}(-x-a_0)e^{ik_F a_0}$.
They are mapped to 
$\phi_+(x)\rightarrow \phi_+(x+a_0)+2\sqrt{2}k_F a_0/r$, where $r$ is
the compactification radius in the $c=1$ theory considered now. 
It must be close to the value of the massless Dirac fermion $1/\sqrt{4\pi}$ 
(see Appendix~\ref{app1}). 
Vertex operators with $\phi_+$ are therefore forbidden, 
except for those where $k_F$ becomes some
special value. These prohibition rules are actually formed 
in the above counting of vertex operators. 
One can confirm that U(1) and one-site translational symmetries 
are maintained in the Hamiltonian (\ref{c1}).
(\rnum{3}) Because the correlation functions 
of massive fields decay exponentially, 
no long-range interaction terms emerge in all the cumulants.
For instance, the correlation lengths of 
two-point functions for $\xi_\alpha^3$ or $\psi$ are at most
the order of $(ma_0)^{-1}\times a_0$, where $m$ has 
the order of the Haldane gap ($\cong 0.41J$),~\cite{Kato} 
and $ma_0$ ($\sim Ja_0$) must be a constant in the present scaling limit. 
In fact, the correlation length of the spin-$1$
AF Heisenberg chain is only about six times as large as
$a_0$.~\cite{Kato} The connected correlation functions of Ising fields
$\sigma_3$ and $\mu_3$ also decay exponentially.~\cite{Wu-Mc}
(\rnum{4}) From (\rnum{3}), roughly speaking, 
the expansion can be thought of as a $J_{\perp}/J$ expansion.~\cite{Wang}

From all the considerations below Eq.~(\ref{eq42}), when the
rung coupling is sufficiently weak, the bosonized Hamiltonian
(\ref{eq42}), possibly with vertex operators of antisymmetric fields
only, could be adopted as an effective theory under a strong uniform
field $H(>m)$.  
Following a standard prescription, we perform a
Bogoliubov transformation,
\begin{eqnarray}
\label{eq44}
\phi'_\pm= \frac{1}{\sqrt{K_\pm}}\,\phi_\pm, && \Pi'_\pm= \sqrt{K_\pm}\,\Pi_\pm,  
\end{eqnarray}
where $K_\pm=\sqrt{A/B_\pm}$, and the canonical relations are conserved: 
$[\phi_\pm(x),\Pi_\pm(y)]=[\phi'_\pm(x),\Pi'_\pm(y)]=i\delta(x-y)$. 
From the view of new boson fields $(\phi'_\pm,\Pi'_\pm)$, 
interaction terms,
\begin{subequations}
\label{eq45}
\begin{eqnarray}
\cos(\sqrt{8\pi}\phi_-)&=&\cos(\sqrt{8\pi K_-}\phi'_-),\label{eq45-1}\\
\cos(\sqrt{2\pi}\theta_-)&=&\cos\Big(\sqrt{\frac{2\pi}{K_-}}\theta'_-\Big),\label{eq45-2}\\
\cos(\sqrt{8\pi}\theta_-)&=&\cos\Big(\sqrt{\frac{8\pi}{K_-}}\theta'_-\Big),\label{eq45-3}
\end{eqnarray}
\end{subequations}
have the scaling dimensions $2K_-$,$1/(2K_-)$ and $2/K_-$,
respectively. The parameters $J_\perp$ and $\lambda$ 
yield a deviation from $K_\pm=1$. It, however, would be quite small in
the weak rung-coupling region. The most relevant term, hence, is 
$\cos(\sqrt{2\pi/K_-}\theta'_-)$, and it locks the phase field
$\theta'_-$. Therefore, the $\phi'_-$ mode 
always becomes massive once the rung coupling $J_\perp$ enters into the system. 
Other interactions $\cos(2\sqrt{8\pi}\phi_{-,L(R)})$ and 
$\cos[\sqrt{2\pi}(\phi_{-,L(R)}+3\phi_{-,R(L)})]$
have a conformal spin, and it is reported
that such fields may engender non-trivial effects.~\cite{Gogo,Ner} 
However, they would not be as powerful as recovering any massless 
$\phi'_-$ modes in the present case. On the other hand, for the symmetric part, 
the linear term $C\partial_x\phi_+$ is absorbed into the quadratic part
by the shift $\phi_+\to \phi_+ +C x/(2 B_+)$. 
From these results, we conclude that for a strong uniform field ($H> m$), 
only the $\phi'_+$ mode remains massless and a $c=1$ phase is
realized, irrespective of the sign of $J_{\perp}$.

Finally, we note the limitations and the validity of the methods used here.
The fermion band width in the effective theory (\ref{eq36}) is expected
to be smaller than or equal to $O(J)$. 
Hence, when the uniform field becomes too strong, such as $H\agt J$, it
is doubtful whether the theory (\ref{eq36}) is valid or not. 
Furthermore, for such a strong-field case, it may be necessary to take into
account other low-energy excitations.  
Meanwhile, the cumulant expansion is not reliable well 
when $|J_\perp|$ reaches $O(J)$. 
For the derivation of Eq.~(\ref{eq42}), 
we used the assumption that the Dirac fermion bands is half-filled.
Removing it does not influence the main results presented in this paragraph.
It merely changes parameters in Eq.~(\ref{eq42}) a bit.

\subsection{\label{sec3-3-A} GS phase diagram of the uniform-field case}
The strategy of the preceding subsection is not suitable for determining
the lower critical uniform field, while we have already known that 
the lowest-excitation-gap profile of
the spin-1 ladder is given by Fig.~\ref{fig11}. Furthermore, the lowest
excitations must consist of a spin-1 magnon triplet [this expectation is
trusted at least for the strong AF or FM rung-coupling regions. 
Moreover, from our effective theory (excitations of fermions
$\xi_{L,R}^a$ are interpreted as spin-1 magnon excitations) and
the NLSM analysis,~\cite{Sene} it would be also true for 
the weak rung-coupling region.]
Therefore, the gap profile in Fig.~\ref{fig11} itself is equivalent to
the shape of the lower critical uniform field in the space
$(J_\perp,H)$. [As $H$ is increased, one of the spin-1 magnon bands goes
down linearly with $H$, as a result of the Zeeman splitting. See the band
$\epsilon_-(k)$.]
   
\begin{figure}
\scalebox{0.35}{\includegraphics{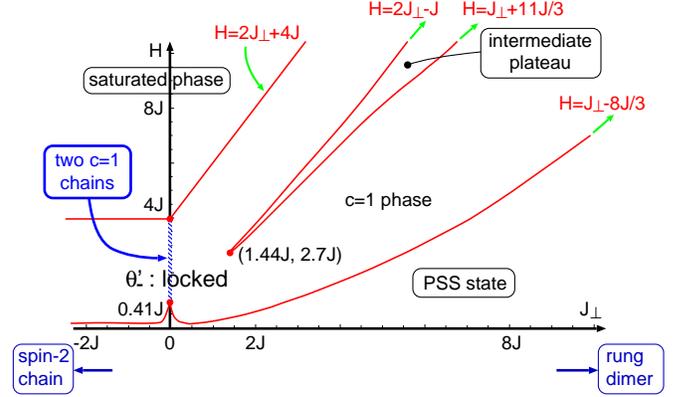}}
\caption{Schematic GS phase diagram of the spin-$1$ AF ladder
 (\ref{eq1}) with the uniform Zeeman term (\ref{eq2-1}).}
\label{fig12}
\end{figure}

Taking into account the above lower critical field and predictions 
in Secs.~\ref{sec2} and \ref{sec3-3}, 
we can finally draw the whole GS phase diagram of the spin-$1$ AF ladder
(\ref{eq1}) with the uniform field as in Fig.~\ref{fig12}.
The sharp form of the intermediate plateau area is one of the
natures of the BKT transition: the correlation length outside of the
critical phase in the BKT transition is still anomalously long, and it 
inversely indicates that the excitation gap, which is proportional to
the width of the plateau in the present case, grows considerably slowly. 
Features of the spin-$1$ GS phase diagram are the phase
boundary near the decoupled point and the existence of the
intermediate plateau. On the other hand, the $c=1$ universality in the
critical phase is common to both spin-$\frac{1}{2}$ and spin-$1$ cases. 

\subsection{\label{sec3-4} No-field case}
This subsection specifically addresses situations in which 
spin-$1$ ladders have no external fields.

\subsubsection{\label{sec3-4-1} String order parameters}
Here, we attempt to evaluate string-type order parameters in spin-$1$ 
systems within the field-theoretical description. 

First, we investigate that of the single spin-$1$ chain, Eq.~(\ref{eq25}).
The estimation of the nonlocal part 
$\hat {\cal O}_{\text{ex}}^a \equiv \exp(i\pi\sum_{n=j+1}^{k-1}S_n^a)$ 
(note $\hat {\cal O}_{\text{ex}}^{a\,\,\dag}=\hat {\cal O}_{\text{ex}}^a$)
can be done similarly to the bosonization technique in
Ref.~\onlinecite{Kim}, where the string parameters of spin-$\frac{1}{2}$ ladders
(\ref{eq3}) were calculated.
In the continuum limit, $\hat {\cal O}_{\text{ex}}^a$ is approximated as
\begin{eqnarray}
\label{eq51}
\hat {\cal O}_{\text{ex}}^a\approx
\exp\Big(\pm i\pi\int_{x_j+a_0}^{x_k-a_0} dy\,\,\big[J_L^a(y)+J_R^a(y)\big]\Big),
\end{eqnarray}
where $x_{k(j)}=k(j)\times a_0$ and the staggered part of the spin
density is dropped. Constructing the boson theory with the scalar field 
$\phi$ from two Ising systems $(\sigma_{a+1},\sigma_{a+2})$, we can
translate Eq.~(\ref{eq51}) as
\begin{eqnarray}
\label{eq54}
\hat {\cal O}_{\text{ex}}^a &\approx&\exp\Big(\mp i\sqrt{\pi}
\int dy\partial_y\phi(y)\Big)\\
&\sim & \exp\left[\mp i\sqrt{\pi}(\phi(x_k)-\phi(x_j))\right]\nonumber\\
&=& \left\{\cos(\sqrt{\pi}\phi(x_k))
\mp i \sin(\sqrt{\pi}\phi(x_k))\right\}\nonumber\\
&&\times\left\{\cos(\sqrt{\pi}\phi(x_j))
\pm i \sin(\sqrt{\pi}\phi(x_j))\right\}\nonumber\\
&\sim & 
\left\{\mu_{a+1}(x_k)\mu_{a+2}(x_k) \mp \sigma_{a+1}(x_k)\sigma_{a+2}(x_k)\right\}
\nonumber\\
&&\times\left\{\mu_{a+1}(x_j)\mu_{a+2}(x_j) 
\pm \sigma_{a+1}(x_j)\sigma_{a+2}(x_j)\right\},
\nonumber
\end{eqnarray}
where we used Eqs.~(\ref{a4}), (\ref{b7}), and (\ref{b9}).
The remaining problem is just two edge spins $S_j^a$ and $S_k^a$. 
[This problem does not appear in the calculation of Eqs.~(\ref{eq3}).]
At least in the Haldane phase where $\langle \sigma_a\rangle=0$, the
staggered parts of the edge spins could not contribute to ${\cal O}^a$.
While, the product of $\hat {\cal O}_{\text{ex}}^a$ and 
edge-spin uniform parts can be evaluated using OPE rules: 
$J_{L,R}^a\times \sigma_{a+1}\sigma_{a+2}\sim \mu_{a+1}\mu_{a+2}$
and $J_{L,R}^a\times \mu_{a+1}\mu_{a+2}\sim \sigma_{a+1}\sigma_{a+2}$.
Therefore, an expected field-theoretical form of ${\cal O}^a$ is
written as ${\cal O}^a\sim \langle\hat {\cal O}_{\text{ex}}^a\rangle 
\sim \langle\mu_{a+1}\mu_{a+2} \rangle^2
+\langle\sigma_{a+1}\sigma_{a+2} \rangle^2$. The derivation of 
this form, however, has some subtle aspects, which are mainly
attributable to the continuous-field (coarse-grained) scheme. 
A similar difficulty is also present
in the estimation of Eqs.~(\ref{eq3}).~\cite{Kim,Naka-Topo} 
To eliminate it, some ideas that are independent of field theories are necessary. 
Actually, Nakamura resolves such an ambiguity of the field-theoretical
expressions of Eqs.~(\ref{eq3}) using a symmetry cleverly.~\cite{Naka-Topo} 
For our spin-$1$ chain case, ${\cal O}^a \neq 0$ in the Haldane phase,
whereas ${\cal O}^a = 0$ in the dimerized phase (see Fig.~\ref{fig8}).
Counting on this fact, we can propose an appropriate form,
\begin{eqnarray}
\label{eq56}
{\cal O}^a &\sim& \langle \mu_{a+1}\mu_{a+2}\rangle^2,
\end{eqnarray}
which may imply that two edge spins follow the rule selecting the 
disorder-field portion from the exponential part 
$\hat {\cal O}_{\text{ex}}^a$. The formula (\ref{eq56}) has the same
form as the field-theoretical form of Eqs.~(\ref{eq3}). That similarity must
be one reflection of the fact that the FM-rung spin-$\frac{1}{2}$ AF ladder 
is tied with the spin-$1$ AF chain smoothly. 
Furthermore, it also reminds us that ${\cal O}^a$ is exactly mapped into 
an FM order parameter by a nonlocal unitary transformation.~\cite{Ko-Ta}

Our new proposal (\ref{eq56}) can also tell us the behavior of 
${\cal O}^a$ in the vicinity of the TB chain.
A scaling argument (or the exact solution for the 2D Ising
model) leads to $m\sim(\delta+1)$ and 
$\langle \mu_a\rangle\sim (\delta+1)^{1/8}$ near the chain. 
Therefore, we predict that the critical behavior,
\begin{eqnarray}
\label{eq56-1}
{\cal O}^a\sim(\delta+1)^{1/2},
\end{eqnarray}
occurs in the Haldane phase close to the TB chain. 

We next examine the spin-$1$ AF ladder. 
We denote two string order parameters of chains 1 and 2 
as ${\cal O}_1^a$ and ${\cal O}_2^a$, respectively.
In Ref.~\onlinecite{Todo}, the quantum Monte Carlo simulation shows
that: (\rnum{1}) a new string parameter 
$\langle \hat {\cal O}_1^a\hat {\cal O}_2^a\rangle$ 
is always finite for the AF-rung side; 
(\rnum{2}) the string order parameter of each single chain vanishes, 
once an AF rung coupling is attached in the system; 
and (\rnum{3}) in the AF-rung side, 
$\langle \hat {\cal O}_1^a\hat {\cal O}_2^a\rangle$ decreases until
$J_\perp\sim 0.4J$ and then grows monotonically until 
$J_\perp\to\infty$ (the rung dimer) (see Fig. 6 in
Ref.~\onlinecite{Todo}).
From these results, it is expected that 
the new string parameter is a quantity characterizing the PSS state.
We discuss how the field theories reproduce these, and what they can
predict. Supposing that Eq.~(\ref{eq56}) is applicable even for the spin-$1$
ladder, we have
\begin{subequations}
\label{eq56-2}
\begin{eqnarray}
\langle \hat {\cal O}_1^a\hat {\cal O}_2^a \rangle &\sim &
\langle\mu_{a+1}\tilde{\mu}_{a+1}\mu_{a+2}\tilde{\mu}_{a+2}\rangle^2\nonumber\\
&\sim&\langle\cos(\sqrt{\pi}\phi_{a+1})\cos(\sqrt{\pi}\phi_{a+2})\rangle^2
\label{eq56-2a}\\
&\sim&\langle\cos(\sqrt{\pi}\Phi)\cos(\sqrt{\pi}\tilde{\Phi})\rangle^2.
\label{eq56-2b}
\end{eqnarray}
\end{subequations}
In Eq.~(\ref{eq56-2a}), the boson $\phi_a$ is made from 
the two Ising systems $(\sigma_a,\tilde{\sigma}_a)$.
In Eq.~(\ref{eq56-2b}), bosons $\Phi$ and $\tilde{\Phi}$ are made from 
$(\sigma_{a+1},\sigma_{a+2})$ and
$(\tilde{\sigma}_{a+1},\tilde{\sigma}_{a+2})$, respectively.
Equations~(\ref{eq56-2a}) and (\ref{eq56-2b}) can be evaluated by a 
bosonized effective theory (\ref{e1}) [or (\ref{e2})] and another theory
(\ref{e3}) plus (\ref{e4}), respectively.  
The semiclassical analysis for Eqs.(\ref{e1}) or (\ref{e2}) predicts 
that $\phi_a$ is locked to the point $\phi_a=0$ for the single-chain
case ($J_\perp=0$), and at that time Eq.~(\ref{eq56-2a}) has a finite value. 
This is in agreement with the fact that
$\langle \hat {\cal O}_1^a\hat {\cal O}_2^a\rangle=
({\cal O}_1^a)^2\neq 0$ is realized at the decoupled point.
While, for a weak (but finite) rung-coupling case, the effective theory
(\ref{e2}) possesses a new potential proportional to 
$J_\perp\sin(\sqrt{\pi}\phi_a)$, as well as the mass potential 
proportional to $m\cos(\sqrt{4\pi}\phi_a)$.  
Their combination must vary the locking point from $\phi_a=0$ to 
a finite and small value irrespective of the sign of $J_\perp$. 
Therefore, we reach the same conclusion as the first content of
(\rnum{3}), and predict that the decrease of 
$\langle \hat {\cal O}_1^a\hat {\cal O}_2^a\rangle$
also occurs in the FM-rung side: the new string parameter would have 
a cusp structure like the gap in Fig.~\ref{fig11}.
Similarly, let us also perform the semiclassical analysis for another
theory, (\ref{e3}) plus (\ref{e4}). It predicts that $\Phi$, $\tilde{\Phi}$, and
$\phi_a$ are all locked to zero at the decoupled point. The rung
coupling engenders a new potential (\ref{e4}) proportional to 
$J_\perp\cos[\pi(\Theta-\tilde{\Theta})]$, in which $\Theta$ and
$\tilde{\Theta}$ are dual fields of $\Phi$ and $\tilde{\Phi}$,
respectively. This potential tends to fix $\Theta$ and
$\tilde{\Theta}$ instead of $\Phi$ and $\tilde{\Phi}$. 
Fixing $\Theta$ means that the fluctuation of $\Phi$ is large because $\Phi$
and $\Theta$ are a canonical conjugate pair (see Appendix~\ref{app1}). 
Therefore, assuming that $\Theta-\tilde{\Theta}$ is locked in the
low-energy limit as the rung coupling is finite, we can
predict that the rung coupling makes the string parameter of the single
chain $\langle \hat {\cal O}^a\rangle \sim \langle
\cos(\sqrt{\pi}\Phi)\rangle^2$ become zero. 
This result consistent with the content (\rnum{2}). 
Moreover, it implies that $\langle \hat {\cal O}^a\rangle$ also
vanishes in the FM-rung side. Note that fixing $\Theta-\tilde{\Theta}$
does not mean $\langle \hat {\cal O}_1^a\hat {\cal O}_2^a\rangle
\rightarrow 0$ because it can be rewritten as 
$\langle \hat {\cal O}_1^a\hat {\cal O}_2^a\rangle\sim
\langle \cos[\sqrt{\pi}(\Phi+\tilde{\Phi})]\rangle+
\langle \cos[\sqrt{\pi}(\Phi-\tilde{\Phi})]\rangle$ and  
$\langle \cos[\sqrt{\pi}(\Phi+\tilde{\Phi})]\rangle$ is expected not to
have a large fluctuation.

We see that spin-$1$ and spin-$\frac{1}{2}$ string parameters take quite
similar field-theoretical expressions. 
However, one should bear in mind that the effective theory for the 
spin-$\frac{1}{2}$ ladder is different from that of the spin-$1$ ladder:
the former is two coupled sine-Gordon-like chains;~\cite{Sel} 
the latter is three coupled sine-Gordon-like chains in Eq.~(\ref{e1}).

\subsubsection{\label{sec3-4-2} In the vicinity of the TB point}
In this paragraph, we briefly consider the spin-1 ladder, in which two
spin-1 chains are located near the TB point (Fig.~\ref{fig8}), through
the perturbative RG technique. In the effective theory of such a chain, 
parameters $m$ and $\lambda$ are much smaller than those of the spin-1
Heisenberg chain. Furthermore, as mentioned already, $m\sim(\delta+1)$
is realized. For the region where $m$, $\lambda$, and $J_\perp$ are
considerably smaller than $J$, the perturbative RG method based on the
TB fixed point becomes a reliable tool to investigate the low-energy
physics.

We construct one-loop RG equations for coupling constants~\cite{noteRG} 
applying the OPE technique.~\cite{Cardy,B-F,Allen2} 
First, we consider all relevant and marginal terms around 
the fixed point (\ref{eq26}). They are summarized in Table~\ref{tab3}, where
we classify them into nine operators $\{{\cal O}_j:j=1,\dots,9\}$ to
render each operator invariant under the spin rotational transformation 
(see Appendix~\ref{app3}) and the interchange of two chains. 
Moreover, we introduced energy operator (mass term) 
$\varepsilon_a(z, \bar z)=i\xi_L^a\xi_R^a$. 
Operators ${\cal O}_{2,3,4,9}$ are generated dynamically in the RG flow, 
even though they are not present initially in the action~(\ref{eq34}).
\begin{table*}
\caption{\label{tab3}Operators in the RG procedure. The sign $G_j(L=0)$
 means the initial value of each running coupling constant 
in the RG flow. The parameter $\alpha_0$ 
is the cut-off parameter defined in Eq.~(\ref{eq46}).}
\begin{ruledtabular}
\begin{tabular}{lcccc}
Operators  & Scaling dimension $x_j$ & $G_j(L=0)$ \\ \hline 
\hline
${\cal O}_1 = \sum_{a=1}^3\varepsilon_a+\tilde{\varepsilon}_a$ & 1
&  $m\times \alpha_0 \times\frac{\pi}{v}$\\
${\cal O}_2 = (\sum_{a=1}^3\varepsilon_a)
(\sum_{b=1}^3\tilde{\varepsilon}_b)$ & 2& 0\\
${\cal O}_3 = \sum_{a>b}^3
2(\varepsilon_a\tilde{\varepsilon}_b+ \varepsilon_b\tilde{\varepsilon}_a)
+(\xi_L^a\xi_R^b+\xi_L^b\xi_R^a)
(\tilde{\xi}_L^a\tilde{\xi}_R^b+\tilde{\xi}_L^b\tilde{\xi}_R^a)$ & 2& 0\\
${\cal O}_4 = (\xi_L^a\xi_R^b+\xi_R^a\xi_L^b)
(\tilde{\xi}_L^a\tilde{\xi}_R^b+\tilde{\xi}_R^a\tilde{\xi}_L^b)$ & 2& 0\\
${\cal O}_5 = J_R^a J_L^a+\tilde{J}_R^a \tilde{J}_L^a$ & 2& 
$-\lambda\times\frac{\pi}{v}$\\
${\cal O}_6 = J_L^a\tilde{J}_L^a+J_R^a\tilde{J}_R^a$ & 2& 
$J_{\perp}a_0\times\frac{\pi}{v}$\\
${\cal O}_7 = J_R^a\tilde{J}_L^a+J_L^a\tilde{J}_R^a$ & 2& 
$J_{\perp}a_0 \times\frac{\pi}{v}$\\
${\cal O}_8 = \kappa\tilde{\kappa}
\sigma_a\mu_{a+1}\mu_{a+2}\tilde{\sigma}_a\tilde{\mu}_{a+1}
\tilde{\mu}_{a+2}$  & 3/4 & 
$C_1^2J_{\perp}a_0 \times \alpha_0^{5/4}\times\frac{\pi}{v}$\\
${\cal O}_{9} = \kappa\tilde{\kappa}\sigma_1\sigma_2\sigma_3
\tilde{\sigma}_1\tilde{\sigma}_2\tilde{\sigma}_3$ & 3/4& 0\\
\end{tabular}
\end{ruledtabular}
\end{table*}

In the low-energy effective action, 
the dimensionless coupling constants $\{G_j\}$ toward the operators 
$\{{\cal O}_j\}$ can be defined as 
\begin{eqnarray}
\label{eq46}
S_{\text{lad}}=S^*
+\sum_{j=1}^9 G_j\int \frac{dx \,v d\tau}{\pi\alpha_0^{2-x_j}}
\,\,{\cal O}_j,
\end{eqnarray}
where $S^*=S_{\text{TB}}[\xi_L^a,\xi_R^a]
+S_{\text{TB}}[\tilde{\xi}_L^a,\tilde{\xi}_R^a]$ is the fixed-point
action, $\alpha_0$ [$\sim O(a_0)$] is the short-distance
cut-off parameter, and $x_j$ is the scaling dimension of ${\cal O}_j$. 
The RG equations for $\{G_j\}$ are the following: 
\begin{subequations}
\label{eq47}
\begin{eqnarray}
\dot G_{1}&=& G_{1}-\frac{1}{2\pi^2}
(G_{2}+4G_{3}-2G_{5})G_{1}\nonumber\\
&&+\pi (G_{8}^2- G_{9}^2),\label{eq47-1}\\
\dot G_{2}&=& -2G_{1}^2+\frac{2}{\pi^2}G_{2}G_{5}
+\frac{2}{\pi^2}(2G_{5}-G_{7})G_{3}\nonumber\\
&&+\frac{2}{\pi^2}G_{4}G_{7}-3\pi^2 G_{8}^2-\pi^2 G_{9}^2,\label{eq47-2}\\
\dot G_{3}&=&\frac{1}{2\pi^2}(3G_{7}-2G_{5})G_{3}
-\frac{1}{2\pi^2}G_{4}G_{7}+2\pi^2 G_{8}^2,\hspace{0.5cm}\label{eq47-3}\\
\dot G_{4}&=& \frac{1}{2\pi^2}(2G_{2}+G_{3})G_{7}\nonumber\\
&&+\frac{1}{\pi^2}(2G_{5}+G_{7})G_{4}-2\pi^2 G_{8}G_{9},\label{eq47-4}
\end{eqnarray}
\begin{eqnarray}
\dot G_{5}&=& 2G_{1}^2+\frac{4}{\pi^2}G_{2}G_{3}
-\pi^2 (G_{8}^2 - G_{9}^2)\nonumber\\
&&+\frac{1}{\pi^2}
\left(\frac{3}{2}G_{2}^2+G_{3}^2+G_{4}^2+\frac{1}{2}G_{5}^2\right),
\label{eq47-5}\\
\dot G_{6}&=& 0,\label{eq47-6}\\
\dot G_{7}&=& \frac{1}{\pi^2}\left(2G_{2}+G_{4}\right)G_{3}
+\frac{1}{2\pi^2}(3G_{3}^2+G_{4}^2+G_{7}^2)\nonumber\\
&&+2\pi^2\left(G_{8}+G_{9}\right)G_{8},\label{eq47-7}\\
\dot G_{8}&=& \frac{5}{4}G_{8}
-\frac{1}{2\pi^2}G_{4}G_{9}+\frac{G_{7}G_{9}}{4\pi^2}\label{eq47-8}\\
&&+\frac{1}{\pi^2}
\left(\pi G_{1}-\frac{1}{8}G_{2}+\frac{3}{2}G_{3}
-\frac{1}{4}G_{4}+\frac{1}{2}G_{7}\right)G_{8} ,\nonumber\\
\dot G_{9}&=& \frac{5}{4}G_{9}-\frac{3}{2\pi^2}G_{4}G_{8}
+\frac{3}{8\pi^2}G_{7}G_{8}\nonumber\\
&&-\frac{1}{\pi^2}
\left(3\pi G_{1}+\frac{9}{8}G_{2}+\frac{3}{2}G_{3}
-\frac{3}{4}G_{5}\right)G_{9}\label{eq47-9}.
\end{eqnarray}
\end{subequations}
Therein, $\dot G_j=dG_j(L)/dL$ and $L$ is the scaling parameter 
(the infinitesimal scaling transformation is 
$\alpha_0\rightarrow \alpha_0 e^{dL}$).
We adopted a simple circle-type cut off.~\cite{Cardy} 

\begin{figure}
\scalebox{0.4}{\includegraphics{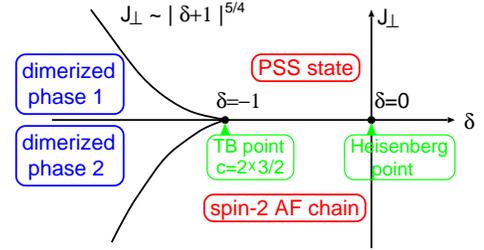}}
\caption{Expected GS phase diagram around two decoupled TB chains ($c=3$).}
\label{fig13}
\end{figure}

Using the RG equations, we discuss the low-energy properties near 
two decoupled TB chains. Close to this fixed point sufficiently, 
we can approximate them as 
\begin{subequations}
\label{eq57} 
\begin{eqnarray}
\dot G_{1}&\approx& G_{1}+\pi (G_{8}^2- G_{9}^2),\label{eq57-1}\\ 
\dot G_{8}&\approx& \Big(\frac{5}{4}+\frac{1}{\pi}G_{1}\Big)G_{8},\label{eq57-2}\\
\dot G_{9}&\approx& \Big(\frac{5}{4}-\frac{3}{\pi}G_{1}\Big)G_{9} \label{eq57-3}.
\end{eqnarray}
\end{subequations}
Couplings $G_1$, $G_8$ and $G_9$ are more relevant than the other 
couplings. The coupling $G_1$ bears a single-chain character, whereas 
both $G_8$ and $G_9$ are the representatives of the rung coupling.
Moreover, we may omit $G_9$ because
its initial value is zero. Under these approximations, 
the treatment of Eqs.~(\ref{eq57}) becomes fairly easier. 
One can find two nontrivial fixed points
$(G_1,G_8)=(-5\pi/4,\pm\sqrt{5/4})$. Let us assume the presence of 
these two fixed points, even though they are not close to the TB 
point $(G_1,G_8)=(0,0)$. Linearization of the approximated RG equations 
around new fixed points indicates that both points are of a divergent type,
thereby implying the existence of phase transitions.
Similarly, the TB point is of a divergent type. Therefore, there exist
two phase transition curves connecting the TB point and a new fixed
point or another. In the vicinity of the TB point where approximations 
$\dot G_1\approx G_1$ and $\dot G_8\approx \frac{5}{4}G_8$ are allowed, 
a conservation law under the  RG transformation, 
$|G_1/G_8^{4/5}|\approx \text{a\,\,constant}$, is realized.
Taking into account this law and recalling that 
$G_1\propto m$, $G_8\propto J_\perp$, and $m\sim (\delta+1)$ 
are realized near the TB point, 
we expect phase transition curves follow:
\begin{eqnarray} 
\pm J_\perp \sim |\delta+1|^{5/4},
\end{eqnarray}
around the TB point.
Consequently, we can draw the GS phase diagram near the two TB chains 
as in Fig.~\ref{fig13}. In this figure, the right side $\delta>-1$ of the 
horizontal (decoupled) line $J_\perp =0$ is probably not
corresponding to any phase transitions. 
In fact, we know that on the Heisenberg line $\delta=0$, 
the point $J_\perp =0$ does not correspond to any transitions. 
Although on the Heisenberg line, the GS of the strong AF-rung limit
(rung dimer) is quite different from one of the FM-rung limit (spin-$2$
AF chain), both AF- and FM-rung sides may belong to the same phase. 
Whereas, it is not sure whether the left side $\delta<-1$ of 
the line $J_\perp =0$ corresponds to a phase transition or not. 
The dimerized phases 1 and 2 must break the
one-site translational symmetry along the chain direction. 
According to the Zamolodchikov's ``$c$ theorem,''~\cite{c-theo}   
two critical curves starting from the TB point ($c=2\times3/2$),
belong to a universality class with $c<3$.

From Fig.~\ref{fig13}, it is believed that the area characterized by the
PSS-state picture (or connected to a spin-2 AF chain) widely expands around
two decoupled Heisenberg chains, in the space $(\delta,J_\perp)$.

\subsection{\label{sec3-5} Staggered-field case}
The low-energy action for the staggered field case is given by
Eq.~(\ref{eq34}) plus Eq.~(\ref{eq35-2}). The latter term 
is only invariant under the U(1) rotation around the spin $z$ axis; 
and it does not possess the SU(2) symmetry.
This partial violation of the symmetry makes each operator
${\cal O}_j$ in Table~\ref{tab3} decoupled to two U(1) invariant parts
through the OPE between ${\cal O}_j$ and the staggered term (\ref{eq35-2}). 
Neither part is invariant under the SU(2) rotation. 
We have to consider more than 21 coupling
constants, to construct the RG equations.
Although we actually constructed the RG, we do not record them here. 
We were unable to extract characteristic contents from them 
because they are extremely complicated. Coupling constants for mass,
rung-coupling and staggered Zeeman terms all grow up just monotonically.
Therefore, the GS in the staggered-field case must have some 
massive excitations. Moreover, two critical curves in the AF-rung side 
(Fig.~\ref{fig7}) do not reach the origin $(J_\perp,H)=(0,0)$.
We infer that the present field-theoretical strategy can not
predict how the curves finish.

\section{\label{sec4} Summary and Discussions}
We explored the spin-$1$ AF ladder (\ref{eq1}) and some of its
extensions.
In the strong AF rung-coupling region, we performed the DPT and  
found that the GS phase diagram of the uniform(staggered)-field case
includes two $c=1$ critical areas (curves). Subsequently, we extended 
these results to the spin-$S$ ladders, and predicted that the spin-$S$
uniform(staggered)-field case has $2S$ $c=1$ areas (curves).
Figure~\ref{fig7} summarizes GS phase diagrams and the magnetization curves. 
The upper critical uniform fields (\ref{eq19}) and
(\ref{eq19-2}) were determined by the spin-wave analysis; 
we saw that, surprisingly, Eq.~(\ref{eq19}) is equal to predictions 
of first-order DPTs (see Table~\ref{tab1}).  
We also proposed the ``RVB'' picture of each plateau
(massive) state.  

On the other hand, we applied field-theoretical methods 
for the weak rung-coupling region.
For the uniform-field case, we employed the non-abelian bosonization
efficiently. Combining the consequences of weak and strong rung-coupling
analyses, we complete the GS phase diagram of the uniform-field case as
in Fig.~\ref{fig12}.
Meanwhile, our field-theoretical approach was not efficient 
for the staggered-field case. 
Therefore, the GS phase diagram is not determined perfectly.
The uncertain area, i.e., the area where the critical curves in
Fig.~\ref{fig7} ($2\alpha$) vanish, is probably located in an
intermediate AF rung-coupling region, which must be distant from both weak and
strong rung-coupling regions.

Using those field theories, 
we revisited and discovered some properties of 1D spin-$1$ systems without
external fields.   
We showed that field theories can describe string order parameters in
spin-$1$ systems, and proposed formulas (\ref{eq56}) and
(\ref{eq56-2}). We hope that these formulas are useful in the search for
some new string-type parameters in spin-$1$ systems.
We also considered the GS phase diagram around two decoupled TB chains 
(Fig.~\ref{fig13}).

\begin{figure}
\scalebox{0.3}{\includegraphics{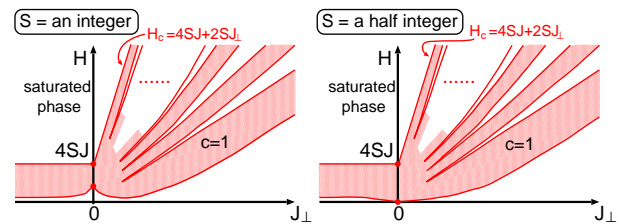}}
\caption{Expected GS phase diagrams of $2$-leg spin-$S$ AF ladder with
 the uniform Zeeman term (\ref{eq2-1}).}
\label{fig15}
\end{figure}

Through the present work, one obtains GS phase diagrams of both
the spin-$\frac{1}{2}$ and spin-$1$ ladders with the uniform 
Zeeman term (\ref{eq2-1}) (see Figs.~\ref{fig1} and \ref{fig12}). 
The former (latter) model consists of two gapless (gapped) spin chains. 
These and our prediction of Fig.~\ref{fig7} enable us to 
expect that phase diagrams of $2$-leg integer-spin and half-integer-spin 
AF ladders are written as in Fig.~\ref{fig15}.
On the other hand, for the staggered-field case, we can give only the
following two predictions about the spin-$S$ ladders. For the $2$-leg
integer-spin ladders, critical curves in Fig.~\ref{fig7}~($2\alpha$)
vanish in a weak AF rung-coupling region. For the half-integer-spin
ladders, Figs.~\ref{fig2} and \ref{fig7} imply that the origin
$(J_\perp,H)=(0,0)$ is perhaps a multicritical point, from which $2S$
critical curves start.

Except for the spin-$\frac{1}{2}$ cases, the bosonization techniques 
for higher spin systems have some subtle and phenomenological aspects.
Establishing more sophisticated bosonizations is an 
interesting but difficult problem that remains for a future work.

Nowadays, several spin-$\frac{1}{2}$ ladder compounds have been 
reported.~\cite{ladder1,SrCuO,Cu2--(NO3)2,Cu2--Cl4,DT-TTF} 
Unfortunately, 
the materials regarded as a spin-$1$ ladder have never been found. 
An organic compound ``BIP-TENO''~\cite{BIP-TENO} 
may be a candidate of spin-$1$ ladders, but its 
magnetic behavior differs from the simple ladder (\ref{eq1}).
Suitable theoretical predictions regarding it have not been constructed.

\begin{acknowledgments}
First, the author would like to thank Masaki Oshikawa for 
a critical reading of this manuscript and fruitful discussions. 
He also thanks Munehisa Matsumoto, Kiyomi Okamoto, and Masaaki Nakamura 
for several useful comments related to spin ladders and quantum phase 
transitions. Furthermore, he gratefully acknowledges Ian Affleck for 
pointing out the importance of estimating $S_{\rm{stag}}$ in Sec.~\ref{sec3-3}.
This work was supported by a 21st Century COE Program at
Tokyo Tech ``Nanometer-Scale Quantum Physics'' by the
Ministry of Education, Culture, Sports, Science and Technology.
\end{acknowledgments}

\appendix

\section{\label{app1}Abelian Bosonization Rule}
In this appendix, we briefly summarize the Abelian 
bosonization~\cite{Hal,Lec-Aff,Boso,CFT,D-S,Gogo,Lec-Sene,LowD-QFT,2D-QFT} 
used in Secs.~\ref{sec2} and \ref{sec3}, which is deeply related with
the concept of TLL and $c=1$ CFT. (Our notation is similar to 
Refs.~\onlinecite{Gogo} and \onlinecite{Lec-Sene}.) 

Bosonization shows that (1+1)D Dirac fermion models 
are equivalent to a (1+1)D boson field theory.
The main root of this technique lies in the identification between 
the massless Dirac fermion and the massless free scalar boson
field theories. The former Hamiltonian is represented as  
\begin{eqnarray}
\label{a1}
\hat {\cal H}_{\text{Dirac}}&=&\int dx\,
iv(\psi_L^{\dag}\partial_x \psi_L -\psi_R^{\dag}\partial_x \psi_R),
\end{eqnarray}
where $\psi_L(z)$ and $\psi_R(\bar z)$ are, respectively, 
the left and right moving components of the 
Dirac fermion. The sign $v$ denotes the ``light'' velocity.
[In real time formalism, 
$z$ and $\bar z$ means, respectively, $i(vt+x)$ and $i(vt-x)$.] 
The fermions obey the equal-time anticommutation relations 
$\{\psi_\alpha^\dag(x),\psi_\beta(y)\}=\delta_{\alpha\beta}\delta(x-y)$
and $\{\psi_\alpha(x),\psi_\beta(y)\}=0$.
The corresponding massless boson theory has the following
Hamiltonian (here we do not make the terms of the 
zero-mode excitations~\cite{Hal,D-S} clear):
\begin{eqnarray}
\label{a2}
\hat {\cal H}_{\text{Scalar}}&=&\int dx\,\frac{v}{2}
\left[ \Pi^2+(\partial_x \phi)^2\right],
\end{eqnarray}
where $\phi(z,\bar z)=\phi_L(z)+\phi_R(\bar z)$ is the scalar boson
field, $\phi_{L(R)}$ is the left (right) mover of $\phi$ 
and $\Pi=\partial_t\phi/v$ is the canonical conjugate of $\phi$. 
The Hamiltonian (\ref{a2}) can be mapped to the same form for 
the dual field $\theta(z,\bar z)=\phi_L-\phi_R$.   
The equal-time commutation relations among boson fields are defined as 
$[\phi_R(x),\phi_R(y)]=-[\phi_L(x),\phi_L(y)]=
\frac{i}{4}\text{sgn}(x-y)$ and $[\phi_L(x),\phi_R(y)]=0$.
Since above two theories have a chiral U(1) symmetry, this method
is called the ``abelian'' bosonization. 

They have the following operator identities: 
\begin{subequations}
\label{a3}
\begin{eqnarray}
\psi_L(z) & = & \frac{\eta_L}{\sqrt{2\pi\alpha}}
\exp\left[-i\sqrt{4\pi}\phi_L(z)\right],\label{a3-1}\\
\psi_R(\bar z) & = & \frac{\eta_R}{\sqrt{2\pi\alpha}}
\exp\left[i\sqrt{4\pi}\phi_R(\bar z)\right],\label{a3-2}
\end{eqnarray}
\end{subequations}
where $\eta_{L,R}$, called Klein factors,~\cite{D-S,Lec-Sene,LowD-QFT,2D-QFT} 
are necessary to guarantee the anticommutation relation
between the left and right movers of the fermions, and they 
hence satisfy $\{\eta_{a},\eta_{b}\}=2\delta_{ab}$. 
The factor $\alpha$ in Eqs.~(\ref{a3}) is a parameter which order is
the inverse of the wave-number cut off of the Dirac fermion.  
The exponential-type operators are
called ``vertex operators.''
Using a point-splitting technique,~\cite{D-S,Lec-Sene} 
one can also bosonize the U(1) currents $J_L$ and $J_R$ as follows:
\begin{subequations}
\label{a4}
\begin{eqnarray}
J_L(z)&=& :\psi_L^{\dag}\psi_L:(z)
=\frac{1}{\sqrt{\pi}}\partial_x\phi_L,\label{a4-1}
\\
J_R(\bar z)&=&:\psi_R^{\dag}\psi_R:(\bar z)
=\frac{1}{\sqrt{\pi}}\partial_x\phi_R,\label{a4-2}
\end{eqnarray}
\end{subequations}
where the symbol $:\,\, :$ stands for the normal-ordered product. 
[Note that the above U(1) currents are different from the SU(2) currents
$J_L^a$ and $J_R^a$ in Sec.~\ref{sec3}. However, if we think of the $c=1$ CFT
here as the $c=1$ part of the $c=3/2$ WZNW model, one component of the
SU(2) currents is proportional to the U(1) current. See Eq.~(\ref{b7}).]

From these relations, one sees that Dirac fermion models even
involving arbitrary interactions can be mapped to a boson theory 
with vertex operators. The RG flow lets several kinds of
interacting Dirac fermion models go to a free boson theory with 
a modified velocity [it is different from $v$ in
Hamiltonians~(\ref{a1}) and (\ref{a2})] and a compactification 
radius $R$ [which determines the period of the boson as 
$\phi=\phi+2\pi R$],~\cite{Lec-Aff,CFT,Lec-Sene} in the low-energy limit.
Nowadays, the terminology ``TLL'' means the low-energy properties of 
such fermion models or the corresponding free boson theories. 
The $c=1$ CFT consists of all the free boson theories
with an arbitrary radius. The radius of the massless Dirac fermion
(\ref{a2}) is $1/\sqrt{4\pi}$.

We mention the scaling dimensions $x=\Delta_L+\Delta_R$ and 
conformal spins $s=\Delta_L-\Delta_R$ of
primary fields in the $c=1$ CFT, where $\Delta_{L(R)}$ is the left (right)
conformal weight. In our notation, the conformal weights 
of the vertex operator $\exp[i(\alpha_L\phi_L+\alpha_R\phi_R)]$ 
is $(\Delta_L,\Delta_R)=(\frac{\alpha_L^2}{8\pi},\frac{\alpha_R^2}{8\pi})$. 
Possible values of $\alpha_L$ and $\alpha_R$ can be determined by
the modular invariance.
The currents $J_L$ and $J_R$ always have weights $(1,0)$ and $(0,1)$,
respectively. In the Dirac fermion (\ref{a1}), 
$\psi_L$ and $\psi_R$ have weights 
$(\frac{1}{2},0)$ and $(0,\frac{1}{2})$, respectively.

\section{\label{app2}Ising Model and $\bol c\bol =\bol 1\bol /\bol 2$ CFT}
We review some facts in terms of the 2D statistical (or 1D transverse) 
Ising model and $c=1/2$ CFT. The latter field theory emerges as 
the effective field theory in Sec.~\ref{sec3}.
 
\subsection{\label{app2-1}Ising model}
Here, we summarize relations between the Ising model on the square 
lattice and its continuum limit. 
(The contents here almost follows Ref.~\onlinecite{Fab}.)

It is well known that both the order-disorder phase transition 
in the 2D Ising model and the quantum phase transition 
in the 1D transverse Ising model~\cite{Cha,Sac} 
belong to the universality of the $c=1/2$ CFT.~\cite{CFT,Gogo,Kog,Boya,I-D}  
These two models are connected with each other via the transfer matrix
method. The latter Hamiltonian is 
\begin{eqnarray}
\label{b1}
\hat {\cal H}_{\text{TI}}=-\sum_j\left[{\cal J}\sigma_j^z\sigma_{j+1}^z 
+h\sigma_j^x \right],
\end{eqnarray}
where $\sigma_j^a$ is $a$ component of Pauli matrices settling in site
$j$, and $h$ is the transverse field. The critical point lies in 
$h={\cal J}$: if $0\leq h<{\cal J}$ ($h > {\cal J}$), the order
parameter satisfies $\langle \sigma_j^z\rangle \neq 0$ ($= 0$).
Here, let us introduce the disorder operator $\mu_{j+1/2}^a$ on the dual lattice
$\{j+1/2\}$ as 
\begin{subequations}
\label{b2}
\begin{eqnarray}
\mu_{j+1/2}^z=\prod_{p=1}^j\sigma_p^x, && 
\mu_{j+1/2}^x=\sigma_j^z\sigma_{j+1}^z,\label{b2-1}\\
\sigma_j^z=\prod_{p=0}^{j-1}\mu_{p+1/2}^x,&& 
\sigma_j^x=\mu_{j-1/2}^z\mu_{j+1/2}^z,\label{b2-2}
\end{eqnarray}
\end{subequations}
where $\mu_{j+1/2}^{x,y,z}$ obey the same commutation relations 
as Pauli matrices. 
From Eqs.~(\ref{b2}), we have 
\begin{eqnarray}
\label{b3}
\hat {\cal H}_{\text{TI}}[\{\sigma_j^a\};{\cal J},h]&=&
\hat {\cal H}_{\text{TI}}[\{\mu_{j+1/2}^a\};h,{\cal J}].
\end{eqnarray}
This is called the Kramers-Wannier duality. 
The self-dual point is just the critical one $h={\cal J}$. 
The ordered phase in the original Ising model, where 
$\langle \sigma_j^z\rangle \neq 0$, corresponds to the disordered
phase in the dual model, where $\langle \mu_{j+1/2}^z\rangle= 0$, and
vice-versa. In the vicinity of the critical point, continuum limits
of $\sigma_j^z$ and $\mu_{j+1/2}^z$ are, respectively, associated with 
the order field $\sigma(z,\bar z)$ and disorder field $\mu(z,\bar z)$ 
in the $c=1/2$ CFT, which are identical with $\sigma_a(z,\bar z)$ and 
$\mu_a(z,\bar z)$ respectively in Sec.~\ref{sec3}. 
It is worth while emphasizing that the relation between $\sigma_j^z$ and 
$\mu_{k+1/2}^z$ is nonlocal: these two commute when $j>k$, but
anticommute otherwise.

Real fermion operators can be introduced as 
\begin{eqnarray}
\label{b4}
\eta_j=\sigma_j^z\mu_{j-1/2}^z,
&&\zeta_j=i\sigma_j^z\mu_{j+1/2}^z,
\end{eqnarray}
where $\{\eta_j,\eta_k\}=\{\zeta_j,\zeta_k\}=2\delta_{jk}$,
$\{\eta_j,\zeta_k\}=0$. Inversely Ising operators are written by
fermions as 
\begin{subequations}
\label{b4'}
\begin{eqnarray}
\sigma_j^x=i\zeta_j\eta_j, &&\mu_{j+1/2}^x=-i\zeta_j\eta_{j+1},\label{b4'-1}\\
\sigma_j^z=i\eta_j\prod_{p=1}^{j-1}i\zeta_p \eta_p, &&
\mu_{j+1/2}^z=\prod_{p=1}^{j}i\zeta_p\eta_p. \label{b4'-2}
\end{eqnarray}
\end{subequations}
The Hamiltonian (\ref{b1}) can be described by these fermions, and is
solvable in the fermion language.
In particular, considering the vicinity of the critical point 
$h={\cal J}$, one can transform it to the following 
field-theoretical Hamiltonian:
\begin{eqnarray}
\label{b5}
\hat {\cal H}_{\text{TI}}&\approx&\int dx \,\,i\frac{v}{2}
(\xi_L\partial_x\xi_L-\xi_R\partial_x\xi_R)+im\xi_L\xi_R,
\end{eqnarray}
where $v=2{\cal J}a_0$, $m=2(h-{\cal J})$, $x=j\times a_0$ and 
$a_0$ is the lattice constant. New Majorana (real) fermions 
$\xi_L$ and $\xi_R$ are defined as 
\begin{eqnarray}
\label{b6}
\xi_L=(\eta+\zeta)/\sqrt{2},&&\xi_R=(\zeta-\eta)/\sqrt{2},
\end{eqnarray}
where $\eta(x)=\eta_j/\sqrt{2a_0}$ and $\zeta(x)=\zeta_j/\sqrt{2a_0}$.
At the critical point, the mass term vanishes, and the effective
field theory (\ref{b5}) then becomes a massless Majorana fermion model,
which is just a $c=1/2$ CFT.
The fields $\xi_L$ and $\xi_R$ are corresponding to 
$\xi_L^a$ and $\xi_R^a$ in Sec.~\ref{sec3-2}, respectively.
From Eqs.~(\ref{b4'-2}) and (\ref{b6}), it is obvious that $\xi_{L(R)}$
has a nonlocal relation with $\sigma$ and $\mu$, too.

Two copies of the critical Majorana fermion theories 
are equivalent to the massless Dirac fermion (\ref{a1}), 
i.e., a $c=1/2+1/2=1$ CFT.
Here let us denote the fields of two $c=1/2$ systems, 
the corresponding Dirac fermion and boson theories as 
$(\xi_{L,R}^1,\sigma_1,\mu_1)$, $(\xi_{L,R}^2,\sigma_2,\mu_2)$, 
$\psi_{L,R}$ defined as Eq.~(\ref{eq36}), and $\phi_{L,R}$, respectively. 
The U(1) currents in Eq.~(\ref{a4}) are written by 
Majorana fermions as follows:
\begin{eqnarray}
\label{b7}
J_L=i\xi_L^1\xi_L^2, && J_R=i\xi_R^1\xi_R^2.
\end{eqnarray}
Energy operators $\varepsilon_{1,2}=i\xi_L^{1,2}\xi_R^{1,2}$ are
mapped to  
\begin{eqnarray}
\label{b8}
\varepsilon_1+\varepsilon_2 &=& i(\psi_L^\dag\psi_R-\psi_R^\dag\psi_L)\nonumber\\
&=& i\frac{\eta_L\eta_R}{\pi\alpha}\cos(\sqrt{4\pi}\phi),
\end{eqnarray}
where $\eta_{L,R}$ are Klein factors in Eq.~(\ref{a3}).
In addition, it is believed that order and disorder fields are 
bosonized as~\cite{Gogo,Fab,F-S-Z}
\begin{eqnarray}
\label{b9}
\sigma_1\sigma_2 \sim   \sin(\sqrt{\pi}\phi),&& 
\mu_1\mu_2       \sim   \cos(\sqrt{\pi}\phi), \nonumber\\
\sigma_1\mu_2    \sim   \cos(\sqrt{\pi}\theta),&& 
\mu_1\sigma_2    \sim   \sin(\sqrt{\pi}\theta), 
\end{eqnarray}
where $\theta$ is the dual of $\phi$.

\subsection{\label{app2-2}OPE}
We write down the OPEs among the primary fields 
in the $c=1/2$ CFT:~\cite{CFT,Allen,F-S-Z,Allen2} 
the identity operator $1$, 
the left (right) mover $\xi_{L(R)}$ of the Majorana fermion, 
the energy operator $\varepsilon(z,\bar z)= i\xi_L\xi_R$, 
the order field $\sigma$ and the disorder one $\mu$. 
These are derived~\cite{CFT,F-S-Z} by making use of fusion rules 
and the Abelian bosonization based on the fact that two copies of 
$c=1/2$ CFTs form a massless Dirac fermion model, a $c=1$ CFT.
The results are
\begin{subequations}
\label{b2-0}
\begin{eqnarray}
\xi_L(z)\xi_L(0)&\sim& \frac{1}{2\pi z},\label{b-2-1}\\
\xi_R(\bar z)\xi_R(0)&\sim& \frac{1}{2\pi \bar z},\label{b-2-2}\\
\sigma(z,\bar z)\sigma(0,0)&\sim& \frac{1}{|z|^{1/4}}+\pi|z|^{3/4} 
\varepsilon(0,0),\label{b2-3}\\
\mu(z,\bar z)\mu(0,0)&\sim& \frac{1}{|z|^{1/4}}-\pi|z|^{3/4} 
\varepsilon(0,0),\label{b2-4}\\
\sigma(z,\bar z)\mu(0,0)&\sim&\frac{\sqrt{\pi}}{|z|^{1/4}}
[e^{i\pi/4}z^{1/2}\xi_L(0) \nonumber\\
&& +e^{-i\pi/4}\bar z^{1/2}\xi_R(0)],\label{b2-5}
\end{eqnarray}
\begin{eqnarray}
\mu(z,\bar z)\sigma(0,0)&\sim& \frac{\sqrt{\pi}}{|z|^{1/4}}
[e^{-i\pi/4}z^{1/2}\xi_L(0) \nonumber\\
&&+e^{i\pi/4}\bar z^{1/2}\xi_R(0)],\label{b2-6}\\
\xi_L(z)\sigma(0,0)&\sim& \frac{e^{i\pi/4}}{2\sqrt{\pi}z^{1/2}}
\mu(0,0),\label{b2-7}\\
\xi_R(\bar z)\sigma(0,0)&\sim& \frac{e^{-i\pi/4}}
{2\sqrt{\pi}\bar z^{1/2}}\mu(0,0),\label{b2-8}\\
\xi_L(z)\mu(0,0)&\sim& \frac{e^{-i\pi/4}}
{2\sqrt{\pi}z^{1/2}}\sigma(0,0),\label{b2-9}\\
\xi_R(\bar z)\mu(0,0)&\sim&\frac{e^{i\pi/4}}
{2\sqrt{\pi}\bar z^{1/2}}\sigma(0,0)\label{b2-10}.
\end{eqnarray}
\end{subequations}
As a reflection of nonlocal natures among $\xi_{L(R)}$, $\sigma$ and
$\mu$, some OPEs have a branch cut.
The OPE (\ref{b2-5}) indicates that the product of the order and
disorder fields must have a fermionic property.
Following Ref.~\onlinecite{Allen2}, in the main text, 
we often use a rule that a disorder field anticommutes with other
disorder fields (when we consider some copies of $c=1/2$ CFTs) and
fermion fields, but commutes with itself and order fields.
This is responsible for the improvement (\ref{eq32-2}).

As mentioned in Sec.~\ref{sec3}, 
three copies of $c=1/2$ CFTs are equivalent to the level-$2$ SU(2) WZNW
model. Using the definition of the SU(2) currents (\ref{eq30}), 
OPEs (\ref{b2-1}) and (\ref{b2-2}), we obtain the following OPEs
among the currents: 
\begin{eqnarray}
J_L^a(z)J_L^b(0)&\sim& \frac{\delta_{ab}}{4\pi^2 z^2}
+i\epsilon_{abc}\frac{J_{L}^c(0)}{2\pi z},\\
J_R^a(\bar z)J_R^b(0)&\sim& \frac{\delta_{ab}}{4\pi^2 \bar z^2}
+i\epsilon_{abc}\frac{J_R^c(0)}{2\pi \bar z}.
\end{eqnarray}

\section{\label{app4}SU(2) Symmetry in spin systems and the WZNW Model}
We consider what transformation for the effective field theories 
(\ref{eq26}) or (\ref{eq34}) 
is corresponding to the global spin rotational transformation 
for spin-$1$ systems.

The latter can be represented by a vector rotation form as  
\begin{eqnarray}
\label{d2}
\vec S_j &\rightarrow& 
T_z(-\varphi_3)T_y(-\varphi_2)T_z(-\varphi_1)\vec S_j\equiv {\cal R}\vec S_j, 
\end{eqnarray}
where $\vec S_j = {}^T(S_j^x,S_j^y,S_j^z)$ and $T_a(\varphi)$ stands for 
the 3D rotation about $a$ axis by angle $\varphi$ [an SO(3) matrix]. 
For example, $T_z$ is defined as
\begin{eqnarray}
\label{d3}
T_z(\varphi)&=&
\left(
\begin{array}{lll}
\cos\varphi & \sin\varphi & 0 \\
-\sin\varphi & \cos\varphi & 0 \\
0    &  0  &    1
\end{array}
\right). 
\end{eqnarray}
From Eq.~(\ref{eq32}), the spin rotation on the lattice is interpreted,
in the WZNW model, as 
\begin{subequations}
\label{d4}
\begin{eqnarray}
\vec J_{L,R} &\rightarrow & {\cal R}\vec J_{L,R}, \label{d4-1}\\
\vec I &\rightarrow & {\cal R}\vec I, \label{d4-2}
\end{eqnarray} 
\end{subequations}
where $\vec J_\alpha= {}^T(J_\alpha^1,J_\alpha^2,J_\alpha^3)$ and 
$\vec I={}^T({\cal G}_1,{\cal G}_2,{\cal G}_3) 
\propto {}^T(\sigma_1\mu_2\mu_3,\sigma_2\mu_3\mu_1,\sigma_3\mu_1\mu_2)$. 
The transformation for the spin uniform part (\ref{d4-1}) is reproduced 
by the following transformation for Majorana fermions:
\begin{eqnarray}
\label{d5}
\vec \xi_{L,R} &\rightarrow & {\cal R}\vec \xi_{L,R},
\end{eqnarray}
where $\vec \xi_\alpha\equiv {}^T(\xi_\alpha^1,\xi_\alpha^2,\xi_\alpha^3)$.
One can confirm that the effective theory for the Heisenberg chain
[Eq.~(\ref{eq26}) plus Eq.~(\ref{eq33})] is invariant under the rotation
(\ref{d5}). Especially, the level-$2$ SU(2) WZNW model (\ref{eq26}) is 
invariant both under the rotation of left movers 
$\vec\xi_L\rightarrow {\cal R}\vec \xi_L$ and under that of right movers
$\vec\xi_R\rightarrow {\cal R}\vec \xi_R$. These two symmetries
must correspond to the chiral SU(2) symmetry of the WZNW model. 
Reversely, both terms in Eq.~(\ref{eq33}) violate the chiral symmetry,
and are invariant just under the ``diagonal'' rotation. 

Next, we focus on the rotation of the spin staggered part (\ref{d4-2}).
When one represents the action of the WZNW model by using the matrix
field $g$,~\cite{Gogo,CFT} its global SU(2) symmetry means that 
the action is invariant under 
\begin{eqnarray}
\label{d6}
g&\rightarrow & VgV^{\dag},
\end{eqnarray}
where $V$ is a SU(2) matrix. A natural expectation is 
that the transformation (\ref{d6}) leads to the rotation (\ref{d4-2}).
Following this idea, in fact, one can verify that 
if the matrix $V$ is parametrized as follows:
\begin{eqnarray}
\label{d7}
V &=& 
\left(
\begin{array}{cc}
e^{-i(\varphi_3+\varphi_1)/2}\cos\frac{\varphi_2}{2}    \\
-ie^{i(\varphi_3-\varphi_1)/2}\sin\frac{\varphi_2}{2}  
\end{array}
\right.\nonumber\\
&&\hspace{0.5cm}
\left.
\begin{array}{cc}
-ie^{-i(\varphi_3-\varphi_1)/2}\sin\frac{\varphi_2}{2}  \\
e^{i(\varphi_3+\varphi_1)/2}\cos\frac{\varphi_2}{2} 
\end{array}
\right),
\end{eqnarray}
the explicit correspondence between (\ref{d4-2}) and (\ref{d6}) 
appears. From a rotation (\ref{d4-2}), one also finds 
the spin rotation (\ref{d2}) does not affect 
${\cal G}_1=\sigma_1\sigma_2\sigma_3$.
Proving that the action (\ref{eq26}) and operators ${\cal O}_j$ in
Table~\ref{tab3} are invariant under the spin rotation
is an easy work.

We touch the rotation around spin $z$ axis 
$\vec S_j\to T_z(-\varphi)\vec S_j$. It does not affect the $z$ component
of the spin, and provides the transformations 
$\xi_\alpha^1\to \cos\varphi \xi_\alpha^1-\sin\varphi \xi_\alpha^2$ and 
$\xi_\alpha^2\to \sin\varphi \xi_\alpha^1+\cos\varphi \xi_\alpha^2$. 
In the Dirac fermion picture, defined by Eq.~(\ref{eq37}), these U(1)
transformations are 
\begin{eqnarray}
\label{d8}
\psi_{L,R}&\to & \exp(i\varphi)\psi_{L,R}.
\end{eqnarray}
This is used in Sec.~\ref{sec3-3}.

\begin{widetext}
\section{\label{app3}Effective Hamiltonian in Sec. \ref{sec3-2}}
The effective Hamiltonian for the spin-$1$ ladder (\ref{eq1}) 
with a uniform field (\ref{eq2-1}) under the condition $H>m$ is 
\begin{eqnarray}
\label{c1}
\hat{\cal H}_{\text{eff}}&=& \int dx 
\Big[\hat H_{\text{free}}(L,R,\xi_\alpha^3,\psi) 
+ \hat H_{\text{free}}(\tilde{L},\tilde{R},\tilde{\xi}_\alpha^3,\tilde{\psi}) 
+\hat H_{\text{int}}^1(L,R,\tilde{L},\tilde{R})
+\hat H_{\text{int}}^2(L,R,\tilde{L},\tilde{R};\xi_\alpha^3,\tilde{\xi}_\alpha^3)
\nonumber\\
&&+\hat H_{\text{int}}^3(L,R,\tilde{L},\tilde{R};\psi,\tilde{\psi})
+\hat H_{\text{int}}^4(\xi_\alpha^3,\tilde{\xi}_\alpha^3;\psi,\tilde{\psi})
+\hat H_{\text{int}}^5(\psi,\tilde{\psi}) +\hat H_{\text{stag}}(\sigma,\mu) \Big],
\end{eqnarray}
where
\begin{subequations}
\label{c2}
\begin{eqnarray}
\hat H_{\text{free}}&=& 
iv'( L^{\dag}\partial_x L-R^{\dag}\partial_x R)
+\psi^{\dag}(-\frac{v^2}{2m}\partial_x^2+m+H)\psi
+i\frac{v}{2}(\xi_L^3\partial_x\xi_L^3-\xi_R^3\partial_x\xi_R^3)
+ i m \xi_L^3\xi_R^3, \label{c2-1}
\end{eqnarray}
\begin{eqnarray}
\hat H_{\text{int}}^1 &=& 
-\frac{\lambda}{4}U_+^2 U_-^2\{(R^{\dag}R-R R^{\dag})^2
+(L^{\dag}L-L L^{\dag})^2
+(R\rightarrow \tilde{R})^2+(L\rightarrow \tilde{L})^2\}\nonumber\\
&&-\frac{\lambda}{4}(U_+^4 +U_-^4)
\{(R^{\dag}R-R R^{\dag})(L^{\dag}L-L L^{\dag})
+(R\rightarrow \tilde{R})(L\rightarrow \tilde{L})\}\nonumber\\
&&-\lambda U_+^2 U_-^2\{R^{\dag}R L L^{\dag}
+L^{\dag}L R  R^{\dag}+(R,L\rightarrow \tilde{R},\tilde{L})\}\nonumber\\
&&+\frac{J_{\perp}a_0}{4}(U_+^2+U_-^2)^2
\{(R^{\dag}R-R R^{\dag})(R\rightarrow \tilde{R})
+(L^{\dag}L-L L^{\dag})(L\rightarrow \tilde{L})\}\nonumber\\
&&+\frac{J_{\perp}a_0}{4}(U_+^2+U_-^2)^2 
\{(R^{\dag}R-R R^{\dag})(\tilde{L}^{\dag}\tilde{L}-\tilde{L}\tilde{L}^{\dag})
+(R\rightarrow \tilde{R})(\tilde{L}\rightarrow L)\}\nonumber\\
&&+4J_{\perp}a_0 U_+^2 U_-^2 [R L^{\dag}\tilde{L}\tilde{R}^{\dag}
+L R^{\dag}\tilde{R}\tilde{L}^{\dag}],\label{c2-5}\\
\hat H_{\text{int}}^2
&=& i \lambda U_+ U_-
\{[R^{\dag}R-R R^{\dag}+L^{\dag}L-L L^{\dag}]\xi_L^3\xi_R^3
+[R\rightarrow \tilde{R}+L\rightarrow \tilde{L}]
\tilde{\xi}_L^3\tilde{\xi}_R^3\}\nonumber\\
&&+J_{\perp}a_0 \{[U_{+}^2(R^{\dag}\tilde{R}+R\tilde{R}^{\dag})
+U_-^2(L^{\dag}\tilde{L}+L\tilde{L}^{\dag})]\xi_R^3\tilde{\xi}_R^3
+[U_-^2(R^{\dag}\tilde{R}+R\tilde{R}^{\dag})
+U_+^2(L^{\dag}\tilde{L}+L\tilde{L}^{\dag})]\xi_L^3\tilde{\xi}_L^3\}\nonumber\\
&&-i J_{\perp}a_0U_+U_-\{[R^{\dag}\tilde{R}-R\tilde{R}^{\dag}
+L^{\dag}\tilde{L}-L\tilde{L}^{\dag}]\xi_L^3\tilde{\xi}_R^3
+[\tilde{R}^{\dag}R-\tilde{R}R^{\dag}+\tilde{L}^{\dag}L
-\tilde{L}L^{\dag}]\tilde{\xi}_L^3\xi_R^3\},\label{c2-3}\\
\hat H_{\text{int}}^3 
&=& \frac{\lambda}{8}(U_+^2+U_-^2)
\{(\psi^{\dag}\psi-\psi\psi^{\dag})
[R^{\dag}R-R R^{\dag}+L^{\dag}L-L L^{\dag}]
+(\psi\rightarrow \tilde{\psi})
[R\rightarrow \tilde{R}+L\rightarrow \tilde{L}]\}\nonumber\\
&&+\frac{\lambda}{2}U_+ U_-\{\psi^{\dag}\psi
(R^{\dag}R+L^{\dag}L)+\psi\psi^{\dag}(R R^{\dag}+L L^{\dag})
+(R,L,\psi\rightarrow \tilde{R},\tilde{L},\tilde{\psi})\}\nonumber\\
&&-\frac{J_{\perp}a_0}{4}(U_+^2+U_-^2)
\{(\psi^{\dag}\psi-\psi \psi^{\dag})
[\tilde{R}^{\dag}\tilde{R}-\tilde{R}\tilde{R}^{\dag}
+\tilde{L}^{\dag}\tilde{L}-\tilde{L}\tilde{L}^{\dag}]
+(\tilde{R},\tilde{L},\psi\rightarrow R,L,\tilde{\psi})\}\nonumber\\
&&+\frac{J_{\perp}a_0}{2}(U_+-U_-)^2
\{\psi^{\dag}\tilde{\psi}^{\dag}(R^{\dag}\tilde{L}^{\dag}+L^{\dag}\tilde{R}^{\dag})
+\psi^{\dag}\tilde{\psi}(R^{\dag}\tilde{R}+L^{\dag}\tilde{L})
+{\rm h.c.}\},\label{c2-4}\\
\hat H_{\text{int}}^4
&=& i\frac{\lambda}{2}\{(\psi^{\dag}\psi-\psi\psi^{\dag})\xi_L^3\xi_R^3
+(\psi\rightarrow\tilde{\psi})\tilde{\xi}_L^3\tilde{\xi}_R^3\}\nonumber\\
&&+\frac{J_{\perp}a_0}{2}(\psi^{\dag}\tilde{\psi}-\tilde{\psi}^{\dag}\psi)
(\xi_R^3\tilde{\xi}_R^3 +\xi_L^3\tilde{\xi}_L^3)
-i \frac{J_{\perp}a_0}{2}(\psi^{\dag}\tilde{\psi}+\tilde{\psi}^{\dag}\psi)
(\xi_L^3\tilde{\xi}_R^3+\tilde{\xi}_L^3\xi_R^3),\label{c2-6}\\
\hat H_{\text{int}}^5
&=&-\frac{\lambda}{16}\{(\psi^{\dag}\psi-\psi^{\dag}\psi)^2
+(\psi\rightarrow\tilde{\psi})\}+\frac{J_{\perp}a_0}{4}
(\psi^{\dag}\psi-\psi\psi^{\dag})(\psi\rightarrow\tilde{\psi}),\label{c2-7}\\
\hat H_{\text{stag}}&=&
C_1^2 J_{\perp} a_0\,\,\kappa\tilde{\kappa}\sigma_a\mu_{a+1}\mu_{a+2}
\tilde{\sigma}_a\tilde{\mu}_{a+1}\tilde{\mu}_{a+2}.\label{c2-8}
\end{eqnarray} 
\end{subequations}
\end{widetext}
We define a new velocity $v'=(H^2-m^2)^{1/2}v/H$. In the interaction
terms $\hat{\cal H}_{\rm int}^{1-5}$, Dirac fermions $L$ and $R$ stand
for $L(-x)$ and $R(-x)$, respectively.
From the field-theoretical point of view, biquadratic terms such as 
$(L^{\dag}L-L L^{\dag})^2$ should be interpreted as a product of two
point-splitting terms. For example, $(L^{\dag}L-L L^{\dag})^2$ means 
$\lim_{\delta\rightarrow 0}(L^{\dag}L-L L^{\dag})(-x)
\times(L^{\dag}L-L L^{\dag})(-x+\delta)$.

\section{\label{app5}Abelian Bosonization in the Spin-1 Ladder}
The low-energy action for the spin-$1$ AF ladder is given in Eq.~(\ref{eq34}).
If we make the boson theory with a scalar field $\phi_a$ from two
Ising systems $\sigma_a$ and $\tilde{\sigma}_a$ ($a=1,2,3$), 
we can obtain a bosonized Hamiltonian for the action~(\ref{eq34})
through several relations in Appendices~\ref{app1} and ~\ref{app2}. 
It has been already given in Ref.~\onlinecite{Allen}. The result is 
\begin{eqnarray}
\label{e1}
\hat {\cal H}_{\text{lad}}&=&\sum_{a=1}^3\int dx 
\Big\{\frac{v}{2}[\Pi_a^2+(\partial_x \phi_a)^2]
-\frac{m}{\pi\alpha}\cos(\sqrt{4\pi}\phi_a)\nonumber\\
&&+B_1\big[\cos(\sqrt{4\pi}\phi_a)\cos(\sqrt{4\pi}\phi_{a+1})\nonumber\\
&&+\cos(\sqrt{4\pi}\theta_a)\cos(\sqrt{4\pi}\theta_{a+1})\big]\nonumber\\
&&+B_2\big[\Pi_a\Pi_{a+1}+\partial_x \phi_a \partial_x \phi_{a+1}\big]\nonumber\\
&&+B_3\big[\sin(\sqrt{4\pi}\phi_a)\sin(\sqrt{4\pi}\phi_{a+1})\nonumber\\
&&+\sin(\sqrt{4\pi}\theta_a)\sin(\sqrt{4\pi}\theta_{a+1})\big]\\
&&+B_4\,\,\sin(\sqrt{\pi}\phi_a)\cos(\sqrt{\pi}\phi_{a+1})
\cos(\sqrt{\pi}\phi_{a+2})\Big\},\nonumber
\end{eqnarray}
where $B_1\propto\frac{2\lambda}{(2\pi\alpha)^2}$, 
$B_{2,3}\propto\frac{J_\perp a_0}{2\pi}$ and 
$B_4\sim C_1^2J_\perp a_0$. We can not consider the Klein factors
correctly, because the formulas (\ref{b9}) are not perfect.  
However, here let us assume that it is allowed. Actually, the same
Hamiltonian in Ref~\onlinecite{Allen} (though its notation is different from
ours) surely seems to work well. 
  
Like Ref.~\onlinecite{Allen}, we take the following mean-field prescription 
for $\hat {\cal H}_{\text{lad}}$. (\rnum{1}) We neglect $B_{1,2,3}$
terms, and leave only the free part (fermion kinetic and mass terms) and  
the most relevant interaction $B_4$ term. 
(\rnum{2}) For the weak AF rung-coupling case, 
the mass potential $-\cos(\sqrt{4\pi}\phi_a)$ will be
still dominant, and therefore the field $\phi_a$ is locked near the
point $\phi_a=0$. This argument allows the following approximation: 
$\cos(\sqrt{\pi}\phi_a) \to B$ (a constant).    
Through (\rnum{1}) and (\rnum{2}), 
the Hamiltonian $\hat {\cal H}_{\text{lad}}$ is reduced to 
\begin{eqnarray}
\label{e2}
\hat {\cal H}_{\text{lad}}^{\text{MF}}&=&\sum_{a=1}^3\int dx 
\Big\{\frac{v}{2}[\Pi_a^2+(\partial_x \phi_a)^2]\\
&&-\frac{m}{\pi\alpha}\cos(\sqrt{4\pi}\phi_a)
+B^2B_4\sin(\sqrt{\pi}\phi_a)\Big\}.\nonumber
\end{eqnarray}
This is three copies of a double-sine-Gordon model.~\cite{Del-Mus}
For details of this mean-field argument, see Ref.~\onlinecite{Allen}.

Next, we consider the case that three boson
fields $\phi_a$, $\Phi$ and $\tilde{\Phi}$ are composed of 
$(\sigma_a,\tilde{\sigma_a})$, $(\sigma_{a+1},\sigma_{a+2})$ and  
$(\tilde{\sigma}_{a+1},\tilde{\sigma}_{a+2})$, respectively.
Like Eq.~(\ref{e1}), the fermion free part in the total Hamiltonian 
is mapped to 
\begin{eqnarray}
\label{e3}
\hat {\cal H}_{\text{free}}&=&\int dx 
\frac{v}{2}\Big[(\Pi_a^2+(\partial_x \phi_a)^2)
+(\Pi_{\Phi}^2+(\partial_x \Phi)^2)\nonumber\\
&&+(\tilde{\Pi}_{\tilde{\Phi}}^2+(\partial_x \tilde{\Phi})^2)\Big]
-\frac{m}{\pi\alpha}\Big[\cos(\sqrt{4\pi}\phi_a)\nonumber\\
&&+\cos(\sqrt{4\pi}\Phi)+\cos(\sqrt{4\pi}\tilde{\Phi})\Big],
\end{eqnarray}
where $\Pi_\Phi$ and $\tilde{\Pi}_{\tilde{\Phi}}$ are canonical conjugated
momenta of $\Phi$ and $\tilde{\Phi}$, respectively. The most relevant
rung-coupling term $J_\perp a_0 C_1^2 \kappa\tilde{\kappa}\sigma_a\mu_{a+1}\mu_{a+2}
\tilde{\sigma}_a\tilde{\mu}_{a+1}\tilde{\mu}_{a+2}$ is mapped as follows:
\begin{eqnarray}
\label{e4}
&\sim& J_\perp\big[\sin(\sqrt{\pi}\phi_a)
\cos(\sqrt{\pi}\Phi)\cos(\sqrt{\pi}\tilde{\Phi})\nonumber\\
&&+\cos(\sqrt{\pi}\phi_a)
\big\{\cos(\sqrt{\pi}\Theta)\cos(\sqrt{\pi}\tilde{\Theta}) \nonumber\\
&&+\sin(\sqrt{\pi}\Theta)\sin(\sqrt{\pi}\tilde{\Theta}) \big\}\big]\nonumber\\
&\rightarrow& J_\perp B \cos\big[\sqrt{\pi}(\Theta-\tilde{\Theta})\big].
\end{eqnarray}
On the right side of the arrow $\rightarrow$, we performed the same 
mean-field approximation used in Eq.~(\ref{e2}): 
$\cos(\sqrt{\pi}\phi_a)\rightarrow B$ and
$\sin(\sqrt{\pi}\phi_a)\rightarrow 0$.

\bibliography{apssamp}

\begin{thebibliography}{l}

\bibitem{ladder1}E. Daggoto and T. M. Rice, Science {\bf 271}, 618
	(1996); T. M. Rice, Z. Phys. B {\bf 103}, 165 (1997).

\bibitem{SrCuO}M. Azuma, Z. Hiroi, M. Takano, K. Ishida and Y. Kitaoka,
	Phys. Rev. Lett. {\bf 73}, 3463 (1994); K. Kojima, A. Keren,
	G. M. Luke, B. Nachumi, W. D. Wu, Y. J. Uemura, M. Azuma and
	M. Takano, {\em ibid}. {\bf 74}, 2812 (1995).
\bibitem{Cu2--(NO3)2}Z. Honda, Y. Nonomura and K. Katsumura,
	J. Phys. Soc. Jpn. {\bf 66}, 3689 (1997). 
\bibitem{Cu2--Cl4}
	G. Chabuoussant, M. -H. Julien, Y. Fagot-Revurat, M. Hanson, 
	L. P. L\'evy, C. Berthier, M. Hovati\'c and O. Piovesana,
	Eur. Phys. J. B {\bf 6}, 167 (1998).
\bibitem{DT-TTF}D. Ar\v con, A. Lappas, S. Margadonna, K. Prassides,
	E. Ribera, J. Veciana, C. Rovira, R. T. Henriques and
	M. Almeida, Phys. Rev. B {\bf 60}, 4191 (1999).
\bibitem{BIP-TENO}K. Katoh, Y. Hosokoshi, K. Inoue and T. Goto,
	J. Phys. Soc. Jpn, {\bf 69}, 1008 (2000); T. Goto,
	M. I. Bartrashevich, Y. Hosokoshi, K. Kato and K. Inoue, Physica
	B {\bf 294-295}, 43 (2001); T. Sakai, N. Okazaki, K. Okamoto,
	K. Kindo, Y. Narumi, Y. Hosokoshi, K. Kato, K. Inoue and
	T. Goto, {\em ibid}. {\bf 329-333}, 1203 (2003); 
	Phys. Status Solidi B {\bf 236}, 429 (2003).

\bibitem{Rev-Heis}E. Manousakis, Rev. Mod. Phys. {\bf 63}, 1 (1991).

\bibitem{O-A}M. Oshikawa and I. Affleck, Phys. Rev. Lett. {\bf 79}, 2883 (1997).
\bibitem{A-O}I. Affleck and M. Oshikawa, Phys. Rev. B {\bf 60}, 1038 (1999).
\bibitem{Nomu}M. Tsukano and K. Nomura, J. Phys. Soc. Jpn. {\bf 67},
	302 (1998). 
\bibitem{Z-R-M}A. Zheludev, E. Ressouche, S. Maslov, T. Yokoo,
	S. Raymond and J. Akimitsu, Phys. Rev. Lett. {\bf 80}, 3630
	(1998).
\bibitem{Mas-Zhe}S. Maslov and A. Zheludev, Phys. Rev. B {\bf 57}, 68
	(1998); Phys. Rev. Lett. {\bf 80}, 5786 (1998).
\bibitem{Erco}E. Ercolessi, G. Morandi, P. Pieri and M. Roncaglia,
	Phys. Rev. B {\bf 62}, 14860 (2000).
\bibitem{Wang}Y. -J. Wang, F. H. L. Essler, M. Fabrizio and
	A. A. Nersesyan, Phys. Rev. B {\bf 66}, 024412 (2002).
\bibitem{M-O}M. Sato and M. Oshikawa, Phys. Rev. B {\bf 69}, 054406
	(2004).

\bibitem{Sene}D. S\'en\'echal, Phys. Rev. B {\bf 52}, 15319 (1995).
\bibitem{Allen}D. Allen and D. S\'en\'echal, Phys. Rev. B {\bf 61}, 12134 (2000).
\bibitem{Todo}S. Todo, M. Matsumoto, C. Yasuda and H. Takayama,
	Phys. Rev. B {\bf 64}, 224412 (2001). 

\bibitem{NLSM}For example, see F. D. M. Haldane, Phys. Rev. Lett. {\bf
	50}, 1153 (1983), E. Fradkin, {\em Field Theories of Condensed
	Matter Systems} (Addison-wesley, New York, 1991), 
	A. Auerbach, {\em Interacting
	Electrons and Quantum Magnetism} (Springer-Verlag, New York,
	1994) and Ref.~\onlinecite{Rev-Heis}. 

\bibitem{M-comm}M. Matsumoto (private communication, 2004).
\bibitem{Matsu}M. Matsumoto, S. Todo, M. Nakamura, C. Yasuda and
	H. Takayama, Physica B {\bf 329-333}, 1010 (2003). 

\bibitem{note1}In the paper, a RVB state means a linear combination of
	the tensor products, each of which is constructed by
	a dimer covering (the dimer is the singlet of a
	spin-$\frac{1}{2}$ pair). ``Short-range'' implies that all 
	the bonds consist of two nearest neighboring sites.  

\bibitem{Witt}E. Witten, Comm. Math. Phys. {\bf 92}, 455 (1984). 
\bibitem{Aff}I. Affleck, Nucl. Phys. B {\bf 265}, 409 (1986). 
\bibitem{A-H}I. Affleck and F.D.M. Haldane, Phys. Rev. B {\bf 36}, 5291 (1987).
\bibitem{Tsv}A. M. Tsvelik, Phys. Rev. B {\bf 42}, 10499 (1990).




\bibitem{Sel}D. G. Shelton, A. A. Nersesyan and A. M. Tsvelik,
	Phys. Rev. B {\bf 53}, 8521 (1996).
\bibitem{White1}S. R. White, R. M. Noack and D. J. Scalapino,
	Phys. Rev. Lett. {\bf 73}, 886 (1994). 
\bibitem{White2}S. R. White, Phys. Rev. B {\bf 53}, 52 (1996).
\bibitem{Nishi}Y. Nishiyama, N. Hatano and M. Suzuki,
	J. Phys. Soc. Jpn. {\bf 64}, 1967 (1995).
\bibitem{S-D}G. Sierra and M. A. Mart\'in-Delgado, Phys. Rev. B {\bf
	56}, 8774 (1997).
\bibitem{Kim}E. H. Kim, G. F\'ath, J. S\'olyom and D. J. Scalapino,
	Phys. Rev. B {\bf 62}, 14965 (2000).

\bibitem{Cab1}D. C. Cabra, A. Honecker and P. Pujol, Phys. Rev. Lett.
	{\bf 79}, 5126 (1997).
\bibitem{Cab2}D. C. Cabra, A. Honecker and P. Pujol, Phys. Rev. B {\bf
	58}, 6241 (1998). 
\bibitem{Mila}F. Mila, Eur. Phys. J. B {\bf 6}, 201 (1998).
\bibitem{Fu-Zh}A. Furusaki and S. C. Zhang, Phys. Rev. B {\bf 60}, 1175
	(1999).
\bibitem{Hi-Fu}T. Hikihara and A. Furusaki, Phys. Rev. B {\bf 63},
	134438 (2001).

\bibitem{Gogo}A. O. Gogolin, A. A. Nersesyan and A. M. Tsvelik, {\em
	Bosonization and Strongly Correlated Systems} (Cambridge Univ.
	Press, Cambridge, England, 1998).
\bibitem{CFT}P. D. Francesco, P. Mathieu and D. S\'en\'echal, {\em
	Conformal\ Field Theory} (Springer-Verlag, New York, 1997).

\bibitem{P-T}V. L. Pokrovsky and A. L. Talapov, Phys. Rev. Lett. {\bf
	42}, 65 (1979).
\bibitem{Sch}H. J. Schulz, Phys. Rev. B {\bf 22}, 5274 (1980).
\bibitem{Ch-Gi}R. Chitra and T. Giamarchi, Phys. Rev. B {\bf 55}, 5816
	(1997).
\bibitem{B-M}S. M. Bhattacharjee and S. Mukherji, J. Phys. A {\bf 31}, 
	L695 (1998).

\bibitem{Hal}F. D. M. Haldane, Phys. Rev. Lett. {\bf 47}, 1840 (1981);
	J. Phys. C {\bf 14}, 2585 (1981). 
\bibitem{Lec-Aff}I. Affleck, in {\em Champs, Cordes et Phenomenes
	Critiques; Fields, Strings and Critical Phenomena}, edited by
	E. Br\'ezin and J. Zinn-Justin (Elsevier, Amsterdam, 1989),
	p. 564.
\bibitem{Boso}{\em Bosonization}, edited by M. Stone (World
	Scientific, Singapore, 1994).
\bibitem{D-S}J. v. Delft and H. Schoeller, Ann. Phys. (Leipzig) {\bf 4}, 225
	(1998).
\bibitem{Lec-Sene}D. S\'en\'echal, cond-mat/9908262.
\bibitem{LowD-QFT}H. J. Schulz, G. Cuniberti and P. Pieri, in {\em Field
	Theories for Low-Dimensional Condensed Matter Systems}, edited
	by G. Morandi {\em et al}. (Springer-Verlag, New York, 2000). 
\bibitem{2D-QFT}E. Abdalla, M. Cristina, B. Abdalla and K. D. Rothe,
	{\em Non-perturbative Methods in 2 Dimensional Quantum Field
	Theory}, 2nd ed. (World Scientific, Singapore, 2001).

\bibitem{Ide}K. Ide, M. Nakamura and M. Sato, 58th Annual Meeting of
	the Physical Society of Japan, 2003.
\bibitem{level}K. Nomura and A. Kitazawa, cond-mat/0201072.
\bibitem{Naka}M. Nakamura and J. Voit, Phys. Rev. B {\bf 65}, 153110
	(2002); M. Nakamura and S. Todo, Phys. Rev. Lett. {\bf 89},
	077204 (2002).





\bibitem{Oka}K. Okamoto, N. Okazaki and T. Sakai, J. Phys. Soc. Jpn. 
	{\bf 70}, 636 (2001).
\bibitem{note2}For example, see Refs.~\onlinecite{Cab2}, \onlinecite{Alca} and
	\onlinecite{H-F}.
\bibitem{Sakai}T. Sakai, K. Okamoto, K. Okunishi and M. Sato, 
	J. Phys: Condens. Matter {\bf 16}, S785 (2004). 
	This paper has a mistake. In Eq.~(7), $J_1^{\text(cr)}=0.491$ 
	should be replaced with $J_1^{\text(cr)}=0.695$.
\bibitem{Oka2}K. Okamoto, K. Okunishi and T. Sakai, in preparation.
\bibitem{note4}In this paper, we say the phase transition is of a BKT type,
	only when it can be regarded as the transition of the $(1+1)$D 
	sine-Gordon model that the marginal relevant sin term becomes 
	marginal irrelevant or vice-versa. Since in a particle language,
	the chemical potential changes when the magnetization (or
	uniform field) is changed, the effective theory near the
	transition occurring with increasing (decreasing) the
	magnetization will become a misfit sine-Gordon model rather
	than a simple sine-Gordon model.  
\bibitem{OYA}M. Oshikawa, M. Yamanaka and I. Affleck, Phys. Rev. Lett. 
	{\bf 78}, 1984 (1997).
\bibitem{Alca}F. C. Alcaraz and A. L. Malvezzi, J. Phys. A {\bf 28},
	1521 (1995).
\bibitem{K-B-I}V. E. Korepin, N. M. Bogoliubov and A. G. Izergin, {\em
	Quantum Inverse Scattering Method and Correlation Functions}
	(Cambridge Univ. Press, Cambridge, England, 1993).
\bibitem{M-Taka}M. Takahashi, {\em Thermodynamics of One-Dimensional
	Solvable Models} (Cambridge Univ. Press, Cambridge, England,
	1999).
\bibitem{E-A-T}S. Eggert, I. Affleck and M. Takahashi,
	Phys. Rev. Lett. {\bf 73}, 332 (1994).
\bibitem{note00}This is reliable near the level-crossing line
	considered. 
\bibitem{note5}Of course, it is possible to obtain the result
	(\ref{eq18}) from the formula (\ref{eq16}). 
\bibitem{note6}The bosonization can lead to the same gap behavior, too,
	although the spin-wave analysis is simpler and exact. 
	
\bibitem{note7}As in Ref.~\onlinecite{note00}, 
	each effective model is valid around the
	corresponding level-crossing line of the two-spin problem. 
	This gap behavior thus is reliable only when $H$ is close to 
	$H^{\text{cr}}$ enough.

\bibitem{L-Z}S. Lukyanov and A. Zamolodchikov, Nucl. Phys. B {\bf
	493}, 571 (1997).
\bibitem{Luk}S. Lukyanov, Mod. Phys. Lett. A {\bf 12}, 2543 (1997).
\bibitem{H-F}T. Hikihara and A. Furusaki, Phys. Rev. B {\bf 69}, 064427
	(2004); cond-mat/0310391.

\bibitem{Sier}G. Sierra, J. Phys. A {\bf 29}, 3299 (1996);
	cond-mat/9610057.

\bibitem{schul}Besides these two, Schulz's method based on the Abelian
	bosonization [see H. J. Schulz, Phys. Rev. B {\bf 34}, 6372
	(1986)] are often used in high spin systems. However, its
	starting point, the spin-$\frac{1}{2}$ XY chain, is far from the
	SU(2) Heisenberg point considered now.
\bibitem{I-O}T. Inami and S. Odake, Phys. Rev. Lett, {\bf 70}, 2016 (1993).
\bibitem{Kita}A. Kitazawa and K. Nomura, Phys. Rev. B {\bf 59}, 11358 (1998).
\bibitem{Tak}L. Takhtajan, Phys. Lett. {\bf A87}, 479 (1982).
\bibitem{Bab}J. Babujian, Phys. Lett. {\bf A90}, 479 (1982);
	Nucl. Phys. B {\bf 215}, 317 (1983).
\bibitem{note9}In Refs.~\onlinecite{S-J-G} and \onlinecite{F-Su},
	the fine reviews are found.
\bibitem{Sol}J. S\'olyom, Phys. Rev. B {\bf 36}, 8642 (1987).
\bibitem{Ken}T. Kennedy, J. Phys.: Cond. Matter {\bf 2}, 5737 (1990).
\bibitem{F-S}G. F\'ath and J. S\'olyom, Phys. Rev. B {\bf 44}, 11836
	(1991); J. Phys.: Condens. Matter {\bf 5}, 8983 (1993); 
	Phys. Rev. B {\bf 47}, 872 (1993); B {\bf 51}, 3620 (1995).
\bibitem{B-X-G}R. J. Bursill, T. Xiang and G. A. Gehring, J. Phys. A 
	{\bf 28}, 2109 (1995).
\bibitem{S-J-G}U. Schollw\"ock, Th. Jolicoeur and T. Garel, Phys. Rev. B
	{\bf 53}, 3304 (1996).
\bibitem{SMKYM}A. Schmitt, K-H. M\"utter, M. Karbach, Y. Yu and
	G. M\"uller, Phys. Rev. B {\bf 58}, 5498 (1998).
\bibitem{G-J-S}O. Golinelli, Th. Jolicoeur and E. S. S\o rensen,
	Eur. Phys. J. B {\bf 11}, 199 (1999).
\bibitem{F-Su}G. F\'ath and A. S\"uto, Phys. Rev. B {\bf 62}, 3778
	(2000).
\bibitem{Nomu2}K. Nomura, J. Phys. Soc. Jpn. {\bf 72}, 476 (2003).
\bibitem{L-S-T}A. L\"auchli, G. Schmid and S. Trebst, cond-mat/0311082.
\bibitem{Lai-Sut}C. K. Lai, J. Math. Phys. {\bf 15}, 1675 (1974);
	B. Sutherland, Phys. Rev. B {\bf 12}, 3795 (1975). 

\bibitem{A-M}F. C. Alcaraz and M. J. Martins, J. Phys. A: Math. Gen. {\bf
	21} L381 (1988); {\em ibid}, 4397 (1988).
\bibitem{AGSZ}I. Affleck, D. Geppner, H. J. Shultz and T. Zimann,
	J. Phys. A {\bf 2}, 511 (1989).
\bibitem{Avd}L. V. Avdeev, J. Phys. A {\bf 23}, L485 (1990).
\bibitem{I-M}C. Itoi and H. Mukaida, J. Phys. A {\bf 27},
	4695 (1994).

\bibitem{I-K}C. Itoi and M.H. Kato, Phys. Rev. B {\bf 55}, 8295 (1997).
\bibitem{AKLT}I. Affleck, T. Kennedy, E.H. Lieb and H. Tasaki,
	Phys. Rev. Lett. {\bf 59}, 799 (1987); Commun. Math. Phys. {\bf
	115}, 477 (1988). 

\bibitem{Aff3}I. Affleck, Phys. Rev. Lett. {\bf 56}, 2763 (1986).
\bibitem{Zam-Fat}A. B. Zamolodchikov and V. A. Fateev, Yad. Fiz. {\bf
	43}, 1031 (1986) [Sov. J. Nucl. Phys. {\bf 43}, 657 (1986)].
\bibitem{C-P-R}D. C. Cabra, P. Pujol and C. vonReichenbach, Phys. Rev. B
	{\bf 58}, 65 (1998).

\bibitem{note11}The most relevant term $g(x)$ can not appear in the
	action because it is forbidden by the one-site translational
	symmetry. Furthermore, for example, the product $g(x)g(x+a_0)$ is
	relevant, too. However, it is translated into fermionic operators 
	through the operator product expansions of the $c=1/2$ CFT 
	[see Eq.~(\ref{eq29}) and Appendix~\ref{app2-2}].
\bibitem{note11'}Generally, the effective theory, derived from a lattice
	model as a continuum limit, has higher symmetries than the
	lattice model. Hence, sometimes symmetries of the lattice do not 
	uniquely associate with one of the effective theory. In our
	case, there may exist more suitable correspondences than those
	we adopt here.
\bibitem{Taka}M. Takahashi, Phys. Rev. Lett. {\bf 62}, 2313 (1989).
\bibitem{S-T}T. Sakai and M. Takahashi, Phys. Rev. B {\bf 42}, R1090
	(1990); J. Phys. Soc. Jpn. {\bf 60}, 760 (1991); {\bf 60}, 3615 
	(1991); Phys. Rev. B {\bf 43}, 13383 (1991).  
\bibitem{Aff2}I. Affleck, Phys. Rev. B {\bf 41}, 6697 (1990); {\bf 43}, 
	3215 (1991).
\bibitem{S-A}E. S. S\o rensen and I. Affleck, Phys. Rev. Lett. {\bf 71},
	1633 (1993).
\bibitem{K-F}R. M. Konik and P. Fendley, Phys. Rev. B {\bf 66}, 144416
	(2002).
\bibitem{Fath}G. F\'ath, Phys. Rev. B {\bf 68}, 134445-1 (2003).
\bibitem{Kato}S. Todo and K. Kato, Phys. Rev. Lett. {\bf 87}, 047203
	(2001).
\bibitem{Wu-Mc}T. T. Wu, B. McCoy, C. A. Tracy and E. Barouch,
	Phys. Rev. B {\bf 13}, 316 (1976). 
\bibitem{Ner}A. A. Nersesyan, A. O. Gogolin and F. H. L. E\ss ler,
	Phys. Rev. Lett. {\bf 81}, 910 (1998).

\bibitem{noteRG}Note that the RG Equations do not contain
	renormalizations of fields and velocities. 
\bibitem{Cardy}J. L. Cardy, {\em Scaling and
	Renormalization in Statistical Physics} (Cambridge Univ. Press, 
	Cambridge, England, 1996).
\bibitem{B-F}L. Balents and M. P. A. Fisher, Phys. Rev. B {\bf 53},
	12133 (1996).
\bibitem{Allen2}D. Allen and D. S\'en\'echal, Phys. Rev. B {\bf 55}, 299
	(1997).
\bibitem{Naka-Topo}M. Nakamura, Physica B {\bf 329-333} 1000 (2003).
\bibitem{Ko-Ta}M. Kohmoto and H. Tasaki, Phys. Rev. B {\bf 46}, 3486
	(1992).
\bibitem{c-theo}A. B. Zamolodchikov, Pis'ma Zh. \'Eksp. Teor. Fiz. {\bf
	43}, 565 (1986) [JETP Lett. {\bf 43}, 730 (1986)];
	Yad. Fiz. {\bf 46}, 1819 (1987) [Sov. J. Nucl. Phys. {\bf 46},
	1090 (1987)].


\bibitem{Fab}M. Fabrizio, A. O. Gogolin and A. A. Nersesyan,
	Nucl. Phys. B {\bf 580}, [FS], 647 (2000).
\bibitem{Cha}B. K. Chakrabarti, A. Dutta and P. Sen, {\em Quantum Ising
	Phases and Transitions in Transverse Ising Models},
	(Springer-Verlag, Berlin, 1996).
\bibitem{Sac}S. Sachdev, {\em Quantum Phase Transitions}, (Cambridge
	Univ. Press, Cambridge, 2001).
\bibitem{Kog}J. B. Kogut, Rev. Mod. Phys. {\bf 51}, 659 (1979).
\bibitem{Boya}D. Boyanovsky, Phys. Rev. B {\bf 39}, 6744 (1989).
\bibitem{I-D}C. Itzykson and J-M. Druffe, {\em Statistical Field
	Theory}, Vols.1 and 2, (Cambridge Univ. Press, Cambridge, 1991). 
\bibitem{F-S-Z}P. D. Francesco, H. Saleur and J. B. Zuber,
	Nucl. Phys. B {\bf 290}, 527 (1987).
\bibitem{Del-Mus}G. Delfino and G. Mussardo, Nucl. Phys. B {\bf 516},
	675 (1998).


\end{thebibliography}

\end{document}